\documentclass[rmp,twocolumn,preprintnumbers,amsmath,amssymb,superscriptaddress]{revtex4}

\usepackage{graphicx}
\usepackage{dcolumn}
\usepackage{bm}
\usepackage{url,hyperref} 

\begin{document}

\title{Fano resonances in nanoscale structures}%
\author{Andrey E. Miroshnichenko}
\email{aem124@rsphysse.anu.edu.au}
\affiliation{Nonlinear Physics Centre and Centre for Ultra-high bandwidth Devices for Optical Systems (CUDOS),\\ Research School of Physics and Engineering, Australian National University, Canberra ACT 0200, Australia}

\author{Sergej Flach}
\affiliation{Max-Planck-Institut f\"ur Physik komplexer Systeme, N\"othnitzer Str. 38, D-01187 Dresden, Germany}

\author{Yuri S. Kivshar}
\affiliation{Nonlinear Physics Centre and Centre for Ultra-high bandwidth Devices for Optical Systems (CUDOS),\\ Research School of Physics and Engineering, Australian National University, Canberra ACT 0200, Australia}

\begin{abstract}
Modern nanotechnology allows to scale down various important devices (sensors, chips, fibres, etc), and, thus, 
opens up new horizons for their applications. The efficiency of most of them is based on fundamental 
physical phenomena, such as transport of wave excitations and resonances. 
Short propagation distances make phase coherent processes of waves important.
Often the scattering of waves involves propagation along different paths, 
and, as a consequence, results in interference phenomena, where constructive interference corresponds to resonant enhancement 
and destructive interference to resonant suppression of the transmission. Recently, a variety of experimental and theoretical 
work has revealed such patterns in different physical settings. The purpose of this Review is to relate  
resonant scattering to Fano resonances, known from atomic physics. One of the main features of the Fano 
resonance is its asymmetric line profile. The asymmetry originates from a close coexistence of resonant  transmission and resonant 
reflection, and can be reduced to the interaction of a discrete 
(localized) state with a continuum of propagation modes. 
We will introduce the basic concepts of Fano resonances, explain their geometrical and/or dynamical origin, and review
theoretical and experimental studies for light propagation in photonic devices, charge transport through quantum dots, 
plasmon scattering in Josephson junction networks, and matter wave scattering in ultracold atom systems, among others.
\end{abstract}

\maketitle
\tableofcontents

\section{Historical remarks}

One of the important diagnostic tools in physics is scattering of radiation (waves) by matter. 
It allows to investigate properties of matter and to control the radiation. 
For example, Rydberg spectral lines (1888) of the hydrogen atom allowed Niels 
Bohr to deduce his model of an atom (1913), which layed the basis of quantum mechanics. Later, ~\textcite{bh:zphn:35} 
observed that some of the Rydberg spectral atomic lines exhibit sharp asymmetric profiles in the absorption. It was 
Ugo ~\textcite{uf:nc1:35} who suggested the first theoretical explanation of this effect and suggested a formula 
(also known as the Beutler-Fano formula) which predicts the shape of spectral lines based on a superposition principle from 
quantum mechanics. The complexity of the physical phenomena was encapsulated in a few key parameters, which made 
this formula a workhorse in many fields of physics, including nuclear, atomic, molecular, and condensed matter physics. 
According to Fano: {\em "the Beutler spectra showed unusual intensity profiles which struck me as reflecting interference 
between alternative mechanisms of excitation"}~\cite{uf:cc:77}. The interpretation provided by Fano of these "strange 
looking shapes" of spectral absorption lines is based of the interaction of a discrete excited state of an atom with a continuum 
sharing the same energy level, which results in interference phenomena. The first paper with the derivation of the line-shape 
formula ~\cite{uf:nc1:35}, was published in 1935, when Ugo Fano was a young postdoctoral fellow in the group of Enrico Fermi. 
Fano has acknowledged the influence of his teacher on the derivation of this key result.
The second much more elaborated paper~\cite{uf:PR:61} became one of the most important publications in the physics of the XX century, 
rated between the first three most relevant works published in The Physical Review~\cite{sr:arxiv:04}, with over 
5300 citations by now (October 2008).
{\em "The paper appears to owe its success to accidental circumstances, such as the timing of its publication and some successful 
features of its formulation. The timing coincided with a rapid expansion of atomic and condensed matter spectroscopy, both optical 
and collisional. The formulation drew attention to the generality of the ingredients of the phenomena under consideration. In fact, 
however, the paper was a rehash of work done 25 years earlier \ldots"}~\cite{uf:cc:77,avab:arxiv:08}.
In his pioneering papers, Ugo Fano introduced an important new ingredient of matter-radiation interaction in atomic physics, making him 
a key player in XX century physics. This was  also acknowledged by the 
Fermi Award in 1995 for "his seemingly formal use of fundamental theory" 
leading to  "the underpinning of a vast variety of practical results which developed naturally from this understanding".

\begin{figure}
\includegraphics[width=0.8\columnwidth]{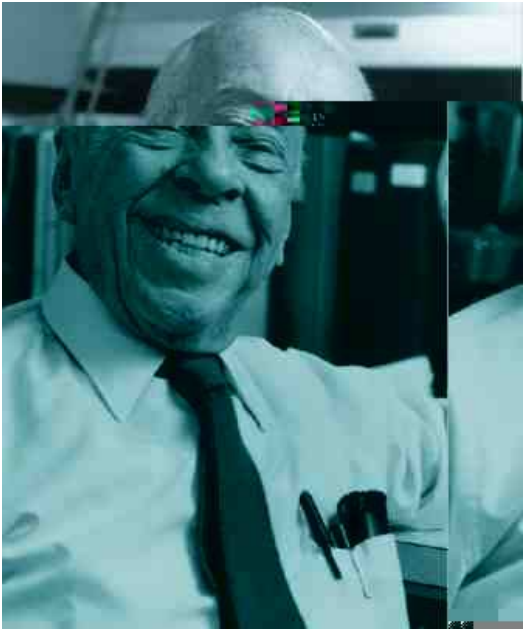}
\caption{\label{fig:UFano}
Ugo Fano (1912-2001) - "outstanding interpreter of how radiation interacts with atoms and cells" ~\cite{cwc:Nature:01}, 
and much more (this Review).
}
\end{figure}

Remarkably, the first observation of the asymmetric line-shapes can be traced back to the discovery made by Wood in 
1902, namely, the presence of unexpected narrow bright and dark bands in the spectrum of an optical reflection grating 
illuminated by a slowly varying light source~\cite{rww:PRSLA:02}. Wood was astounded to see that under special illumination 
conditions the grating efficiency in a given order dropped from maximum to minimum illumination, within a wavelength range 
not greater than the distance between the sodium lines. These rapid variations of intensities of the various diffracted 
spectral orders in certain narrow frequency bands were termed {\em anomalies}, since the effects could not be explained by 
the conventional grating theory~\cite{rww:PR:35}. The first theoretical treatment of these anomalies is due to Lord~\textcite{lr:PRSA:07}. 
His "dynamical theory of the grating" was based on an expansion of the scattered electromagnetic field in terms of outgoing 
waves only. This theory correctly predicted the wavelength (Rayleigh wavelengths) at which anomalies occurred. However, one 
of the limitations of Rayleigh's approach was that it yields a singularity at the Rayleigh wavelength, and, therefore, 
does not give the shape of the bands associated with the anomaly. Fano tried to overcome this difficulty in a series of 
papers~\cite{uf:PR:36,uf:PR:37,uf:ap:38,uf:JOSA:41} by assuming a grating consisting of lossy dielectric material, 
and suggesting that anomalies could be associated with the excitation of a surface wave along the grating.
The resonant excitation of leaky surface waves near the grating, which occurs when a suitable phase matching between the 
incident plane wave and the guided wave is satisfied, leads to a strong enhancement of the field near the grating 
surface~\cite{ahaao:AO:65,msjpvjmv:PRB:03,fjgda:RMP:07}. As it was pointed out by \textcite{msjpvjmv:PRB:03}, the observed 
asymmetric profiles can be fitted by the Fano formula with very good accuracy.
Thus, the interaction of excited leaky modes with an incoming radiation leads to similar interference phenomena as in 
absorption by Rydberg atoms, where a leaky mode can be associated with a discrete state, and the incoming radiation with a continuum. 
These examples reveal the universality of Fano's approach in describing the origin of asymmetric line-shapes in terms of 
interference phenomena, regardless of the nature of the constituting waves, as well as in 
predicting both the position and the width of the resonance.

Similar asymmetric profiles were observed in various other systems and settings. 
But sometimes it is not obvious to determine the origin of the interference. 
In the present survey paper, we provide a very general explanation of appearance 
of the Fano resonances in various physical systems based on a simple model, which sheds light on the origin of the 
interference phenomena, which is well along the lines of 
Steven Weinberg: "our job in physics is to see things simply, to understand many complicated phenomena in a 
unified way, in terms of a few simple principles." (1979 Nobel Prize Lecture).



\section{The Fano resonance \label{sec:fano_res}}

\subsection{Two oscillators with a driving force}

\begin{figure}
\includegraphics[width=\columnwidth]{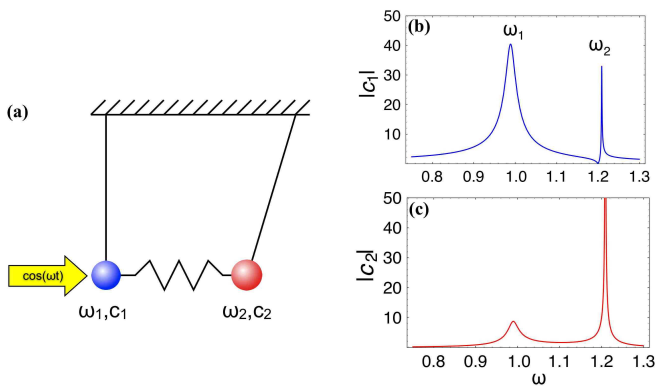}
\caption{\label{fig:fano_res1}
(Color online)
Resonances of parametrically driven coupled oscillators.
(a) Schematic view of two coupled damped oscillators with a driving force applied to one of them; (b) the resonant dependence of the amplitude of the forced oscillator $|c_1|$, and (c) the coupled one $|c_2|$. 
There are two resonances in the system. The forced oscillator exhibits resonances with symmetric and 
asymmetric profiles near the eigenfrequencies $\omega_1=1$ and $\omega_2=1.2$ (b), respectively. The 
second coupled oscillator responds only with symmetric resonant profiles (c). Adapted from ~\textcite{ysjamscsk:PS:06}.
}
\end{figure}

Usually, a resonance is thought to be an enhancement of the response 
of a system to an external excitation at a particular frequency. It is referred to as the resonant frequency, or natural 
frequency of the system. One of the simplest examples is a harmonic oscillator with periodic forcing.
When the frequency of the driving force is close to the eigenfrequency of the oscillator, the amplitude of the latter 
is growing towards 
its maximal value. Often many physical systems may also exhibit the opposite phenomenon, when their response is suppressed if 
some resonance condition is met (which lead even to the term {\sl antiresonance}).
This can be illustrated by using two weakly coupled underdapmed harmonic oscillators, where one of them is driven by a periodic 
force [see Fig.~\ref{fig:fano_res1}(a)]. In such a system, in general, there are two resonances located close to eigenfrequencies 
$\omega_1$ and $\omega_2$ of the oscillators~\cite{ysjamscsk:PS:06}. One of the resonances of the forced oscillator 
demonstrates the standard enhancement of the amplitude near its eigenfrequency $\omega_1$, while the other resonance exhibits an
unusual sharp suppression of the amplitude near the eigenfrequency of the second oscillator $\omega_2$ [see Fig.~\ref{fig:fano_res1}(b,c)]. 
The first resonance is characterized by a symmetric profile, described by Lorentzian function, and known as a Breit-Wigner 
resonance~\cite{gbew:PR:36}. The second resonance is characterized by an asymmetric profile.

\subsection{Light and atoms}


\begin{figure}
\includegraphics[width=\columnwidth]{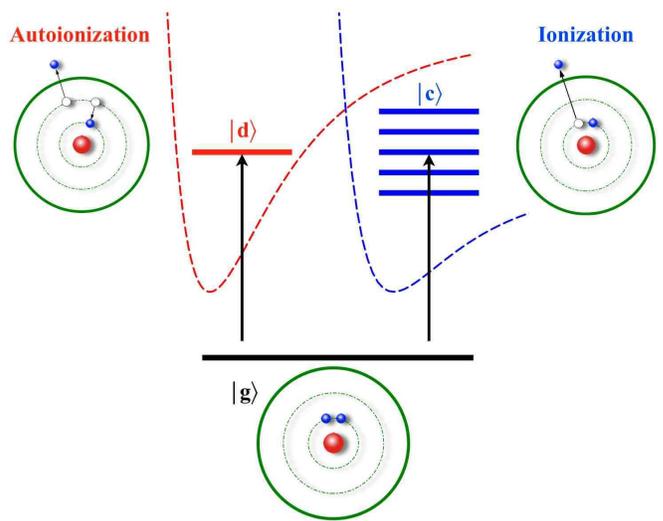}
\caption{\label{fig:fano_res2}
(Color online) 
Fano resonance as a quantum interference of two processes - direct ionization of a deep inner-shell electron and 
autoionization of two excited electrons followed by the Auger effect.
This process can be represented as a transition from the ground state of an atom $|g\rangle$ either to a discrete 
excited autoionizing state $|d\rangle$ or to a continuum $|c\rangle$.
Dashed lines indicate double excitations and ionization potentials.
}
\end{figure}

The second resonance was described for the first 
time by~\textcite{uf:nc1:35,uf:PR:61}, when being 
attracted by unusual sharp peaks in the absorption spectra of noble gases observed by~\textcite{bh:zphn:35}. 
The nature of the asymmetry was established with the theory of configuration by ~\textcite{uf:PR:61}. 
The photoionization of an atom can go along various ways. The first, straightforward one, is the excitation 
of the inner-shell electron above the ionization threshold $A+\hbar\nu\rightarrow A^++e$. Another possibility is to excite the atom 
into some quasi-discrete level, which can spontaneously ionize by ejecting an electron into the continuum 
$A+\hbar\nu\rightarrow A^*\rightarrow A^++e$. Such levels were named autoionizing ones after \textcite{ags:rpp:38}. 
In other words, the autoionized state is a bound state of an atom with the energy above the first ionizing threshold. 
Autoionization is one of the most fundamental electron-electron correlation phenomena, and it is forbidden in the noninteracting 
particle approximation ~\cite{connerade}. 
One of the possible autoionized states is the excitation of two electrons by one photon, 
when the excitation energies of each electron are of the 
same order of magnitude, and the total excitation energy exceeds the atom ionization threshold. The interaction between 
electrons leads to the decay of this state when one electron transfers into a lower state, and the second electron is 
ejected into the continuum, using the energy of the relaxed electron. In spectroscopy this process is known as the Auger 
effect~\cite{pa:cr:25,pa:jp:25,pa:adp:26}. Different types of other autoionizing states are described in~\textcite{smirnov}. 
In general, autoionization can be considered as a mechanism which couples bound states of one channel with continuum 
states of another. Due to the superposition principle of quantum mechanics, whenever two states are coupled by different paths, 
interference may occur.

\begin{figure}
\includegraphics[width=\columnwidth]{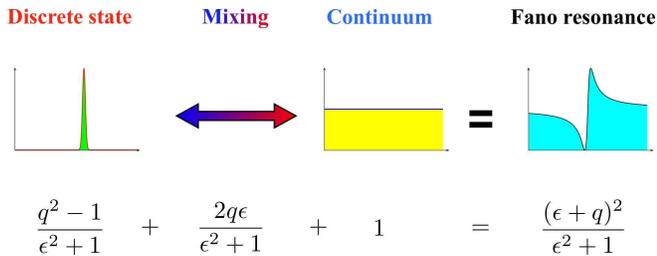}
\caption{\label{fig:fano_res3}
(Color online) 
Illustration of the Fano formula (\ref{eq:Fano}) as a superposition of the Lorentzian lineshape of the discrete 
level with a flat continuous background.
}
\end{figure}

\begin{figure}
\includegraphics[width=\columnwidth]{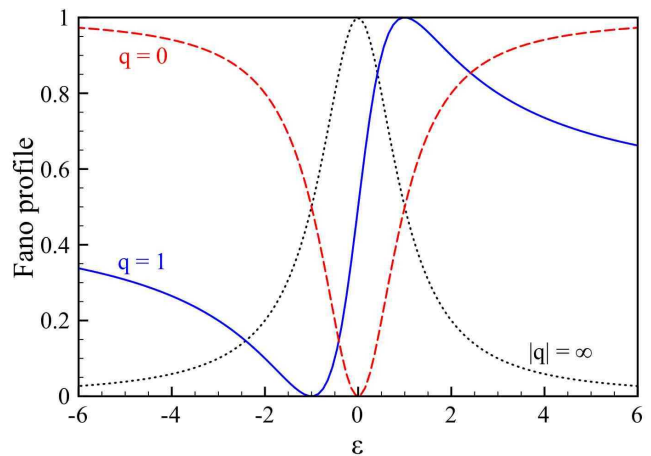}
\caption{\label{fig:Fano_profiles}
(Color online) Normalized Fano profiles (\ref{eq:Fano}) with the prefactor $1/(1+q^2)$ (\ref{eq:extremuma}) for various 
values of the asymmetry parameter $q$.
}
\end{figure}

Fano used a perturbation approach to explain the appearance of asymmetric resonances. 
He considered a so-called prediagonalised  state by putting the coupling between a discrete bound state, which is degenerate 
in energy with a continuum of states, to zero. Such a prediagonalized state may or may not have a 
clear physical analogy, but serves in any case as a convenient
mathematical construction, which allows to solve the problem. 
As a result Fano obtained the formula for the shape of the resonance profile~\cite{uf:nc1:35,uf:PR:61}
of a scattering cross-section
\begin{eqnarray}\label{eq:Fano}
\sigma=\frac{(\epsilon+q)^2}{\epsilon^2+1}
\end{eqnarray}
using a phenomenological shape parameter $q$ and a reduced energy $\epsilon$ defined by $2(E-E_F)/\Gamma$. 
$E_F$ is a  resonant energy, and $\Gamma$ is the width of the auto-ionized state.  
Formula (\ref{eq:Fano}) suggests that there are exactly one maximum and one minimum in the Fano profile
\begin{eqnarray}\label{eq:extremuma}
	\begin{array}{ll}
		\sigma_{\rm min}= 0\;,& {\rm at}\; \epsilon = -q\\
		\sigma_{\rm max}=1+q^2\;,& {\rm at}\; \epsilon = 1/q\;.
	\end{array}
\end{eqnarray}
In his original paper~\textcite{uf:PR:61} has introduced the asymmetry parameter $q$ as a ratio of transition 
probabilities to the mixed state and to the continuum.
In the limit $|q|\rightarrow\infty$ the transition to the continuum is very weak, and the lineshape is entirely 
determined by the transition through the discrete state only with the standard Lorentzian profile of a Breit-Wigner resonance. 
When the asymmetry parameter $q$ is order of unity both the continuum and discrete transition are of the same strength 
resulting is the asymmetric profile (\ref{eq:Fano}), with the maximum value at $E_{\rm max}=E_F+\Gamma/(2q)$ and minimum 
value at $E_{\rm min}=E_F-\Gamma q/2$. 
The case of zero asymmetry parameter $q=0$ is very unique to the Fano resonance and  describes a symmetrical dip, 
sometimes called an anti-resonance (see Fig.~\ref{fig:Fano_profiles}). The main feature of the Fano resonance is the
possibility of destructive interference, leading to asymmetric line 
shapes~\cite{gpralpf:ssc:90,junad:PRB:94,sb:jpamop:00,ab:03,sbbdrhcm:AJP:04,arp:ps:04}.
The actual resonant frequency of the discrete level $E_F$ may 
lie somewhere between the maximum and the minimum 
of the asymmetric profile, and the parameter $q$ defines the relative deviation. In the situation $|q|\rightarrow\infty$ 
the resonant frequency coincides to the maximum of the profile, while in the case $q=0$ the resonant frequency coincides 
to the minimum. Ror $q=1$ it is located exactly at half distance between the minimum and maximum 
(see Fig.~\ref{fig:Fano_profiles}).

Due to recent advances in the generation of ultrashort attosecond pulses ~\textcite{mwjbfkmd:PRL:05}
theoretically investigated the possibility to observe the buildup of 
Fano resonances in time by using attosecond streaking techniquies . Excitation by an ultrashort pump 
pulse opens two interfering paths from the ground state to the continuum, which are then studied by a weak probe pulse. 
After the characteristic time of the autoionizing level, the transient coupling to the resonant state starts to "burn a hole" 
in the energy distribution of the initial Gaussian. This method may become very useful in determining both coherent and incoherent 
pathways to ionization.

The Fano formula (\ref{eq:Fano}) was successfully  used to fit and explain various experimental 
data~\cite{Fano_Festschrift,wm:jesrp:98,sbbdrhcm:AJP:04,jasuf:PRL:63,uf:PR:64,uf:PR:65,ufjwc:PR:65,ufjwc:RMP:68,jkjl:PRL:70,uhjkjl:PRL:70,ufcml:PRL:73,ksdegsobwtard:PRA:73,derdms:PRA:74,adbjpl:JCP:76,sdd:JCP:77,lcdlaf:PRB:77,lacetmjw:PRA:78,sfsljpat:PR:79,sndpl:PRA:79,dfhsm:JCP:79,yy:PRB:81,jgmrhhjg:PRL:84,llrkrlj:PRL:84,dah:PRA:85,ejggrshgg:PRB:85,lnojww:PRB:85,ubtpesbsgew:PRA:86,jasjew:JCP:87,ku:PRA:87,amgb:PRB:89,aneeetbep:PRA:90,mcnsvc:PRL:91,cwpwl:JCP:91,kmkutnki:PRA:92,kswsjrs:PRL:92,dptpmj:JCP:93,isfm:PRA:94,junad:PRB:94,plr:JCP:94b,plr:JCP:94a,plr:JCP:95,awsabszseg:PRB:95,usmamsgdsc:PRB:95,tdomdvjaccgm:PRB:95,kahyms:PRL:96,sbpkmvmsmdsc:PRL:97,gwlzwec:PRA:97,cwl:PRA:98,fpmhswdsbd:PRL:99,lphnjh:JCP:00,rrtmobslsihssmbssanb:PRA:01,sg:PRB:02,vkcdppfms:PRA:02,egbucf:SPIE:03,uetfgrmk:PRL:03,vammap:JP:04,mwjbfkmd:PRL:05,pkvbjh:PRA:05,mhjdmk:PRB:06,sjxsjxjlhzz:EL:06,jfavb:PRB:07}, 
thus, revealing the underlying mechanism of the observed resonances in terms of quantum-mechanical interaction between 
discrete and continuous states. In nuclear and atomic physics, interferences are often originating from the interaction of 
open (continuum) and closed (discrete levels) channels ~\cite{hf:AP:58,hf:AP:62}. 
\textcite{akbat:PRA:84} unified the approaches of Fano and Feshbach with
{\em ab initio} calculations and derived a rigorous expression of the asymmetry parameter $q$ ~\cite{akbat:PRA:84}.

There are limitations to the applicability of the Fano formula (\ref{eq:Fano}) ~\cite{connerade}. 
First, it can be applied to describe single, isolated resonances. 
The appearance of more than two propagation pathes will change the profiles.
Second, the width of the discrete level should be narrow enough compared to other resonant
structures in the scattering profile.

In general, the Coulomb interaction between an outgoing electron $e^-$ and a charged ion core $A^+$ during auto-ionization leads to  a
renormalization of the energy levels of the many-electron system. Such a renormalization is known as the quantum defect of Rydberg series. 
To precisely describe  the positions and width of the resonances, a multichannel quantum defect theory was developed 
by~\textcite{mjs:PPS:66} and~\textcite{uf:PRA:70}, which provides a rigorous description of the process. It allows to 
derive all asymptotic quantities such as phase shifts or amplitudes of the auto-ionized levels. Eq. (\ref{eq:Fano}) 
was derived by Fano by neglecting effects due to long-range Coulomb interaction. Still it provides a
physical insight into the auto-ionization process in terms of quantum-mechanical interference of discrete and continuum states.

At the resonance the phase of the scattering wave changes sharply by $\pi$. Thus, the interaction of  
scattering waves will result in constructive and destructive interference phenomena located very close to 
each other, corresponding to a maximum $E_{\rm max}$ and a minimum $E_{\rm min}$ of the transmission (absorption), respectively. 
The width of the resonance is proportional to the distance between them $\Gamma\sim|E_{\rm max}-E_{\rm min}|$. 
In principle, they may be located very close to each other $E_{\rm max}\approx E_{\rm min}$, resulting in a very narrow 
resonance $\Gamma\approx0$, corresponding to a very long-lived quasi-bound state ~\cite{fhsdrh:PRA:75}.
using artificial one-dimensional potentials one can even achieve $\Gamma=0$ ~\cite{jvnew:ZP:29}, as a proof of concept.
By applying Feshbach's theory of resonances to two overlapping 
Fano resonances, \textcite{hfdw:PRA:85b,hfdw:PRA:85a} demonstrated that the interference of several auto-ionizationing levels 
of a Rydberg atom may lead to the formation of bound states in the continuum with anomalously narrow resonances.

\subsection{Light and structured matter}
\begin{figure}
\includegraphics[width=\columnwidth]{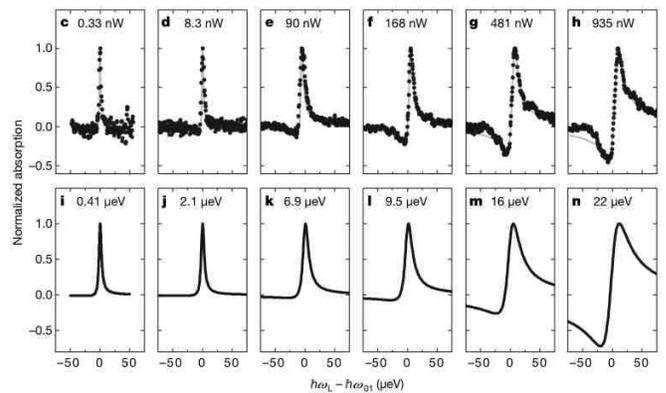}
\caption{\label{fig:nlFano}
Measured (upper row) and calculated (bottom row) absorption spectra of a single quantum dot for various laser powers. 
The absorption profile varies from a symmetrical to an asymmetrical one with increase of the laser power, indicating the 
enhancement of the continuum transition.
From~\textcite{kroner:n:08}. }
\end{figure}

Experiments on the absorption cross-section of a single quantum dot, which is often considered as an artificial atom, 
have revealed that 
the asymmetry parameter $q$ can be continuously tuned with the power of the laser~\cite{kroner:n:08}. In this system, 
the transition rate to the discrete level saturates at high power, while the rate of the continuum transition does not~\cite{wzaoggwb:PR:06}. 
Eventually, the initially weak continuum transition rate will match the saturated transition rate to the discrete level with 
increasing laser power. As a result, a symmetric Lorentzian profile at low power will transform to  an asymmetric Fano 
profile at sufficiently large power (see Fig.~\ref{fig:nlFano}).

In biased semiconductor superlattices the Fano coupling parameter  $\Gamma$ between the discrete state and the 
continuum can be continuously tuned by varying the applied electric field~\cite{cphflmskldmwkk:PRL:98}.
The external bias gives rise to Wannier-Stark states, which interact with 
excitons, and result in asymmetric absorption spectra of 
Wannier-Stark transitions~\cite{khnt:PRB:05,sjxsjxjlhzz:EL:06}. The external bias determines the energy spacing of a 
Wannier-Stark subband, and, thus, controls the effective coupling between the discrete states and the continua.  It allows 
to study the dephasing dynamics of the Fano resonance.

In general, the asymmetry parameter $q$ is not restricted to be only real. In systems with broken time reversal symmetry 
transition amplitudes to the discrete level and to the continuum may become complex, and so does the asymmetry parameter. 
The Fano resonance in such systems can be studied by analyzing the dynamical response. In particular, \textcite{ommhkimk:jetpl:05} 
found that the time-dependent reflection of light a bismuth single crystal after the excitation by an ultrashort laser pulse 
exhibits Fano asymmetric profiles in the Fourier transform of a time-periodic signal. They demonstrated that the asymmetric parameter 
varies periodically with the time delay between pump and probe pulses. The breaking of time reversal symmetry is indicated 
by the change of the sign of the asymmetry parameter.

Asymmetric lineshapes were also observed in Raman spectra of heavily doped 
semiconductors~\cite{hjjdpjtdg:pr:67,fctafmc:pr:73,mibrntmc:ssc:73,fbkp:pss:75,mcjbrmc:pr:78,vmrb:pr:02} and high-$T_c$ 
superconductors~\cite{fbtccm:prl:90,mflairstay:prl:98,mflstay:pr:00,ovmkkkssn:pr:00}. Although, almost any asymmetric 
profile of these spectra can be fitted by the Fano 
formula~\cite{fcafmc:ssc:73,mcfctaf:pr:74,cardona:book:83,jmmc:pr:85,vibacmcctstp:jpcm:97,kzjczygczhxhssc:pr:01,mhjdmk:prcmmp:06,jdjimh:prl:06,ksj:apl:07,vyaavalvgvig:prcmmp:07}, 
a suitable theory for a quantitative description of these cases is still lacking. The general qualitative understanding is 
that the absorbed photon can initiate two kinds of processes. The first one is the  inter- or intra-band electronic 
transition from the ground state to the continuum. The second process is the transition to an intermediate state followed 
by a one-phonon Raman emission and electron transition to either the initial ground state or to the excited donor state. 
Thus, the interference of two processes may in principle result in the Fano resonance.


\subsection{Atoms and atoms}

\begin{figure}[t]
\begin{center}
\includegraphics[width=\columnwidth]{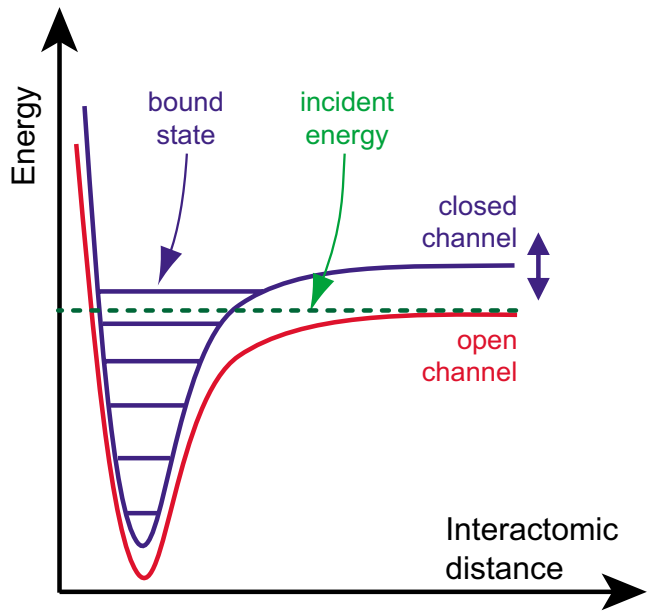}
\end{center}
\caption{ (Color online)
The two-channel model for a Feshbach resonance. Atoms which are prepared in the open channel,
undergo a collision at low incident energy. In the course of the collision, the open channel
is coupled to the closed channel. When a bound state of the closed channel has an energy
close to zero, a scattering resonance occurs. 
From ~\textcite{ibjdwz:RMP:08}.}
\label{fig:Bloch3}
\end{figure}

When two atoms collide with each other a quasi-bound state can be formed, which is characterized 
by a complex energy $E=E_F+i\Gamma$. In scattering theory this quasi-bound state is called a resonance since it possesses 
a finite life-time $\hbar/\Gamma$. The quasi-bound state is formed due to the excitation and sharing of electrons, 
and can interpreted as an interaction between discrete and continuous states [see Fig.~\ref{fig:fano_res2}(b)]. 
In a similar manner, the observed asymmetric resonances in pre-dissociation 
~\cite{adbjpl:JCP:76,rcjrhve:JCP:94,brlstgpoktrkplrjd:PRL:01,mljchddrrl:PRL:06,apzhws:PRA:07}(or fragmentation) 
of molecules were explained by ~\textcite{okr:JCP:33} in terms of auto-ionization.
The concept was introduced by ~\textcite{hf:AP:58} in the context of reactions forming a compound nucleus.
A Feshbach resonance in a two-particle collision appears whenever a bound state in a closed
channel is coupled resonantly with a scattering continuum of an open channel~\cite{ibjdwz:RMP:08}.
The scattered particles are temporarily captured in the quasibound state, and the associated
long time delay gives rise to a Breit-Wigner-type resonance in the scattering cross section (see Fig.\ref{fig:Bloch3}).

A series of recent studies was devoted to the explicit calculation of scattering states
for one-dimensional chains with two interacting bosons or fermions
\cite{mgrwwpsammw:JPB:07,nnrpkm:PRAR:08,nnrpkm:PRA:08,mvdp:JPB:09}.
These systems allow for two-particle continuum states, but also for bound states of two particles.
Tuning the Bloch wave number, the bound state dissolves with the two-particle continuum.
However, its trace inside the continuum remains, leads to a $\pi$ phase shift of the scattering phase,
and to corresponding Fano or Feshbach resonances in the scattering length.
Notably in these problems a clear notion of resonant transport is absent, since there is no difference between
a probe beam and a target due to indistinguishability of the two particles.

Efimov predicted that a three-body quantum system can support weakly bound states (trimer) under 
conditions when none of the three constituting pairs are bound~\cite{vne:plb:70,vne:sjnp:71}. Efimov trimer states 
appear in the limit where the two-body interaction is too weak to support a two-body bound state (dimer). 
Such trimer states should exist regardless of the nature of the two-body interaction, and, thus, are generic 
in few-body systems. 
Recently, the first experimental observation of Efimov states has been reported in ultracold cesium 
trimers~\cite{tkmmpwjgdccbeadlkpajhnrg:n:06}, by measuring the three-body recombination process Cs+Cs+Cs$\rightarrow$Cs$_2$+Cs.
The fingerprint of Efimov trimers in this system appears as a resonant enhancement and suppression of three-body collisions 
as a function of the two-atom interaction strength~\cite{tkmmpwjgdccbeadlkpajhnrg:n:06,bdechg:n:06}, with typical asymmetric 
profiles. ~\textcite{imarprvsb:PRL:06} explained this asymmetric response in terms of a Fano resonance, suggesting 
that the asymmetry can be used as a diagnostic tool for the Efimov effect.


\section{\label{sec:fano_model} Modeling: complex geometries}
One possibility to model a Fano resonance is to choose the geometry of a given system in such a way,
that (at least) two scattering pathes are available.
In this Section we will consider the basic geometries which will do the job, and discuss several extensions.

\subsection{Fano-Anderson model}
One of the simplest models which describes the physics and the main
features of the Fano resonance is the Fano-Anderson
model~\cite{mahan}, which mimics the energy level structures [see Fig.~\ref{fig:fano_res2}(a)] of the model proposed by~\textcite{uf:PR:61}.
In a simplified version ~\cite{aemsfmsfysk:PRE:05} it can be described by the following
Hamiltonian
\begin{eqnarray}\label{eq:fano_model1}
H=C\sum\limits_n (\phi_n\phi^{*}_{n-1}+c.c.)+ E_F|\psi|^2+V_F(\psi^{*}\phi_0+ c.c.) ,
\end{eqnarray}
where the asterisk denotes complex conjugation. This model
describes the interaction of two subsystems. One is a linear discrete chain with the complex field amplitude
$\phi_n$ at site $n$ and nearest-neighbor coupling with strength $C$.
This system supports propagation of plane waves with dispersion $\omega_k=2C\cos k$. The second subsystem
consists of a single Fano state $\psi$ with the 
energy $E_F$. The interaction between these two subsystems is
given by the coupling coefficient $V_F$ between the state $\psi$ and one site of the discrete chain $\phi_0$.
A propagating wave may directly pass through the chain, or instead visit the Fano state, return back and continue
with propagation. These two pathes are the ingredients of the Fano resonance.

The lattice Hamiltonian (\ref{eq:fano_model1}) generates the following differential equations:
\begin{eqnarray}\label{eq:fano_model2}
\mathrm{i}\dot{\phi}_n&=&C(\phi_{n-1}+\phi_{n+1})+ V_F\psi\delta_{n0}\;,\nonumber\\
\mathrm{i}\dot{\psi}&=&E_F\psi+V_F\phi_0.
\end{eqnarray}
With the ansatz
\begin{equation}\label{eq:fano_model3}
\phi_n(\tau) =A_n\mathrm{e}^{-\mathrm{i}\omega \tau}, \;\;\; \psi(\tau
)=B\mathrm{e}^{-\mathrm{i}\omega\tau}\;,
\end{equation}
we obtain a set of algebraic equations for the amplitudes:
\begin{eqnarray}\label{eq:fano_model4}
\omega A_n&=&C(A_{n-1}+A_{n+1})+ V_FB \delta_{n0}\;,\nonumber\\
\omega B&=&E_FB+V_F A_0\;.
\end{eqnarray}
For a scattering problem, the system (\ref{eq:fano_model4}) should be solved for frequencies chosen
from the propagation band $\omega=\omega_k$
with the following boundary conditions
\begin{eqnarray}\label{eq:fano_model5}
A_n = \left\lbrace %
\begin{array}{lc}
  I\mathrm{e}^{\mathrm{i}kn}+\rho\mathrm{e}^{-\mathrm{i}kn},& n<0,\\
  \tau\mathrm{e}^{\mathrm{i}kn}, & n> 0,
 \end{array}
 \right.
\end{eqnarray}
where $I$, $r$, and $t$ have the meaning of the incoming, reflected
and transmitted wave amplitudes, respectively.

\begin{figure}
\includegraphics[width=\columnwidth]{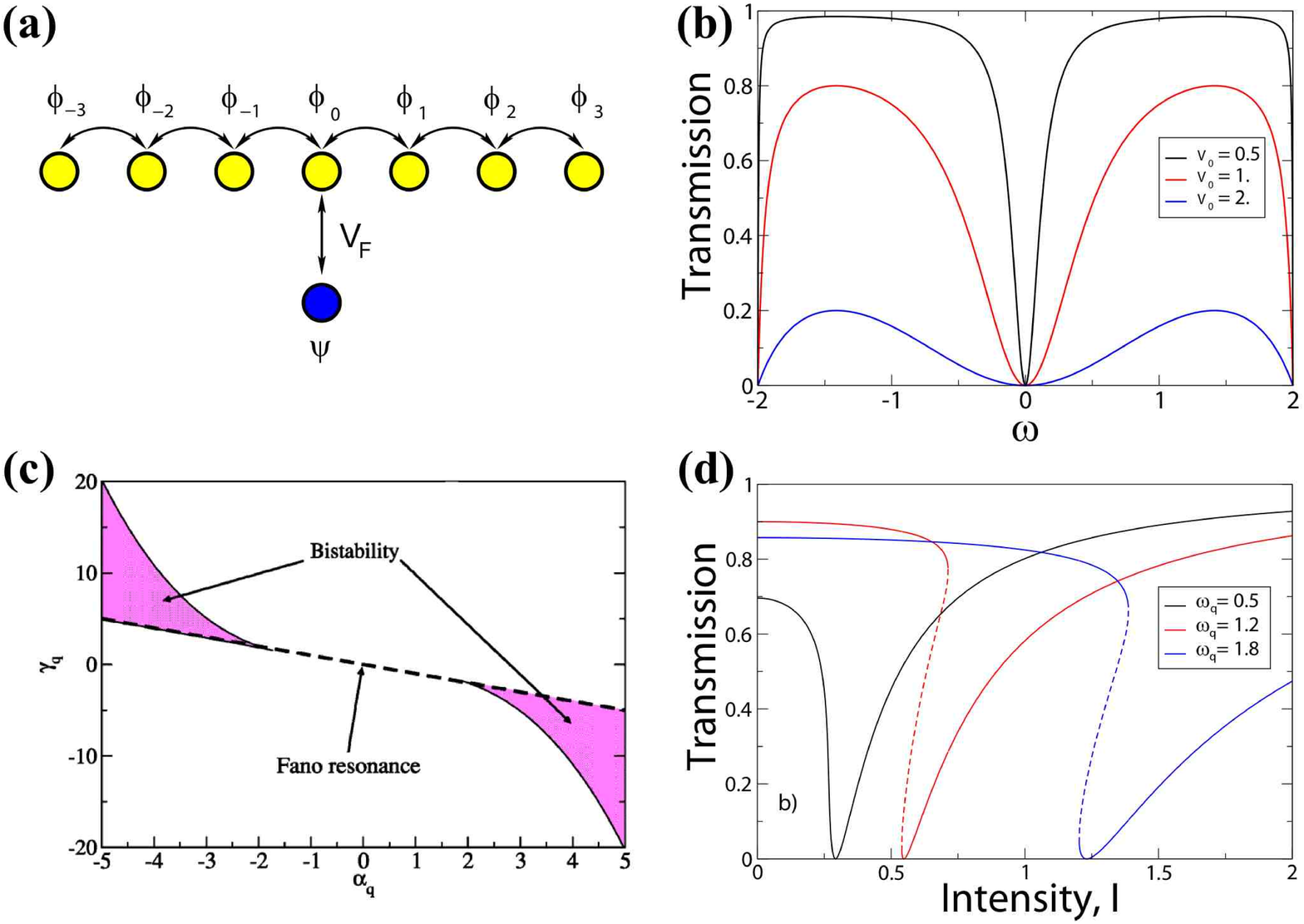}
\caption{\label{fig:fano_model1}
(Color online) Fano-Anderson model as a discrete one-dimensional system with a single side-coupled Fano state defect.
(a) The array of yellow circles corresponds to a linear chains, and the isolated blue circle is the Fano state. Arrows 
indicate the coupling between different states;
(b) Transmission coefficient (\ref{eq:fano_model8}) for various values of the coupling coefficient $V_F$. Other
parameters are $C=1$, and $E_F=0$;
(c) Areas of bistability of the nonlinear Fano resonance (dashed line) in the parameter space $(\alpha_k,\gamma_k)$;
(d) Nonlinear transmission coefficient versus input intensity for various frequencies $\omega_k$ for $C=1$, 
$V_F=0.8$, $E_F=0$ and $\lambda=1$. Regions of bistability are indicated by dashed lines, corresponding to unstable solutions. 
Adapted from \textcite{aemsfmsfysk:PRE:05}.
}
\end{figure}

From (\ref{eq:fano_model4}) it follows
\begin{eqnarray}\label{eq:fano_model6}
B=\frac{V_F A_0}{\omega_k-E_F}\;,
\end{eqnarray}
and finally
\begin{eqnarray}\label{eq:fano_model7}
\omega_k A_n=C(A_{n-1}+A_{n+1})+\frac{V_F^2}{\omega_k-E_F}A_0\delta_{n0}\;.
\end{eqnarray}
The main resulting action of the Fano state is that the strength of the effective scattering potential 
$V_F^2/(\omega_k-E_F)$ resonantly depends on the frequency of the incoming wave $\omega_k$. If $E_F$
lies inside the propagation band of the linear chain $|E_F|<2C$, the scattering potential will become infinitely large for 
$\omega_{k_F}=E_F$, completely blocking propagation. Therefore meeting the resonance condition
leads to a resonant suppression of the transmission, which is the main feature of the Fano resonance.

The transmission coefficient $T=|\tau/I|^2$ can be computed by using the transfer matrix approach~\cite{ptblbh:PRB:99}, 
and expressed in the following form~\cite{aemsfmsfysk:PRE:05}
\begin{equation}\label{eq:fano_model8}
T=\frac{\alpha_k^2}{\alpha_k^2+1},
\end{equation}
where
\begin{equation}\label{eq:fano_model9}
\alpha_k=c_k(E_F-\omega_k)/V_F^2,
\;\; c_k=2C\sin k.
\end{equation}
Transmission vanishes at $\omega_k=E_F$. 
The expression of the transmission coefficient (\ref{eq:fano_model8}) corresponds to the Fano formula (\ref{eq:Fano})
wht $q=0$, 
where $\alpha_k$ corresponds to the dimensionless energy, and $E_F$ is the resonant frequency.
The Fano state is an additional degree of freedom which allows
waves propagating in the chain to interfere with those propagating 
through the discrete state. 

The width of the resonance is defined as
\begin{eqnarray}\label{eq:fano_model10}
\Gamma=\frac{V_F^2}{C\sin k_F}\;,
\end{eqnarray}
where $k_F$ is the wavenumber at the resonance, $E_F =
\omega_{k_F}$.
The width of the resonance is proportional to the square of the coupling strength $V_F^2$.

The Fano-Anderson model (\ref{eq:fano_model1}) is  perhaps the simplest one-dimensional model, 
which shows up with a Fano resonance. Since its asymmetry parameter $q=0$, the location of the maximum in the 
Fano profile is tuned to infinity. The essence of the Fano resonance - destructive interference - is therefore
not encapsulated in an asymmetric scattering profile with both a maximum and a minimum. It is the minimum
which is generated by interference along several propagation pathes.
Due to its analytical simplicity the model may serve as a guideline for the analysis 
of more complicated physical models. There are many variations of this 
model~\cite{aemysk:PRE:05,pbdcpsatav:Chaos:05,rbdcpsatav:PRE:06,ac:PRB:06} studied recently.

\subsection{Tuning the asymmetry parameter}

\begin{figure}
\includegraphics[width=\columnwidth]{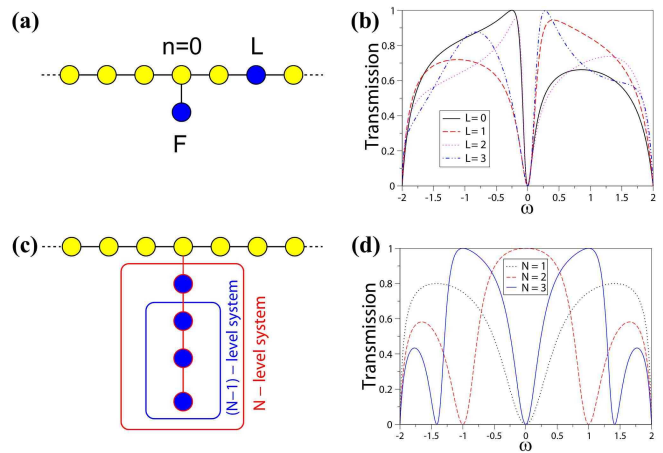}
\caption{\label{fig:fano_model2}
(Color online) Variations of the Fano-Anderson model.
(a) Schematic view of the Fano-Anderson model with an additional defect in the chain.
(b) Transmission coefficient for different distances between the Fano site and the additional defect for 
parameters $C=1$, $V_F=0.5$, $E_F=0$, and $E_L=1$.
(c) Schematic view of the Fano-Anderson model with a locally coupled $N$-defect chainlet.
(d) Transmission coefficient of the $N$-site chainlet model. All sites in the chainlet are identical and with zero 
eigenfrequencies $E_m=0$, and the couplings are $C=V_m=1$.
Adapted from ~\textcite{aemysk:PRE:05}.
}
\end{figure}

The Fano-Anderson model (\ref{eq:fano_model1}) describes the resonant suppression of the transmission with a symmetric 
lineshape ($q=0$), emphasizing the main property of the Fano resonance which is destructive interference (resonant reflection). 
It can be easily extended in order to obtain a nonzero asymmetry parameter $q$ with asymmetric lineshapes, such that 
both resonant suppression and resonant 
enhancement of the transmission will be located close to each other. 
Introducing a defect $E_L\phi_L\delta_{nL}$ in the 
main array (\ref{eq:fano_model2}) [see Fig.~\ref{fig:fano_model2}(a)], both pathes for scattering waves will yield
phase shifts. As a result, both constructive and destructive interference phenomena may coexist, generating asymmetric 
transmission profiles [see Fig.~\ref{fig:fano_model2}(b)].
As observed in Fig.~\ref{fig:fano_model2}(b) the sign of the asymmetry parameter $q$ alternates  
with the distance between the side-coupled defect and the defect in the main array
(which is known as 
$q$-reversal~\cite{bkky:JCP:93})
\begin{eqnarray}\label{eq:fano_model15}
{\rm sign}(\omega_{\;T_{\mathrm max}}-\omega_{\;T_{\mathrm min}})=(-1)^L.
\end{eqnarray}
Note that 
the maximum of the transmission does not need to reach the value $T=1$. This 
incomplete constructive interference is due to additional phase accumulation along the propagation distance between two defects. 
It does not affect the destructive interference, at which strictly $T=0$, confirming that $T=0$ is the
only necessary and sufficient result of 
destructive interference and the Fano resonance.

\subsection{Many resonances}

Consider now a replacement of the Fano site in the Fano-Anderson model by a finite chainlet,
consisting of $N$ coupled sites~\cite{aemysk:PRE:05,pbdcpsatav:Chaos:05} 
[see Fig.~\ref{fig:fano_model2}(c)]. 
If the chainlet is decoupled from the linear discrete chain, the standing waves of the chainlet will
give rise to $N$ eigenfrequencies. Once the chainlet is coupled back to the linear discrete chain,
each of the standing waves will provide with an additional path for a propagating wave,
leading to a variety of interference phenomena.
The finite chainlet could be considered as an approximation 
of a complex $N$-level system, such as a quantum dot, for example. \textcite{aemysk:PRE:05} showed that, in general, 
there are exactly $N$ total reflection ($T=0$) and $N-1$ total transmission ($T=1$) resonances [see Fig.~\ref{fig:fano_model2}(d)]. 
Each frequency of the total reflection corresponds to an eigenfrequency of the chainlet standing wave, and each total transmission 
corresponds to an eigenfrequency of the chainlet with $(N-1)$ sites, indicated in Fig.~\ref{fig:fano_model2}(c). Tt 
resonances some particular eigenstates of the side-coupled chainlet are excited. 

\begin{figure}
\includegraphics[width=\columnwidth]{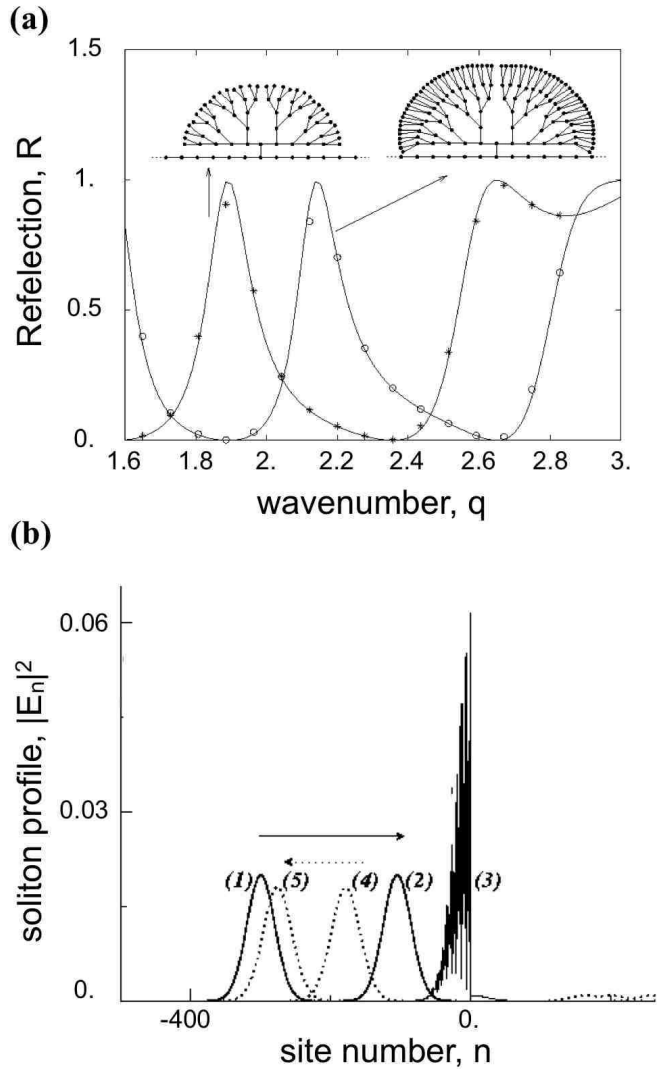}
\caption{\label{fig:fano_model3}
Resonant reflection of a soliton in topological networks
(a) The reflection coefficient versus wavenumber $k$ for two Cayley trees of length $M=5$ (line) and $M=6$ (line) attached 
to the discrete array. Empty circles and stars correspond to direct numerical simulations of the soliton propagation.
(b) Example of the soliton reflection by a Fano-like defect.
From ~\textcite{pbdcpsatav:Chaos:05}.
}
\end{figure}

Many other inhomogeneous networks were considered to design various topological filters~\cite{pbdcpsatav:Chaos:05,rbdcpsatav:PRE:06}. 
One can even plant Cayley trees into a discrete array and gather very well pronounced Fano resonances [see Fig.~\ref{fig:fano_model3}(a)].

\subsection{Nonlinear Fano resonance}

The Fano state amplitude becomes largest
\begin{eqnarray}\label{eq:fano_model11}
|B_{\rm max}|^2=4V_F^2|I|^2/\Gamma^2\;,
\end{eqnarray}
exactly at the resonant value of the wave number $k_F$, and it diverges (and is therefore much
larger than the amplitudes in the chain which are bounded by $I$) in the limit of small coupling strength $V_F$.

Whatever the physical origin of the waves whose scattering is studied, large amplitudes call for 
corrections - either many-body interactions in a quantum setting, or nonlinear response corrections
in a classical setting. Notably these corrections apply in first order only for the Fano state.
Taking the classical setting,
nonlinear Fano resonances~\cite{aemsfmsfysk:PRE:05} were studied by introducing 
nonlinear corrections to the evolution equation for the Fano state only (\ref{eq:fano_model4})
\begin{eqnarray}\label{eq:fano_model12}
\omega B=E_FB+\lambda|B|^2B+V_FA_0\;.
\end{eqnarray}

The nonlinear transmission coefficient can be expressed in the following form~\cite{aemsfmsfysk:PRE:05}
\begin{eqnarray}\label{eq:fano_model13}
T=\frac{x^2}{x^2+1}\;,
\end{eqnarray}
where $x=-\cot\delta(k)$ is a function of the scattering phase $\delta(k)$, and satisfies the cubic equation
\begin{eqnarray}\label{eq:fano_model14}
(x^2+1)(x-\alpha_k)-\gamma_k=0\;,
\end{eqnarray}
with the parameter $\gamma_k=\lambda c_k^3|I|^2/V_F^4$.
The nonlinear Fano resonance condition corresponds to $x=0$ in Eq.(\ref{eq:fano_model14}), which needs the 
condition $\gamma_k=-\alpha_k$ to be satisfied [see Fig.~\ref{fig:fano_model1}(c)].
The transmission coefficient depends not only on the frequency of the incoming wave $\omega_k$, but 
on its intensity $|I|^2$ as well.
The presence of nonlinearity leads to a renormalization of the self-energy of the Fano state, and consequently to an 
intensity-dependent shift of the resonance.
~\textcite{aemsfmsfysk:PRE:05} have shown that the nonlinear Fano resonance exists for any value of the input 
intensity $|I|^2$ [see Fig.~\ref{fig:fano_model1}(c)].
Therefore, nonlinearity allows to tune the location of the Fano resonance by
changing the
intensity of the input waves.
In general, there exist up to three solutions of the cubic Eq.(\ref{eq:fano_model14}), which will 
result in bistable transmission [see Fig.~\ref{fig:fano_model1}(d)]. 


\subsection{Resonant reflection of pulses and solitons}

So far we discussed the scattering of monochromatic plane waves. Consider a pulse instead which is launched towards the scattering region.
The more narrow the pulse is in real space, the broader is its spectral decomposition in Fourier (plane wave) space $k$,
which is characterized by the maximum frequency $\omega_m$ and the spectral width $\Delta \omega$.
Each pulse component in Fourier space $k$ will scatter as discussed above. The spectral width $\Delta \omega$ has to be compared
with the width of a Fano resonance $\Gamma$. If $\Delta \omega \ll \Gamma$, tuning $\omega_m$ into resonance with a
Fano resonance will lead to a practically complete reflection of the pulse. If on the contrary $\Delta \omega \gg \Gamma$,
only a narrow part of the spectral component of the pulse will be reflected, while the rest will be transmitted
with a spectral hole 'burned' into it.

If nonlinearities are added into the propagation channel, they lead to an interaction between the various
plane waves constituting the pulse and may ultimately yield nondispersing solitons.
Their scattering by Fano defects was studied as well
~\cite{aemsfbm:CHAOS:03,uwvvs:PRB:05,pbdcpsatav:Chaos:05,rbdcpsatav:PRE:06}. 
There are two characteristic time scales important for the scattering of solitons. One of them is the time the soliton
resides in the vicinity of 
the defect $\tau_{\rm rs}$, which is inversely proportional to its spectral width $\Delta \omega$ and 
the soliton velocity $v$. The second 
one is set by the nonlinearity. It is the time scale on which the plane wave which constitute the soliton
interact with each other $\tau_{\rm int}$~\cite{aemsfbm:CHAOS:03}. For fast propagating 
solitons the residence time is much smaller than the interaction time $\tau_{\rm rs}\ll\tau_{\rm int}$. 
Then, during the scattering process the soliton can be considered as a set of noninteracting plane waves,
and the results of the above pulse scattering apply
~\cite{aemsfbm:CHAOS:03,pbdcpsatav:Chaos:05,rbdcpsatav:PRE:06} [see Fig.~\ref{fig:fano_model3}(b)].
In the opposite case, when the residence time is much larger than the interaction time 
$\tau_{\rm rs}\gg\tau_{\rm int}$, the nonlinearity-induced mode-mode interaction 
becomes crucial during the scattering process. In general a nonlinear interaction between
many degrees of freedom (modes or plane waves) will lead to chaotic dynamics, and consequently to
a dephasing of individual plane waves. 
Therefore phase coherence will not be maintained during the scattering, and interference effects
will vanish. Therefore the Fano resonance should quickly deteriorate as the soliton parameters
are tuned into the region of validity of the second case. This was numerically confirmed by \textcite{aemsfbm:CHAOS:03}.


\begin{figure}
\includegraphics[width=\columnwidth]{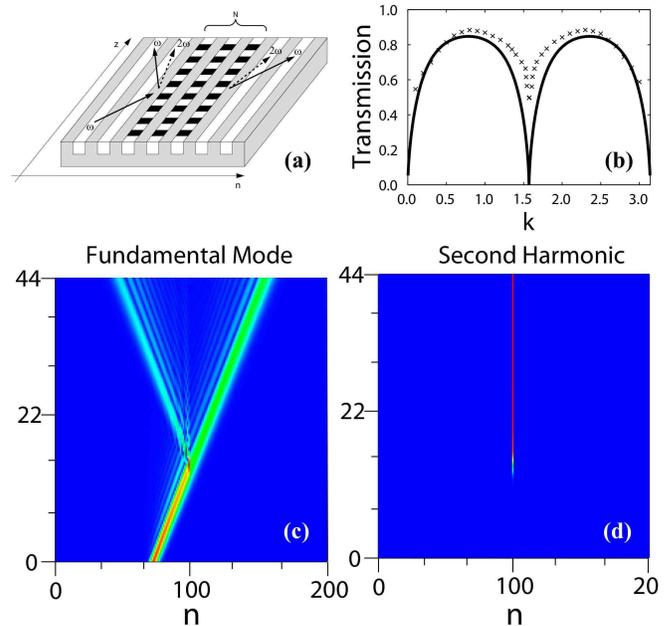}
\caption{\label{fig:fano_model4}
(Color online) Light scattering in an array of channel waveguides with quadratic nonlinearity.
(a) Schematic view of a one-dimensional array of channel waveguides with nonlinear defects, created by periodic poling. 
Arrows indicate the scattering process.
(b) Comparison of the transmission coefficients of plane waves (solid line) and a Gaussian beam (crosses).
Bottom: Example of the Gaussian beam scattering by a single nonlinear defect showing the resonant reflection part of the 
beam at the fundamental frequency (c), and resonant excitation of the second harmonic (d). 
Adapted from ~\textcite{aeyskravmim:OL:05}.
}
\end{figure}

\subsection{Quadratic nonlinearities}

Consider the wave scattering in an array of channel waveguides with quadratic nonlinearity generated by 
periodic poling of several waveguides~\cite{aeyskravmim:OL:05}.
When the matching conditions are satisfied, the fundamental-frequency (FF) mode with frequency $\omega$ can  
parametrically generate a second-harmonic (SH) wave with the frequency $2\omega$ [see Fig.~\ref{fig:fano_model4}(a)], such that a 
structure with several poled waveguides may behave as a nonlinear defect with spatially confined
quadratic nonlinearity~\cite{rirsgistpflymws:PRL:04}.
The waveguide array can be described by a discrete model of weakly coupled linear waveguides with several waveguides having a 
quadratic nonlinear response~\cite{rirsgistpflymws:PRL:04,aeyskravmim:OL:05}, which is very similar to the Fano-Anderson model 
(\ref{eq:fano_model2}). The fundamental mode in this case can be considered as a continuum of propagating states, while the 
generated second harmonic can be either extended or effectively localized depending on the phase matching condition~\cite{aeyskravmim:OL:05}. 
In the latter case the excited second harmonic will act as a discrete state in the continuum, leading to the appearance of a 
Fano resonance in the transmission [see Fig.~\ref{fig:fano_model4}(b)]. Results of the direct numerical simulations of the 
Gaussian beam scattering are in a good agreement with the plane wave analysis  [see Fig.~\ref{fig:fano_model4}(b)]. 
Figures~\ref{fig:fano_model4}(c,d) show the evolution of the fundamental and second harmonic of the Gaussian beam scattering at 
resonance. A part of the fundamental harmonic of the Gaussian beam is resonantly reflected 
by a single nonlinear defect [see Fig.~\ref{fig:fano_model4}(c)].
Since the spectral width of the Gaussian beam is larger than the width of the resonance some part of the beam 
still propagates through the defect. During the scattering the second harmonic is resonantly excited 
[see Fig.~\ref{fig:fano_model4}(d)]. After the scattered
beam parts leave the defect region, the second harmonic persists  in a self-sustained form. 

\section{\label{sec:dnls}Modeling: complex dynamics}

Several propagation pathes and interference phenomena can be generated not only by imprinting complex geometries,
but also by using complex dynamics. Nonlinear wave excitations, e.g. discrete solitons, when scattering small amplitude
waves, generate several propagation pathes purely dynamically. 
The reason is that the scattering potentials are time-dependent (in fact usually time-periodic).
The amplitude and the temporal period can be tuned by controlling the characteristics of the nonlinear excitations
~\cite{wllir:PRB:99,dfmler:PRB:01,ailer:PRA:02}.  
Total resonant reflection was also observed~\cite{pfbrkl:PRB:92}. This is because the time-periodic 
scattering potential generates several harmonics. In general, these harmonics will correspond to 
open and closed propagation channels, respectively. 
The presence of such dynamically generated channels is 
equivalent to a local increase of the spatial dimensionality, discussed in the previous Section. In other words, each new channel 
generates an alternative pathway for the scattering wave to propagate. 
The spectrum of excitations in each additional closed channel may contain discrete (localized) states, which
happen to resonate with the continuum of the original open channel. 
As a result, Fano resonances can be expected, where the Fano state is the discrete state from a dynamically
generated closed channel.


\subsection{Scattering by discrete breathers}
Discrete breathers (DBs) are known as time-periodic and spatially localized solutions of nonlinear wave equations on lattices
~\cite{rsmsa:Nonlin:94,sa:PD:97,sfcrw:PR:98,sfavg:PR:08}. 
They originate from a constructive interplay between nonlinearity and 
discreteness. 
DBs exist independent of the lattice dimension, and are not relying on integrability properties.
In return, these excitations can not freely move through lattices. Therefore, they act as scattering
centers for small amplitude plane waves. Tuning the amplitude of the DB excitation, one tunes its temporal period,
and all other characteristics of the resulting time-periodic scattering potential.
DBs were detected and studied experimentally in interacting Josephson junction networks~\cite{etjjmtpo:PRL:00,pbdaavusfyz:PRL:00}, 
coupled nonlinear optical waveguides~\cite{hseysrmarb:PRL:98}, lattice vibrations in crystals~\cite{bisjabsplgfsapsarbwzwmis:PRL:99}, 
anti-ferromagnetic structures~\cite{utslqeajs:PRL:99}, micro-mechanical cantilever arrays~\cite{msbehajsbidachgc:PRL:03}, 
Bose-Einstein condensates loaded on optical lattices~\cite{betamamtptkpmmko:PRL:04}, and many others \cite{sfavg:PR:08}.

Resonant scattering of plane waves by DBs was 
studied and showed Fano resonances with zero transmission
$T=0$~\cite{sslsk:IJMPB:00,swksk:PD:00,swkwk:PRB:01,sfaemvfmvf:PRL:03,sfaemmvf:Chaos:03,aemmssfmvfavu:PRB:05}. 
Below we will demonstrate the concept using a particular example 
of wave scattering by DBs in the discrete nonlinear Schr\"odinger model (DNLS)~\cite{sfaemvfmvf:PRL:03}.

The equations of motion for the DNLS are given by
\begin{equation}\label{eq:dnls1}
i\dot{\Psi}_n = C(\Psi_{n+1} + \Psi_{n-1}) + |\Psi_n|^2\Psi_n\;,
\end{equation}
where $n$ is an integer labeling the lattice sites, $\Psi_n$ is a
complex scalar variable and $C$ describes the nearest neighbor
interaction (hopping) on the lattice. The last term in
(\ref{eq:dnls1}) is a cubic nonlinearity. For small
amplitude waves $\Psi_n(t) = \epsilon e^{i(\omega_kt-kn)}$ the
dispersion relation
\begin{equation}\label{eq:dnls2}
\omega_k = -2C \cos k
\end{equation}
follows from Eq.(\ref{eq:dnls1}).

The DNLS model supports DB solutions with a single harmonic 
\begin{equation}\label{eq:dnls3}
\hat{\Psi}_n(t) = \hat{A}_n e^{-i\Omega_b t}\;,\;\hat{A}_{|n|
\rightarrow \infty} \rightarrow 0\;,
\end{equation}
where the time-independent amplitude $\hat{A}_n$ can be taken real
valued, and the breather frequency $\Omega_b \neq \omega_k$ is
some function of the maximum amplitude $\hat{A}_0$. The spatial
localization is given by an exponential law $\hat{A}_n \sim
e^{-\lambda |n|}$ where $\cosh \lambda = |\Omega_b|/2C$. Thus the
DB can be approximated by a single-site excitation if
$|\Omega_b| \gg C$. In this case the relation between the
single-site amplitude $\hat{A}_0$ and $\Omega_b$ becomes $\Omega_b
= \hat{A}_0^2$. In the following, the DB amplitudes for $n
\neq 0$ will be neglected, i.e. $\hat{A}_{n\neq 0} \approx 0$,
since $\hat{A}_{\pm 1} \approx (C/\Omega_b)\hat{A}_0\ll A_0$.

\begin{figure}
\includegraphics[width=\columnwidth]{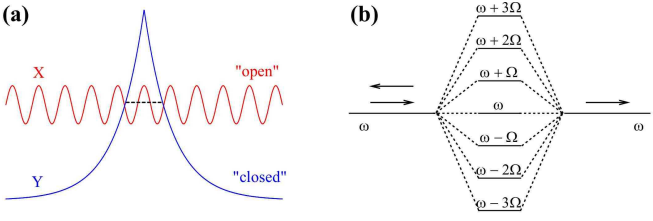}
\caption{\label{fig:dnls_fig1}
(Color online) Time-periodic scattering potentials.
(a) Schematic view of the open channel $X$ and closed channel $Y$ from the Eqs. (\ref{eq:dnls7}-\ref{eq:dnls8}). 
The dashed line indicates the localized state of the closed channel $Y$ inside the open channel $X$;
(b) Schematic view of the virtual states, generated by a possibly infinite 
number of harmonics of the time-periodic scattering potential.
}
\end{figure}

Let us perturb the breather solution with small fluctuations $\phi_n(t)$
\begin{equation}\label{eq:dnls4}
\Psi_n(t) = \hat{\Psi}_n(t) + \phi_n(t)
\end{equation}
and substitute this ansatz into (\ref{eq:dnls1}). 
Linearization in the small fluctuating perturbation leads to
the following set of equations:
\begin{equation}\label{eq:dnls5}
i\dot{\phi}_n = C(\phi_{n+1}+\phi_{n-1}) +\Omega_b \delta_{n,0}
(2\phi_0 + e^{-2i\Omega_b t} \phi_0^*)
\end{equation}
with $\delta_{n,m}$ being the Kronecker symbol.
The DB generates a scattering potential that consists of two parts: a static (dc) one, which depends on the breather 
intensity only $\sim\Omega_b=\hat{A}_0^2$, and a dynamical (ac) one, which depends periodically 
on time $\sim\Omega_be^{-2i\Omega_b t}$.
With the two channel ansatz
\begin{equation}\label{eq:dnls6}
\phi_n(t) = X_n e^{i\omega t} + Y_n^* e^{-i(2\Omega_b +\omega)t}\;,
\end{equation}
Eq. (\ref{eq:dnls5}) is reduced to a set of algebraic equations for the complex channel amplitudes $X_n$ and $Y_n$
\begin{eqnarray}\label{eq:dnls7}
-\omega X_n = C(X_{n+1}+X_{n-1}) +\Omega_b \delta_{n,0} (2X_0 +
Y_0)\;,\;\;\\ 
(2\Omega_b + \omega) Y_n = C(Y_{n+1}+Y_{n-1})
+\Omega_b \delta_{n,0} (2Y_0+X_0)\;.\;\; \label{eq:dnls7+1}
\end{eqnarray}
For propagating the frequency $\omega$ should be chosen from the propagation band $\omega_k$. As a result, 
the channel $X_n$ supports extended waves, while for $Y_n$ channel does not, since the frequency 
$-(2\Omega_b+\omega_q)$ is outside the propagation band $\omega_k$~\cite{sfaemvfmvf:PRL:03} [see Fig.~\ref{fig:dnls_fig1}(a)]. 
Therefore the scattering takes place with an open channel $X_n$ which interacts with a closed channel $Y_n$.

Let us 
consider the  more general set of equations
\begin{eqnarray}\label{eq:dnls8}
-\omega_k X_n = C(X_{n+1}+X_{n-1}) - \delta_{n,0} (V_x X_0 + V_a
Y_0)\;,\;\; \label{nae1}\\ 
(\Omega + \omega_k) Y_n = C(Y_{n+1}+Y_{n-1})
- \delta_{n,0} (V_y Y_0+ V_a X_0)\;,\;\; \label{nae1+1}
\end{eqnarray}
which can be reduced to Eq. (\ref{eq:dnls7}) with the following parameters $\Omega=2\Omega_b$ and $V_x=V_y=2V_a=-2\Omega_b$.
For $V_a=0$ the closed channel $Y_n$ possesses exactly one localized eigenstate
\begin{eqnarray}\label{eq:dnls9}
Y_n=Y{\rm e}^{-\lambda|n|}\;.
\end{eqnarray}
with eigenfrequency
\begin{equation}\label{eq:dnls10}
\omega^{(y)}_L = -\Omega + \sqrt{V_y^2 + 4C^2}\;.
\end{equation}
The transmission coefficient for the general case $V_a \neq 0$ ~\cite{sfaemvfmvf:PRL:03}
\begin{eqnarray}\label{eq:dnls12}
T = \frac{4 \sin ^2 k}{\left( 2\cos k - a -\frac{d^2
\eta}{2-b\eta} \right)^2 + 4\sin ^2 k}\nonumber\;,
\\
a = \frac{-\omega_k + V_x}{C}\;,\;b=\frac{\Omega+\omega_k+ V_y}{C}
\;,\; d = \frac{V_a}{C}\;.
\end{eqnarray}
From Eq. (\ref{eq:dnls12}) it follows that the transmission coefficient vanishes, when the condition
\begin{eqnarray}\label{eq:dnls13}
2-b\eta=0
\end{eqnarray}
is satisfied, which is equivalent to requesting the resonance condition
\begin{eqnarray}\label{eq:dnls14}
\omega_k = \omega^{(y)}_L\;.
\end{eqnarray}
The conclusion is, that total reflection
takes place when a local mode, originating from the closed
$Y$-channel, resonates with the plane wave spectrum $\omega_k$
of  the open $X$-channel. 
The resonance condition is not renormalized by the actual value of $V_a$. 
The existence of local modes which originate from
the $X$-channel for nonzero $V_x$ and possibly resonate with the
closed $Y$-channel is evidently not of any relevance. The
resonant total reflection is a Fano resonance,
as it is unambiguously related to a local state resonating
and interacting with a continuum of extended states. The fact that
the resonance is independent of $V_a$ is due to the local coupling between the Fano state (originating from
the $Y$-channel) and the open channel\index{open channel}, and originates from the
approximative DB solution in the limit $|\Omega_b| \gg C$. Corrections to the DB solution will
increase the range of coupling between the Fano state and the continuum, and correspondingly lead
to a renormalization of the resonance location \cite{sfaemvfmvf:PRL:03}.
Therefore we conclude, that the resonance location is not significantly renormalized,
if the wavelength of the propagating wave is large
compared to the extension of the space region where the coupling between a Fano state and a continuum occurs~\cite{sfaemmvf:Chaos:03}.

If the closed channel is reduced to the localized discrete Fano state $Y$ only, the equations for the amplitudes
take the form
\begin{eqnarray}\label{eq:dnls15}
-\omega X_n&=&C(X_{n-1}+X_{n+1})+ V_a  Y \delta_{n0}\;,\nonumber\\
-\omega Y&=&E_F Y-V_a X_0\;.
\end{eqnarray}
The different signs in front of the coupling between the chain and the Fano state
are due to the fact, that time-periodic scattering potentials correspond to
eigenvalue problems with a symplectic propagator. 
At variance, complex geometries (\ref{eq:fano_model4}) do not leave the grounds
of unitary propagators. Remarkably, these differences in the symmetries of the underlying 
dynamical processes do not alter the final result of destructive interference and Fano resonances.

The above analysis leads to a 
recipe of finding the position of resonances. One first calculates the localized states of closed channels decoupled 
from the open one~\cite{sfaemvfmvf:PRL:03,sfaemmvf:Chaos:03}.
Switching on the coupling again,
Fano resonances will take place exactly at the eigenfrequencies of the localized states for weak coupling. 
For stronger coupling the positions of the resonances will renormalize.
In general, there is an infinite number of harmonics of the DB, which generate an infinite number of closed channels
~\cite{sfaemvfmvf:PRL:03,sfaemmvf:Chaos:03}.
The approach described above is rather generic and can be applied to the scattering through many types of oscillating barriers, 
self-induced (like DBs) or parametrically driven 
(by external forces)~\cite{pfbrkl:PRB:92,wllir:PRB:99,dbmler:PRB:00,dfmler:PRB:01,ailer:PRA:02,swk:PRB:02,sf:PRB:06}. 
All of them produce similar scattering potentials with an open and a number of  closed channels for small amplitude scattering waves. 

\subsection{Light scattering by optical solitons}

\begin{figure}
\includegraphics[width=\columnwidth]{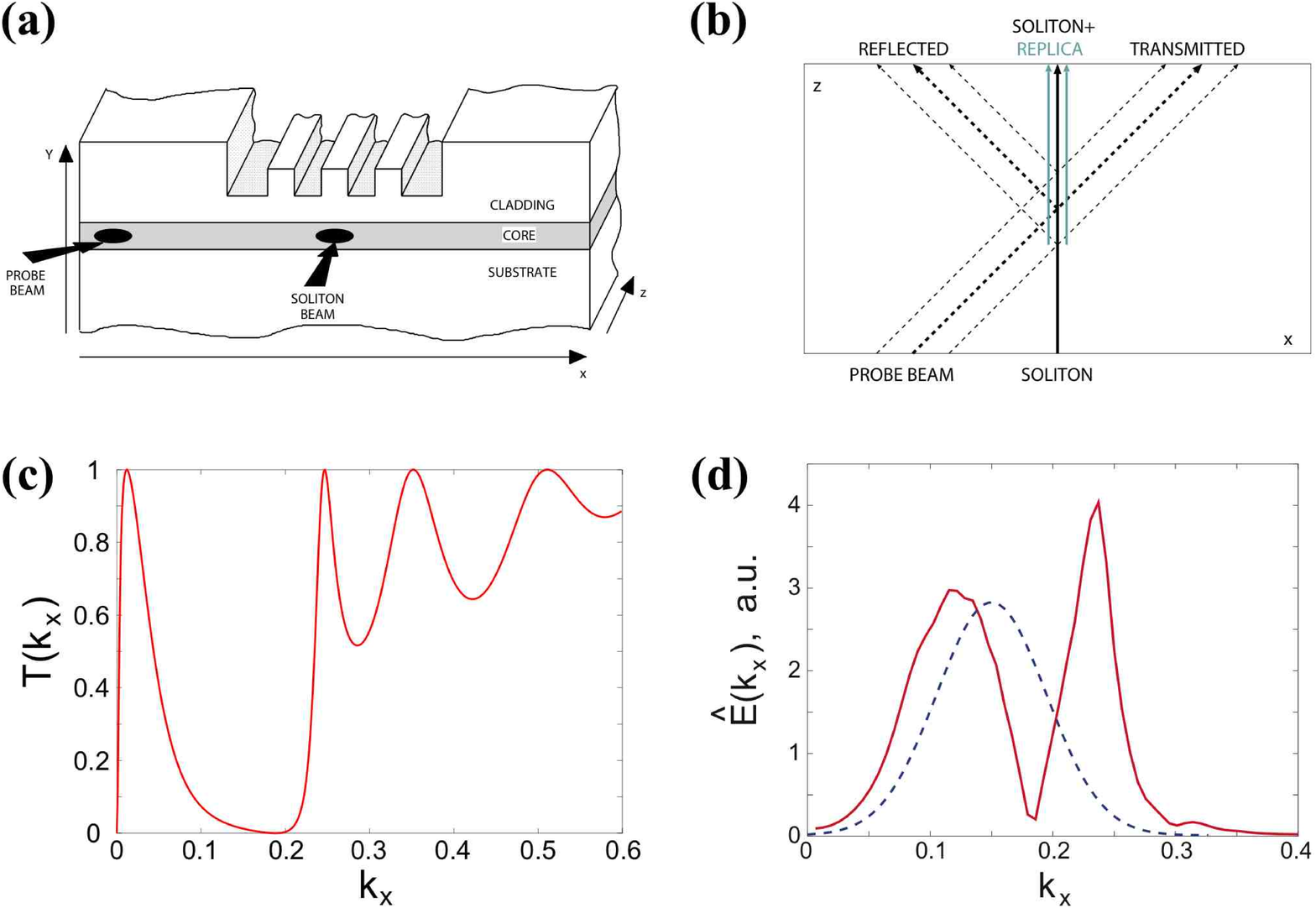}
\caption{\label{fig:dnls_fig2}
(Color online) Light scattering by optical solitons.
(a) Sketch of the scattering setup by an optical soliton in a one-dimensional waveguide array. The soliton beam is sent along 
the $z$-axis, while the probe
beam propagates in the $xz$-plane at some angle to the soliton; (b) top view of the scattering process; (c) transmission 
coefficient vs $k_x$ for plane waves under oblique incidence. There is total suppression of the transmission near 
$k_x\approx0.181$; (d) Fourier spectrum of the incident (dashed line) and transmitted (solid line) beams. 
The suppression of the resonant frequency [see plot (c)] in the spectrum is observed. Adapted from ~\textcite{sfvfavgaem:PRL:05}.
}
\end{figure}
The above concept of scattering by solitary excitations was applied to predict
resonant light scattering by optical solitons in a slab waveguide with an inhomogeneous refractive 
index core~\cite{sfvfavgaem:PRL:05,sfvfavgaem:ps:06}. The soliton is generated in a nonlinear planar waveguide by a laser 
beam injected into the slab along the $z-$direction [see Fig.~\ref{fig:dnls_fig2}(a)].
The soliton beam is confined in the $y-$direction by total internal reflection. The localization in the $x-$direction 
is achieved by a balance between linear diffraction and an instantaneous Kerr-type nonlinearity. The analogy with the 
discussed above scattering problem by time-periodic potentials comes from the possibility to interpret the spatial propagation 
along the $z-$direction as an artificial time ~\cite{agrawal}. Thus, the propagation constant of the soliton can be considered 
as the frequency of the breather. The evolution of the soliton envelope function satisfies the nonlinear Schr\"odinger equation 
(NLS), the continuum analog of Eq. (\ref{eq:dnls1})~\cite{sfvfavgaem:PRL:05}. The analysis of the scattering problem is 
similar to the above discussed one. Figure~\ref{fig:dnls_fig2}(c) shows the dependence of the transmission coefficient for oblique 
incident light for various $k_x$ wavenumbers. It results in a Fano resonance for plane waves at $k_x\approx0.181$, 
where the  transmission coefficient vanishes. This result has been confirmed by direct numerical simulations of a 
propagating small-amplitude 
wavepacket scattered by the optical soliton~\cite{sfvfavgaem:ps:06} [see Fig.~\ref{fig:dnls_fig2}(b)]. The Fourier spectrum of the 
transmitted wavepacket reveals that the resonant wavenumber $k_x\approx0.181$ was filtered out from the initial wavepacket 
[see Fig.~\ref{fig:dnls_fig2}(d)]. Such a spectral hole burning effect can be used as a characteristic  feature for the detection of the 
Fano resonance in an experimental setup.

\subsection{Plasmon scattering in Josephson junction ladders}

\begin{figure}
\includegraphics[width=\columnwidth]{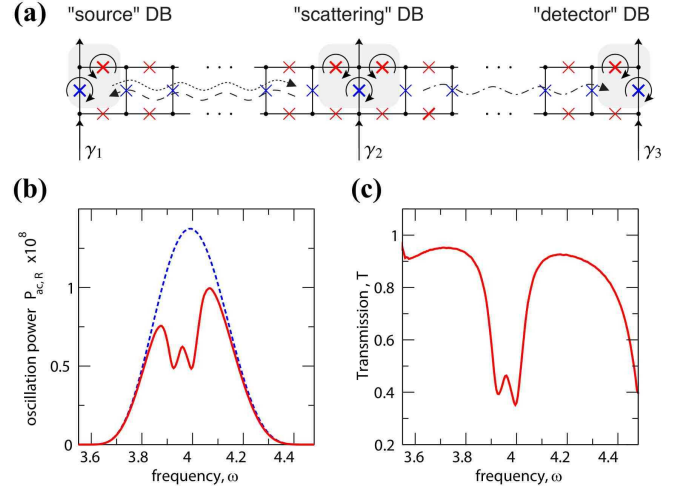}
\caption{\label{fig:dnls_fig3}
(Color online) Plasmon scattering by discrete breathers in Josephson junction ladders.
(a) Schematic setup for the measurement of plasmon scattering with the use of controlled bias currents $\gamma_i$;
(b) Oscillating power $P_{{\rm ac}, R}$ at the right end with (red solid line) and without (blue dashed line) the DB;
(c) Transmission coefficient $T$, derived from (b) by using Eq. (\ref{eq:dnls151}). Adapted from ~\textcite{aemmssfmvfavu:PRB:05}.
}
\end{figure}
Another theoretical prediction concerns 
the plasmon wave scattering by DBs in Josephson junction ladders (JJLs). JJLs are formed by an array of 
small Josephson junctions that are arranged along the spars and rungs of a ladder [see Fig.~\ref{fig:dnls_fig3}(a)].  Each 
junction consists of two small weakly coupled superconducting islands. The dynamical state of a junction is described by the 
phase difference $\phi(t)$ (Josephson phase) of the superconducting order parameters of two neighbouring islands. When the difference 
does not vary in time $\phi(t)=const$, the junction is traversed by a superconducting current only,
with zero voltage drop. Otherwise, the junction is traversed in addition by a resisitive current component
with a nonzero voltage drop $V\propto\dot{\phi}(t)$. It was observed experimentally that JJLs support dynamic localized 
states (DBs)~\cite{etjjmtpo:PRL:00,pbdaavusfyz:PRL:00}. A discrete breather is characterized by a few junctions being in the 
resistive state $\langle\dot{\phi}\rangle\not=0$ while the others reside in the superconducting state $\langle\dot{\phi}\rangle=0$. 
The frequency of a DB is proportional to the average voltage drop across the resistive junctions 
$\Omega_b\propto\langle\dot{\phi}\rangle$. ~\textcite{aemmssfmvfavu:PRB:05} have recently proposed an experimental 
setup to measure Fano resonances in that transmission line. Small amplitude waves are generated in a JJL with open ends by applying 
locally a time-periodic current $\gamma_1(t)=\gamma_{\rm ac}\cos(\omega t)$. The local current  acts as a local parametric drive. 
It excites edge junctions at a frequency $\omega$. This tail extends into the ladder.
To monitor the linear wave propagation in the system ,the time-averaged oscillation power 
$P_{{\rm ac},n}=\langle\dot{\phi}_n^2\rangle$ is measured. The transmission coefficient can be obtained by relating the oscillation 
power at the right boundary with and without an excited DB in the system
\begin{eqnarray}\label{eq:dnls151}
T=\frac{P_{{\rm ac},R}({\rm with\; DB})}{P_{{\rm ac},R}({\rm without\; DB})}\;.
\end{eqnarray}
Figures~\ref{fig:dnls_fig3}(b,c) show the presence of resonant suppression of the transmission coefficient for particular 
frequencies $\omega$. The analysis in Ref.~\cite{aemmssfmvfavu:PRB:05} reveals that they correspond to Fano resonances, 
which originate from localized states of closed channels of the time-periodic scattering potential which is
generated by the DB.

\subsection{Matter wave scattering in Bose-Einstein condensates}

Over the last couple of years, it has been shown that optical
lattices, generated by counter-propagating laser beams and
providing a periodic potential modulation for the atoms,
introduce many interesting and potentially useful effects
by modifying single atom properties and enhancing correlations
between atoms~\cite{ommo:rmp:06}. 
Using about 1000 $^{87}$Rb atoms in a quasi one-dimensional optical lattice, \textcite{betamamtptkpmmko:PRL:04}
obtained a spatially localized Bose-Einstein condensate (BEC) which is an experimental
manifestation of a gap soliton, or a discrete breather.
The solitary state exists due to the atom-atom interaction, which can be tuned
in various ways experimentally.

\begin{figure}[t]
\begin{center}
\includegraphics[width=\columnwidth]{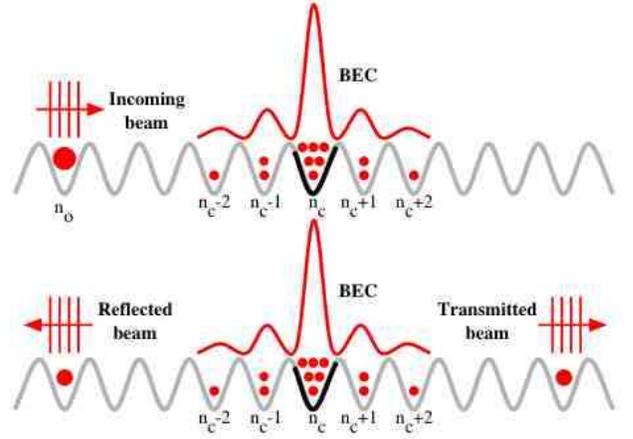}
\end{center}
\caption{(Color online) Scattering scheme in an optical lattice.
The incoming, reflected, and transmitted beams of atoms are
represented as plane waves. The atoms interact only around $n=n_c$,
where the BEC is centered. From~\textcite{ravjbsf:PRL:07}. } \label{BEC_fig1}
\end{figure}

\textcite{ravjbsf:PRL:07} 
considered a BEC on a lattice, where interactions between
atoms are present only in a very localized region (see
Fig.~\ref{BEC_fig1}). Such a situation could be realized experimentally
by combining optical lattices with atom-chip technology
\cite{whphtwhjr:n:01,hjfgsagcz:prl:01} or in optical micro-lens arrays~\cite{rdtmmvwegb:prl:02}. 
The system is described by the discrete nonlinear Schr\"odinger
(DNLS) equation, a classical variant of the Bose-Hubbard
model appropriate for a BEC in a periodic potential in the
tight binding limit~\cite{ommo:rmp:06}. With interactions being present only
on site number $n_c$, it follows
\begin{equation}
\label{BEC_eq1}
i\frac{d\Psi_n}{dt} = - (\Psi_{n+1} + \Psi_{n-1}) - \gamma |\Psi_{n_c}|^2\Psi_{n_c} \delta_{n, n_c},
\end{equation}
where $\Psi_n$ is the complex amplitude of the condensate field at site
n and $-\gamma =U/J$ is the interaction strength on site $n_c$,
where $J$  is the tunneling energy between the lattice cites and $U$ is
on-site interaction energy per atom. 

Equations (\ref{BEC_eq1}) support a localized state $\Psi_n(t) = b x^{|n-n_c|} \exp(-iE_bt)$, where
$x=-\frac{1}{2}(E_b + g)$ with $g=\gamma b^2$, $b$ is
the condensate amplitude and $E_b=-(4+ g^2)^{1/2}$ is the chemical potential.

\begin{figure}[t]
\begin{center}
\includegraphics[width=\columnwidth]{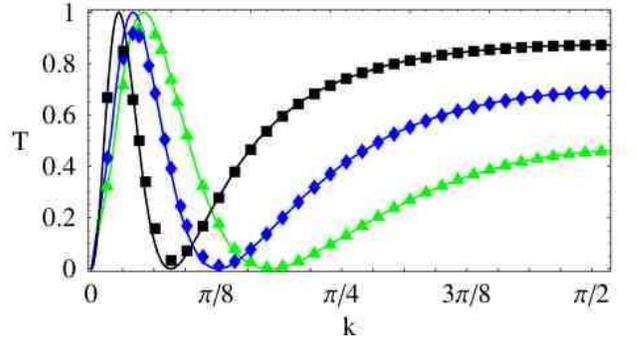}
\end{center}
\caption{(Color online). Transmission T versus momentum k.
Lines: analytic solution, symbols: real time numerical simulations of
Eq.~(\ref{BEC_eq1}) using wave packets for $g=0.36$ (line and boxes),
$g =0.6$ (line and diamonds), and $g=0.9$ (line and triangles).
From~\textcite{ravjbsf:PRL:07}. } \label{BEC_fig2}
\end{figure}

The scattering of propagating atomic matter waves with the energy $E_k = -2\cos k$
by this localized BEC were calculated analytically within the framework of
the Bogolyubov-DeGenne equations \cite{ravjbsf:PRL:07},
The transmission $T(k)$ is shown in Fig.~\ref{BEC_fig2}
for three values of $g$ (solid curves). As $g$ increases, the width and the
position of the resonance increase. Furthermore, the more
localized the BEC becomes, the stronger it reflects the
atom beam off resonance. By tuning the nonlinear parameter $g$, we can thus choose the amount
of the beam which passes through the BEC. Off resonance
(for larger values of $k$), we can select the percentage of the
incoming beam that is transmitted for a defined quasi-momentum.
Therefore, the actual setup can be used as a 100\%
blockade or as a selective filter.

The analytical results have been confirmed by numerical simulation of
Eq. (\ref{BEC_eq1}) with a Gaussian atom beam profile. The results are
shown in Fig.~\ref{BEC_fig2} by the symbols for three different
values of the parameter $g$. The agreement between theory and simulations is
very good.


\section{Light propagation in photonic devices}


Optical microcavity structures are of great interest
for device applications, and many of these structures involve coupling of
one or several cavities to a waveguide. Such waveguide-cavity systems
can naturally exhibit Fano resonances with high quality factors, and they can be used
for optical modulations and switching. The on/off switching functionality
can be realized by shifting the resonant frequency either toward
or away from the signal frequency.

\begin{figure}[t]
\begin{center}
\includegraphics[width=\columnwidth]{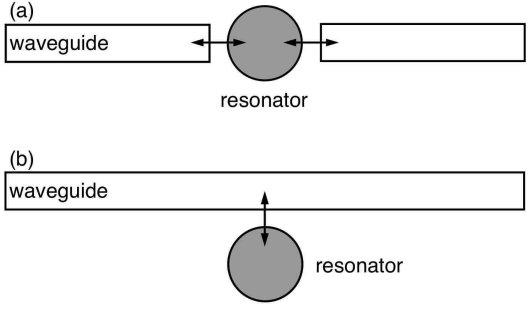}
\end{center}
\caption{Schematic setup for (a) a waveguide directly coupled to a
cavity and (b) a waveguide side-coupled to a cavity.}
\label{ph_fig1}
\end{figure}

The basic geometry of a waveguide-cavity system which demonstrates a sharp Fano resonance has been introduced
and analyzed in Refs.~\cite{hahy:jlt:91,yyrkay:pre:00}. It consists of a waveguide coupled to a cavity (or resonator).
In general, two-port photonic devices based upon waveguide-resonator interaction can be presented in two geometries, as 
shown in Figs.~\ref{ph_fig1}(a,b). The first configuration is based on a direct-coupling geometry~\cite{msmisgjyfjdj:PRE:02}, 
and the second geometry is a waveguide side coupled with a single-mode cavity~\cite{yyrkay:pre:00,mfysfms:APL:03}.
Such structures are tunable by adding cavities with nonlinear response
or by employing an external control. Below, we review the basic properties
of the simplest waveguide-cavity systems, and discuss several
generalizations including all-optical switching structures based on
the concepts of Fano resonances.

\subsection{Green's function formalism}

The Green's function approach~\cite{sfmysk:OL:02,sfmysk:JOSAB:02} allows to obtain
very accurate results in comparison to the time-consuming direct numerical finite-difference time-domain (FDTD)
simulations, even for rather complex geometries of the waveguide-cavity systems.
To derive the corresponding equations, one takes the explicit temporal
dependencies into account which allow one to study the pulse propagation and scattering.

We consider a photonic crystal created by a periodic square lattice of
infinite cylindrical rods parallel to the $z$ axis. We neglect the material dispersion
and assume the dielectric constant $\epsilon(\vec{r})$ to be periodic in two transverse
directions, $\vec{r}=(x,y)$. The evolution of the $E$-polarized electric field propagating
in the $(x,y)$ plane is governed by the scalar wave equation
\begin{eqnarray}\label{eq:eq1}
\nabla^{2}E_{z}(\vec{r},\tau)-\frac{1}{c^{2}}\partial_{\tau}^{2}\left[
\epsilon(\vec{r})E_{z}(\vec{r},\tau)\right]  =0\;,
\end{eqnarray}
where $\nabla^{2}=\partial_{x}^{2}+\partial_{y}^{2}$. We assume that the light field
propagating in such structures can be separated into fast and slow components,
$E_{z}(\vec{r},\tau)=e^{-i\omega\tau}E(\vec{r},\tau|\omega)$, where $E(\vec{r},\tau|\omega)$ is a
slowly varying envelope of the electric field, i.e. $\partial_{\tau}^{2}%
E(\vec{r},\tau|\omega)\ll\omega\partial_{t}E(\vec{r},\tau|\omega)$.
This allows to simplify Eq.~(\ref{eq:eq1}) to the following form
\begin{eqnarray}\label{eq:eq2}
\left[  \nabla^{2}+\epsilon(\vec{r})\left(  \frac{\omega}{c}\right)
^{2}\right]  E(\vec{r},\tau|\omega)\simeq-2i\epsilon(\vec{r})\frac{\omega
}{c^{2}}\frac{\partial E(\vec{r},\tau|\omega)}{\partial\tau}\;.
\end{eqnarray}

Both the straight waveguide and the side-coupled cavity are created by introducing
defect rods into a perfect two-dimensional periodic structure.
Therefore, the dielectric constant can be
represented as a sum of two components, describing the
periodic and defect structures $\epsilon(\vec{r})=\epsilon_{\mathrm{pc}}+\delta\epsilon$.
We employ the  Green's function of the two-dimensional periodic structure without defects,
and rewrite Eq.~(\ref{eq:eq2}) in the integral form
\begin{eqnarray}\label{eq:eq4}
E(\mathbf{x},\tau|\omega)=\int d^{2}\mathbf{y}G(\mathbf{x},\mathbf{y}%
|\omega)\hat{L}E(\mathbf{y},\tau,\omega)\;,
\end{eqnarray}
where we introduce the linear operator
\begin{eqnarray}\label{eq:eq5}
\hat{L}=\left(  \frac{\omega}{c}\right)  ^{2}\delta\epsilon(\vec
{r})+2i\epsilon(\vec{r})\frac{\omega}{c^{2}}\frac{\partial}{\partial\tau}\;,
\end{eqnarray}
and consider the time evolution of the slowly varying envelope
as a perturbation to the steady state.

The defect rods introduced into the periodic structure can formally be
described as follows:
\begin{eqnarray}\label{eq:eq6}
\delta\epsilon(\vec{r})=\sum\limits_{n,m}\left[  {\delta\epsilon^{(0)}_{m,n}+\chi^{(3)}%
|E(\mathbf{x},\tau|\omega)|^{2}}\right]  \theta
(\mathbf{x}-\mathbf{x}_{n,m})\;,
\end{eqnarray}
where we use the $\theta$-function to
describe the position of a defect rod at site $n,m$, with  $\theta
(\mathbf{x})=1$ for $\mathbf{x}$ inside the defect rods, and $\theta(\mathbf{x})=0$
otherwise. $\delta\epsilon^{(0)}_{m,n}$ is the variation of the dielectric constant of
the defect rod  $(m,n)$. Importantly, this approach allows us to incorporate a
nonlinear response in a straightforward manner, which is
assumed to be of the Kerr type being described by the term $\chi^{(3)}|E|^{2}$.

Substituting Eq.~(\ref{eq:eq6}) into the integral equation (\ref{eq:eq4}) and
assuming that the electric field does not change inside the dielectric rods, we can
evaluate the integral at the right hand side of Eq.~(\ref{eq:eq4}) and derive
a set of \textit{discrete nonlinear equations}
\begin{widetext}
\begin{eqnarray}\label{eq:eq7}
i\sigma\frac{\partial}{\partial\tau}E_{n,m}-E_{n,m}+\sum\limits_{k,l} J_{n-k,m-l}(\omega)\large(\delta\epsilon^{(0)}_{k,l}+\chi^{(3)}|E_{k,l}|^2\large) E_{k,l} =0,
\end{eqnarray}
\end{widetext}
for the amplitudes of the electric field $E_{n,m}(\tau|\omega)=E(\mathbf{x}
_{n,m},\tau|\omega)$ calculated at the defect rods. The parameters $\sigma$ and
$J_{k,l}(\omega)$ are determined by using the corresponding integrals of the Green's
function, where the whole information about the photonic crystal dispersion is now hidden
in their specific frequency dependencies, which can be found in Refs.~\cite{sfmysk:PRL:01,sfmaemyskkb:PRE:06}.
In this way, the Green's function needs to be calculated only once
for a given photonic structure, e.g. by employing the approach outlined in Ref.
\cite{ajwjbp:PRB:98}, and then it can be used to study any photonic circuit in that structure.

\begin{figure}[t]
\begin{center}
\includegraphics[width=\columnwidth]{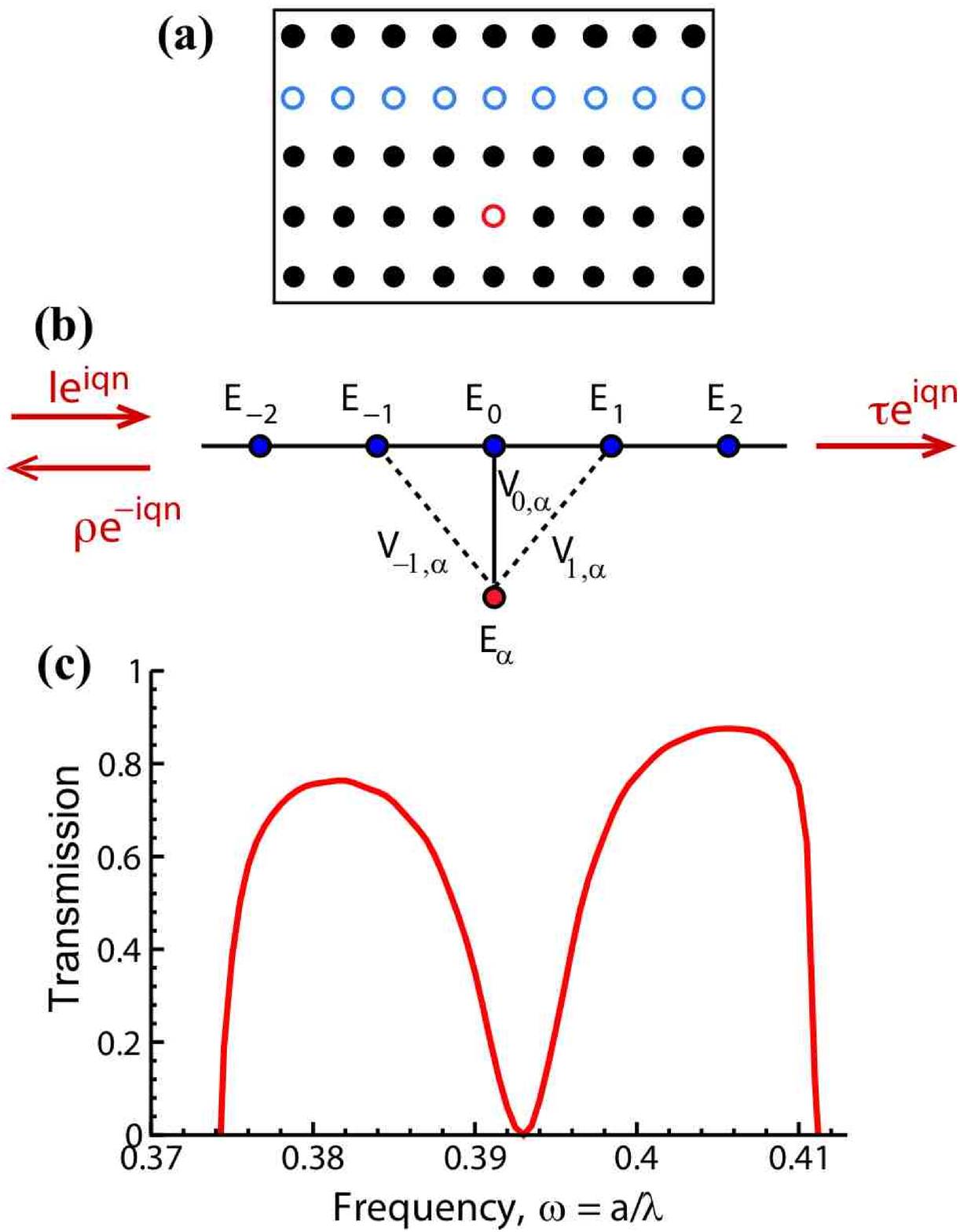}
\end{center}
\caption{Schematic view of (a) photonic crystal waveguide with an
isolated side-coupled cavity, and (b) effective discrete system.
(c) Typical profile of the Fano resonance.}
\label{ph_fig2}
\end{figure}

For the simple system when the photonic crystal has a waveguide side coupled
to a single defect see Fig.~\ref{ph_fig2}(a)], the problem describes a discrete
system studied earlier [see Fig.~\ref{ph_fig2}(b)], and the transmission
shows a Fano resonance [see Fig.~\ref{ph_fig2}(c)], analyzed in details in 
Refs.~\cite{sfmaemyskkb:PRE:06,aemsfmsfysk:PRE:05}.

In a general case, the effective interaction between defect rods
is of long-range nature~\cite{sfmyskras:PRE:00,sfmysk:JOSAB:02}.
However, the coupling strength decays exponentially with
the distance and, as a result, for coupled-resonators optical waveguides
the specific discrete arrays with nearest-neighbor interactions (at $L=1$) give
already an excellent agreement with direct FDTD simulations~\cite{sfmysk:JOSAB:02}.


\subsection{Defects in the waveguide}
The two basic geometries shown in Figs.~\ref{ph_fig1}(a,b) can be further improved by placing partially reflecting elements 
into the waveguides~\cite{sf:APL:02,akbdjovpad:PRB:03}. These elements allow creating sharp and asymmetric response 
line shapes. In such systems, the transmission coefficient can vary from 0\% to 100\% in a frequency range narrower than 
the full width of the resonance itself.

\begin{figure}[t]
\begin{center}
\includegraphics[width=\columnwidth]{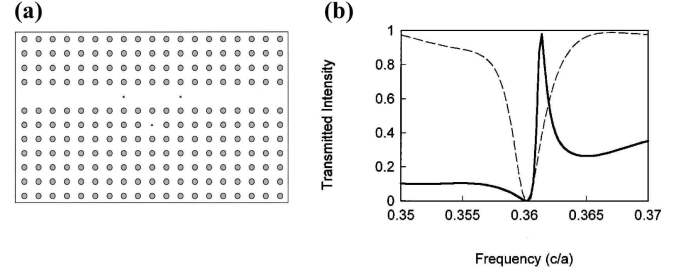}
\end{center}
\caption{
Light propagation in a photonic crystal waveguide with a side-coupled cavity.
(a) Photonic crystal waveguide formed by removing a single row of rods.
Within the line defect there are two smaller rods. A point defect, created by reducing
the radius of a single rod, is placed away from the waveguide.
(b) Transmission spectra through the structure (a) with (solid) and without (dashed)
the two defects in the waveguide. From ~\textcite{sf:APL:02}.}
\label{ph_fig3}
\end{figure}

To illustrate the effect of defects, Fan~\cite{sf:APL:02} simulated the response
of the structure shown in Fig.~\ref{ph_fig3}(a) using a FDTD scheme with perfectly
matched layer boundary conditions. A pulse is excited by a monopole source at one
end of the waveguide. The transmission coefficient is then
calculated by Fourier transforming the amplitude of the
fields at the other end, and is shown as a solid line in Fig.~\ref{ph_fig3}(b).
In comparison,  the transmission spectra for the same structure, but without
the two small cylinders in the waveguide, is shown by a dashed line.

Importantly, no detailed tuning of either the resonant frequency
or the coupling between the cavity and the waveguide is required
to achieve asymmetric line shapes. Also, since the reflectivity of the partially
reflecting elements need not to be large, the underlying physics
here differs from typical coupled-cavity systems, and resembles
Fano resonances involving interference
between a continuum and a discrete level.

\subsection{Sharp bends}
One of the most fascinating properties of photonic crystals is their ability to guide electromagnetic
waves in narrow waveguides created by a sequence of line defects, including light propagation through
extremely sharp waveguide bends with nearly perfect power transmission~\cite{amjcciksprvjdj:PRL:96,syecvhprvjdj:s:98}. It is believed
that the low-loss transmission through sharp waveguide bends in photonic crystals is one of
the most promising approaches to combine several devices inside a compact nanoscale optical
chip.

Interestingly, the transmission through sharp bends in photonic crystal waveguides
can be reduced to a simple model with Fano resonances, where the waveguide bend
hosts a specific localized defect. ~\textcite{aemysk:OE:05}
derived effective discrete equations for two types of the waveguide
bends in two-dimensional photonic crystals and obtained exact analytical
solutions for the resonant transmission and reflection. 

\subsection{Add-drop filters}
\begin{figure}[t]
\begin{center}
\includegraphics[width=\columnwidth]{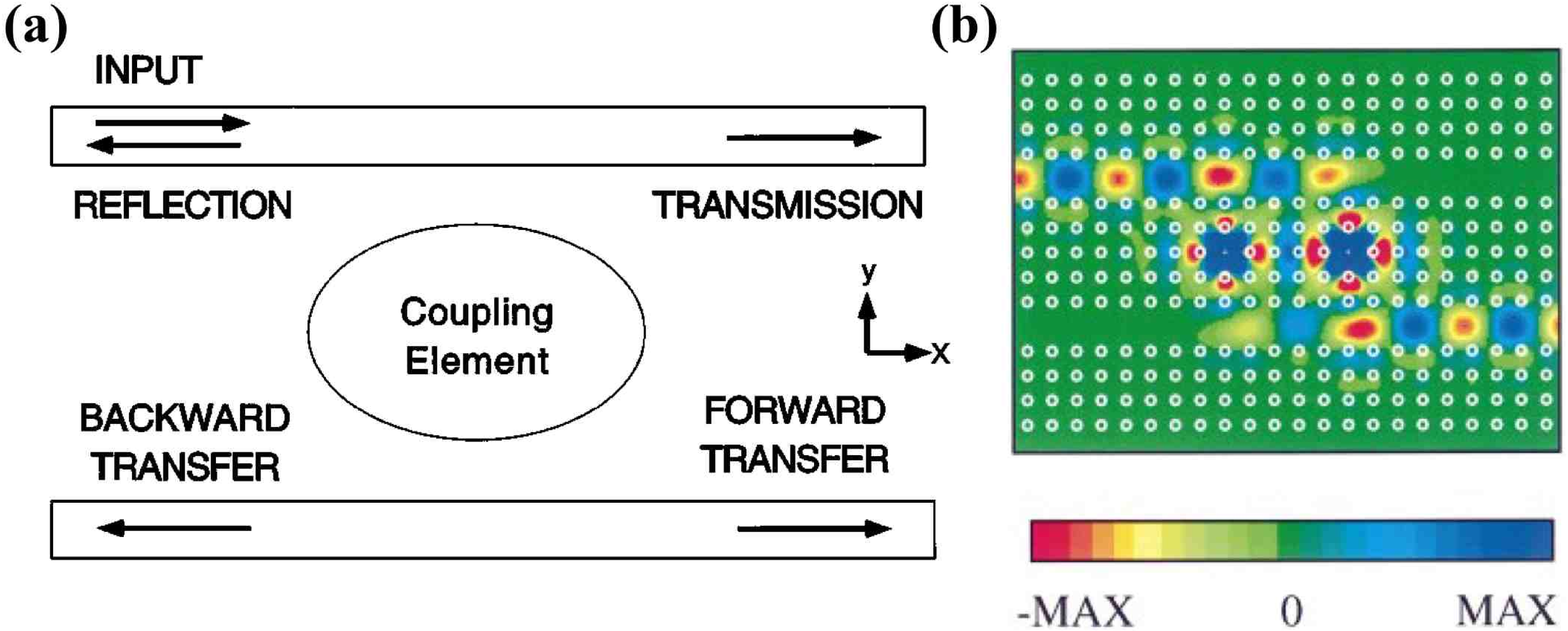}
\end{center}
\caption{
(Color online) Add-drop filter.
(a) Schematic diagram of two waveguides coupled
through an element which supports a localized resonant state.
(b) Electric field pattern of the photonic crystal
at the resonant frequency. The white circles indicate the
position of the rods. From ~\textcite{sfprvjdjhah:PRL:98}.}
\label{ph_fig4}
\end{figure}

Fano resonances can be employed for
a variety of photonic devices based on resonant
tunneling. In particular, if two waveguides interact through a coupling element
which supports a localized mode, a channel add-drop filter can be realized
via the resonant tunneling between the waveguides~\cite{sfprvjdjhah:PRL:98,sfprvjdjmjkcmhah:PRB:99,mscljdjsf:OL:03}.
The schematic diagram of a generic coupled system of this kind is shown in Fig.~\ref{ph_fig4}(a).
At Fano resonance, the propagating state excites the resonant modes,
which in turn decay into both waveguides. The transmitted signal in the first waveguide
is made up of the directly propagating signal and the signal which originates from
the second path which visits the coupling region. In order to achieve complete transfer
from one waveguide to the other one,
these two signal components must interfere destructively. The reflected
amplitude, on the other hand, originates entirely from the second path into the coupling region.
Hence, at least two states in the coupling region are needed
to achieve also destructive interference of backscattered waves in the first waveguide.
With these conditions satisfied, one may resonantly transfer the excitation from the first into the second
waveguide.

This concept was developed by ~\textcite{sfprvjdjhah:PRL:98}
for the propagation of electromagnetic waves in a two-dimensional photonic crystal.
To realize this concept, they used two photonic crystal waveguides and two
coupled single-mode high-$Q$ cavities, as shown in Figure~\ref{ph_fig4}(b).
The photonic crystal is made of a square lattice of
high-index dielectric rods, and the waveguides
are formed by removing two rows of dielectric rods.
The cavities are introduced between the waveguides by
reducing the radius of two rods. 
The resonant states have different symmetry.
An accidental degeneracy, caused by an exact cancellation between the two
coupling mechanisms, is enforced by reducing the dielectric
constant of four specific rods in the photonic crystal. The
cancellation could equally have been accomplished by reducing the size of the rods
instead of their dielectric constant.

Figure~\ref{ph_fig4}(b) shows the field pattern 
at resonance.
The quality factor
is larger than $10^3$. The backward transferred signal is almost
completely absent over the entire frequency range.

This type of four-port photonic crystal systems can be employed for
optical bistability, being particularly suitable for integration with
other active devices on a chip~\cite{mscljdjsf:OL:03}. A similar concept can be employed for the realization
of all-optical switching action in a nonlinear photonic crystal
cross-waveguide geometry with instantaneous Kerr nonlinearity.
There the transmission of a signal can be reversibly switched on and
off by a control input~\cite{mfysfmsjdj:OL:03}.

\subsection{All-optical switching and bistability}

\begin{figure}[t]
\begin{center}
\includegraphics[width=\columnwidth]{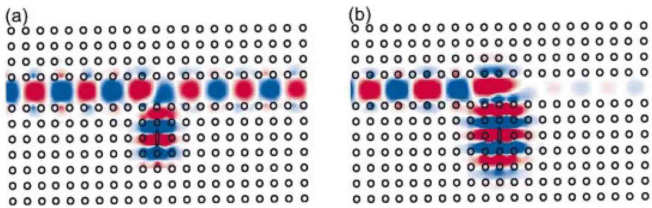}
\end{center}
\caption{
(Color online)
Electric field distributions in a photonic crystal for
(a) high and (b) low transmission states. Red and blue colors represent large positive or negative
electric fields, respectively. The same color scale is used for both panels.
The black circles indicate the positions of the dielectric rods. From ~\textcite{mfysfms:APL:03}.}
\label{ph_fig5}
\end{figure}

A powerful principle that could be explored to implement
all-optical transistors, switches, and logical gates is
based on the concept of optical bistability. The use of photonic crystals
enables the system to be of a size of the order of the wavelength of light, consume
only a few milliwatts of power, and have a recovery and
response time smaller than 1 ps.  Several theoretical and experimental studies explored 
nonlinear Fano resonances for designing optimal bistable switching in
nonlinear photonic crystals
~\cite{sfmysk:JOSAB:02,msmisgjyfjdj:PRE:02,mfysfms:APL:03,arcjfy:PRE:03,sfmaemyskkb:PRE:06,sfmaemysk:oe:07,bmpbrb:oe:08}.
A photonic crystal provides an optimal control over the input and output
and facilitates further large-scale optical integration.

The main idea of using the Fano resonance for all-optical switching and
bistability is quite simple: One should introduce an element with nonlinear response
and achieve nonlinearity-induced shifts of the resonant frequency, as was
discussed above for discrete models. Thus, by employing
{\em nonlinear Fano resonances} we can achieve bistability in many of the device
structures suggested on the photonic-crystal platform.  For example,
for the side-coupled geometry shown in Fig.~\ref{ph_fig1}(b), one could take
advantage of the interference between the propagating wave inside the
waveguide and the decaying wave from the cavity, to greatly
enhance achievable contrast ratio in the transmission between
the two bistable states. This approach was realized by ~\textcite{mfysfms:APL:03}
who demonstrated that such a configuration can generate extremely high contrast between
the bistable states in its transmission with low input
power.

One of the great advantages in using nonlinear photonic-crystal cavities
is the enhancement of nonlinear optical processes, including nonlinear Fano
resonance~\cite{msjdj:NM:04,jbaarpbsgjjdjms:OE:07}. Such an enhancement can be very efficient in the regime
of the slow-light propagation, that was demonstrated experimentally
with the smallest achieved group velocity $c/1000$~
\cite{mnkyasjtctiy:prl:01,rjallfcpbzgmjfpb:oe:05,yavmohfhsjm:n:05,hgtjkrjpewbjpknfhtfklk:prl:05}.
Because of this success, the interest in slow-light applications
based on photonic-crystal waveguides is rapidly growing, and posing 
problems of a design of different types of functional
optical devices which would efficiently operate in the slow-light
regime.

\begin{figure}[t]
\begin{center}
\includegraphics[width=\columnwidth]{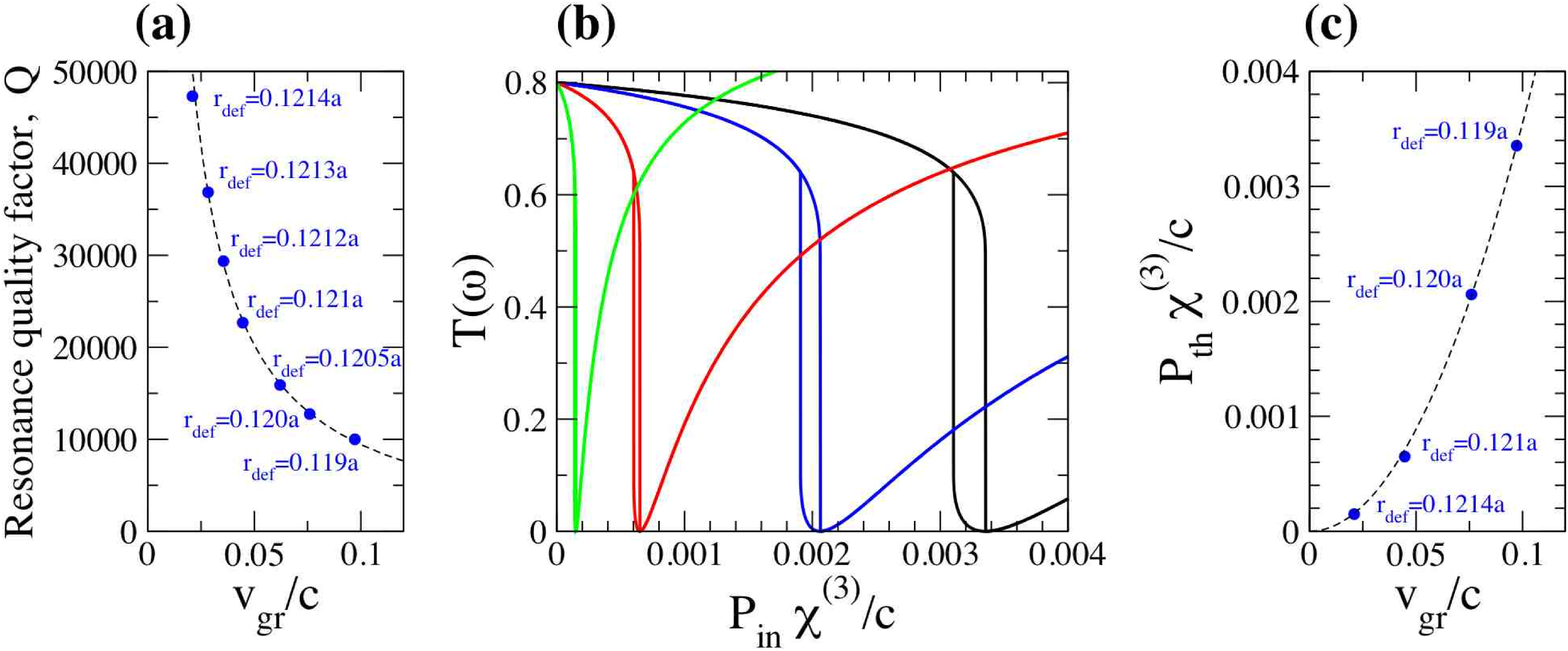}
\end{center}
\caption{
(Color online) Ultra-low all-optical switching in the slow-light regime. 
(a) Quality factor $Q$ vs. group velocity $v_{g}$ at resonance
for the waveguide-cavity structure. (b) Nonlinear bistable
transmission at the frequencies with 80\% of
linear light transmission vs. the incoming light power for different
values of the rod radius; (c) Switch-off bistability
threshold vs. the group velocity  at resonance. From ~\textcite{sfmaemysk:oe:07}.}
\label{ph_fig6}
\end{figure}

Recently,~\textcite{sfmaemysk:oe:07} have studied the resonant
transmission of light through a photonic-crystal waveguide coupled to a nonlinear cavity, and
demonstrated how to modify the structure geometry for achieving
bistability and all-optical switching at ultra-low powers in the
slow-light regime. This can be achieved by placing a side-coupled cavity
between two defects of a photonic-crystal waveguide assuming that
all the defect modes and the cavity mode have the same symmetry. In this structure
the quality factor  grows inversely proportional to the group velocity of light at the
resonant frequency  and, accordingly, the power threshold required for
all-optical switching vanishes as a square of the group velocity (see Fig.~\ref{ph_fig6}).

The numerically obtained dependence $Q(v_{gr})
\sim 1/v_{gr}$ is shown in Fig.~\ref{ph_fig6}(a), and it is in an
excellent agreement with the theoretical predictions. Since the
bistability threshold power of the incoming light in
waveguide-cavity structures scales as $P_{th} \sim 1/Q^2$
\cite{sfmaemyskkb:PRE:06}, one observes a rapid diminishing
of $P_{th} \sim v_{g}^2$ when the resonance frequency approaches the
band edge, as shown in numerical calculations summarized in
Figs.~\ref{ph_fig6}(b,c).

By now, several experimental observations of optical bistability enhanced through
Fano interferences have been reported~\cite{ewscar:APL:07,xychcwwmydlk:APL:07}. In particular, ~\textcite{xychcwwmydlk:APL:07}
employed a high-$Q$ cavity mode ($Q=30 000$) in a silicon photonic crystal and
demonstrated Fano resonance based bistable states and switching with 
thresholds of 185$\mu$W and 4.5 fJ internally stored cavity energy that might be useful
for scalable optical buffering and logic.

\begin{figure}[t]
\begin{center}
\includegraphics[width=\columnwidth]{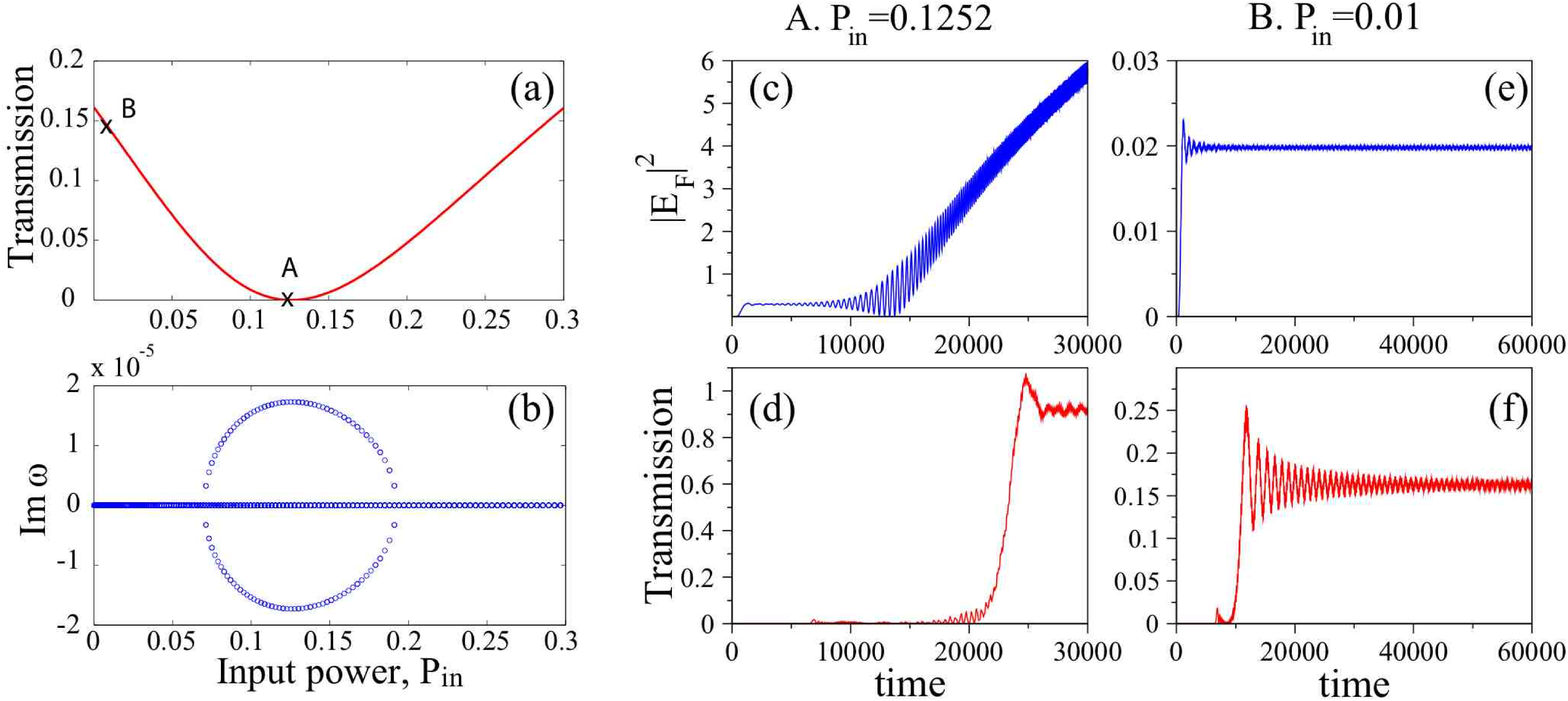}
\end{center}
\caption{
(Color online) 
Dynamical instability of the nonlinear Fano resonance. 
(a) Nonlinear transmission coefficient, and (b) imaginary part of eigenvalues of the stability problem vs input power. 
In the vicinity of  the nonlinear Fano resonance the plane wave excitation becomes dynamically unstable.
Temporal evolution of (c,e) the field inside the side-coupled cavity, and (d,f) the transmission coefficient for two different 
values of the input power values, indicated in plot (a). Near the resonance the dynamics of the field inside the nonlinear 
cavity yields a buildup of a modulation instability in time.
Adapted from ~\textcite{aemykcetprifl:pra:09}.}
\label{fig:Fano_dynamics}
\end{figure}

It is important to note, that the nonlinear Fano resonance shows dynamical instabilities with plane 
wave excitations~\cite{aemykcetprifl:pra:09}. Near the resonance the intensity of the scattered 
wave starts to grow in time, leading to modulational instability, while far from resonance it converges to a steady-state 
solution (see Fig.~\ref{fig:Fano_dynamics}). However, as it was demonstrated by \textcite{aemykcetprifl:pra:09} this 
instability can be suppressed for temporal Gaussian pulses excitations, providing with an effective method 
of recovering the bistable transmission.

\subsection{Overlapping resonances}

A very important effect associated with the Fano resonances
in double-resonator photonic structures can be linked to the electromagnetically-induced transparency
(EIT)~\cite{mfaijpm:RMP:05}. Coupled-resonator-induced transparency (CRIT) structures have been
introduced in 2004~\cite{ddshckafatrrwb:PRA:04,lmabmaasvsi:ol:04,wzws:ijqe:04},
although the early work~\cite{todw:pra:01} suggested already an idea of macroscopic
double-resonator optical system exhibiting the EIT-like effect. Recently, the CRIT effect
has been observed experimentally in the system of two interacting microresonators (glass
spheres of about 400 $\mu$m in diameter) with whispering-gallery modes~\cite{angfsisatr:PRA:05}, in a cavity with at 
least two resonant modes~\cite{jdfsmh:PRA:06},
and in integrated photonic chips with two microring
resonators~\cite{qxssmlpjssfml:PRL:06,mtktrhtm:josa:09}. Providing an
efficiently tunable transparency on an optical chip, such CRIT
devices are considered as a crucial step towards the development of
integrated all-optical chips \cite{rwbdjg:Nature:06}.

\begin{figure}[t]
\begin{center}
\includegraphics[width=\columnwidth]{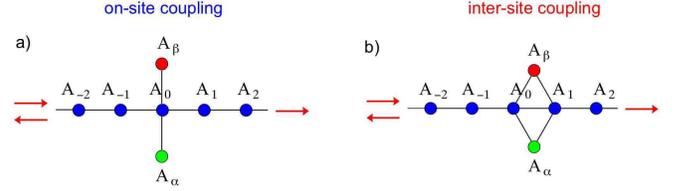}
\end{center}
\caption{
(Color online)
Two types of the geometries of a photonic-crystal waveguide side coupled
to two nonlinear optical resonators. Light transmission and bistability are qualitatively different
for (a) on-site and (b) inter-site locations of the resonator along the waveguide.
Adapted from ~\textcite{sfmaemysk:oe:08}.}
\label{ph_fig7}
\end{figure}

To explain the origin of CRIT resonances, we characterize
the light transmission by the transmission and reflection coefficients
which can be presented in the form
\begin{eqnarray}
\label{tr-Fano-abs} T(\omega)=
\frac{\sigma^{2}(\omega)}{\sigma^{2}(\omega) + 1} \quad \; , \quad
R(\omega)= \frac{1}{\sigma^{2}(\omega)+1} \;,
\end{eqnarray}
where the detuning function $\sigma(\omega)$ may have quite different
type of frequency dependence for different types of waveguide-cavity
structures. Zero transmission (total reflection) corresponds to the condition
$\sigma(\omega)=0$, while perfect transmission (zero reflection) corresponds to
the condition $\sigma(\omega)=\pm \infty$.

For the waveguide-cavity structure shown in Fig.~\ref{ph_fig1}(b), we obtain~\cite{sfmaemyskkb:PRE:06}
\begin{eqnarray}
\label{sigma-one-cavity-strip}
\sigma(\omega) \simeq
\frac{(\omega_{\alpha}-\omega)}{\gamma_{\alpha}} \; ,
\end{eqnarray}
where $\omega_{\alpha}$ is the eigenfrequency of the localized cavity
mode of an isolated cavity $\alpha$. The spectral width $\gamma_{\alpha}$
of the resonance is determined by the overlap integral between the cavity mode and the guided mode
at the resonant frequency.

To find $\sigma(\omega)$ for the two-cavity structure,
one can apply a variety of methods but the simplest approach is based on the transfer-matrix
technique~\cite{sf:APL:02}. When two cavities are separated by
the distance $d = 2\pi m/k(\omega_{t})$, where $k(\omega)$
is the waveguide's dispersion relation, $m$ is any
integer number, and the frequency $\omega_{t}$ is defined below,
and there is no direct coupling between
the cavities, we obtain
\begin{eqnarray}
\label{sigma-two-cavity-d0}
\sigma(\omega) \simeq
\frac{(\omega_{\alpha}-\omega)(\omega_{\beta}-\omega)}{\Gamma
(\omega_{t}-\omega)} \; ,
\end{eqnarray}
with the total resonance width $\Gamma = \gamma_{\alpha}+
\gamma_{\beta}$ and the frequency of perfect transmission
$ \omega_t = (\gamma_{\alpha}\omega_{\beta}+
\gamma_{\beta}\omega_{\alpha})(\gamma_{\alpha} + \gamma_{\beta})^{-1}$,
lying in between the two cavity frequencies, $\omega_{\alpha}$ and
$\omega_{\beta}$, of zero transmission.

\begin{figure}[t]
\begin{center}
\includegraphics[width=\columnwidth]{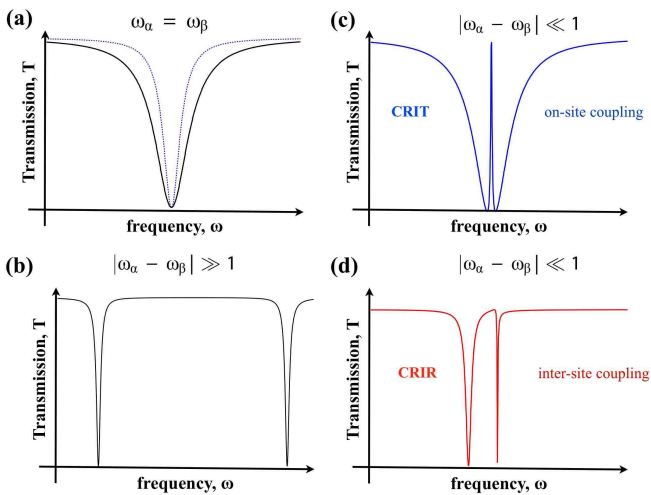}
\end{center}
\caption{(Color online) Typical transmission curves for four different cases (a) two identical side-coupled 
defects $\omega_\alpha=\omega_\beta$ (solid). Transmission for a single side-coupled cavity is shown by a dashed line; 
(b) two side-coupled cavities with strongly detuned eigenfrequencies $|\omega_\alpha-\omega_\beta|\gg1$; (c,d) 
two side-coupled cavities with slightly detuned eigenfrequencies $|\omega_\alpha-\omega_\beta|\ll1$ for (c) 
on-site coupling and (d) inter-site coupling. From ~\textcite{sfmaemysk:oe:08}.}
\label{ph_fig8}
\end{figure}

In the case when the cavities $\alpha$ and $\beta$ are identical,
we obtain a single-cavity resonance and
the only effect of using two cavities is the doubling of the spectral width,
$\Gamma=2\gamma_{\alpha}$, of the resonant reflection line,
as it is illustrated in Fig.~\ref{ph_fig8}(a). However,
introducing even the smallest difference between two cavities leads to the
opening of an extremely narrow resonant transmission line on the background of
this broader reflection line, as it is illustrated in Fig.~\ref{ph_fig8}(c). Indeed, for slightly
different cavities we may rewrite
Eq.~(\ref{sigma-two-cavity-d0}) in the vicinity of the resonant
transmission frequency, $\omega_t=\omega_{\alpha}+\delta\omega/2$, as
$\sigma(\omega) \approx \Gamma_t/(\omega-\omega_t)$, with the line
width  $\Gamma_t=\delta\omega^2/8\gamma_{\alpha}$,
which can easily be controlled by tuning the frequency difference $\delta\omega$.
The quality factor of this transmission line,
$Q_t=\omega_t/2\Gamma_t \approx 4\gamma_{\alpha}\omega_{\alpha}/\delta\omega^2$,
grows indefinitely when $\delta\omega$ vanishes.
As mentioned above, this effect is the all-optical analogue of the electromagnetically-induced transparency
and is now often referred to as the effect of coupled-resonator-induced transparency~\cite{ddshckafatrrwb:PRA:04}.

In contrast, the inter-coupling between two cavities, as shown in Fig.~\ref{ph_fig7}(b)
manifests itself as a qualitatively new effect of coupled-resonator-induced reflection (CRIR):
for small detuning $\delta\omega = \omega_{\beta} - \omega_{\alpha}$, one
of the resonant reflection frequencies shifts very close to the
perfect transmission frequency, $\omega_{t}$, producing a
narrow resonant reflection line, as is illustrated in
Fig.~\ref{ph_fig8}(d). The frequency of this line is always
close to the frequency $\omega_{\alpha}$ of the cavity mode, while
its spectral width is determined by the frequency difference $\delta\omega$,
growing indefinitely as $\delta\omega$ vanishes~\cite{lymsdmkc:OE:06,sfmaemysk:oe:08}.

It should be emphasized that despite such a qualitative difference in their spectral manifestations,
both CRIT and CRIR effects have the same physical origin which can be attributed to the Fano-Feshbach
resonances \cite{hf:AP:58,hf:AP:62,fhm:PR:68} which are known to
originate from the interaction of two or more resonances (e.g., two
Fano resonances) in the overlapping regime, where the spectral widths of resonances
are comparable to or larger than the frequency separation between them.
In a general situation it leads to a drastic deformation of the transmission
spectrum and the formation of additional resonances with sharp peaks.
The Fano-Feshbach resonances are associated with a collective response
of multiple interacting resonant degrees of freedom, and they have numerous
evidences in quantum mechanical systems~\cite{aimirsis:PRB:03,mrfhm:PRA:04}.

Finally, we discuss the interaction between two Fano resonances~\cite{kh:PRB:01,aem:pre:09} which can be employed to 
stop and store light coherently, with an all-optical adiabatic and reversible pulse bandwidth compression 
process~\cite{mfysf:PRL:04,mfywszwsf:PRL:04}. Such a process overcomes the fundamental bandwidth delay constraint in optics 
and can generate arbitrarily small group velocities for any light pulse with a given bandwidth, without any coherent or 
resonant light-matter interaction. The mechanism can be realized in a system consisting of a waveguide side coupled to 
tunable resonators, which generates a photonic band structure that represents a classical EIT 
analogue~\cite{mfywszwsf:PRL:04,bmpbrb:josa:05}.


\subsection{Guided resonances in photonic crystal slabs}

\begin{figure}[t]
\begin{center}
\includegraphics[width=\columnwidth]{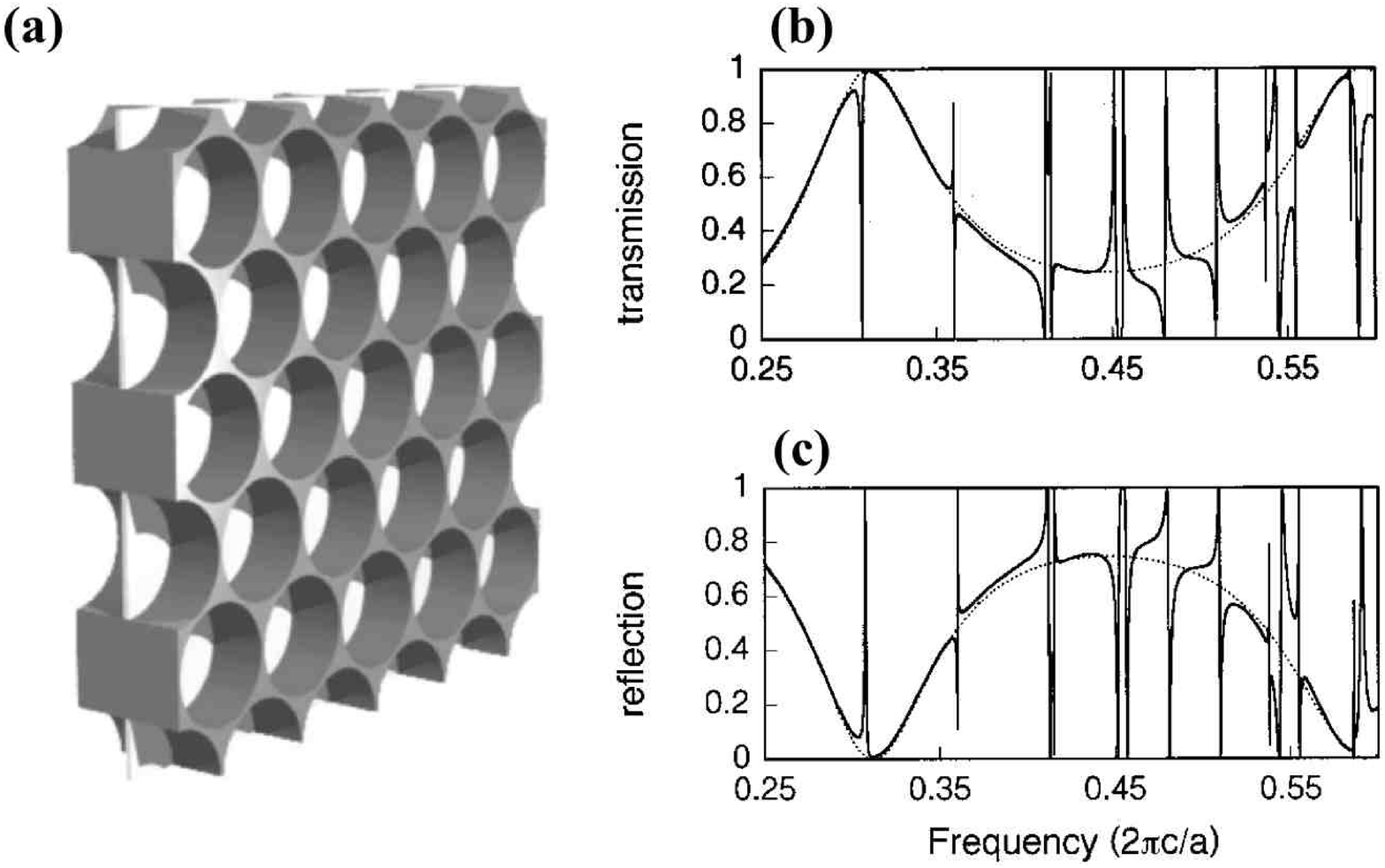}
\end{center}
\caption{
Light scattering by photonic crystal slabs.
(a) Geometry of the photonic-crystal film.
(b) Transmission and (c) reflection spectra. The solid
lines are for the photonic crystal structure, and the
dashed lines are for a uniform dielectric slab with a frequency-dependent
dielectric constant. Adapted from~\textcite{sfjdj:PRB:02}.}
\label{ph_fig9}
\end{figure}

Scattering of light by photonic crystal slabs leads to
another class of Fano resonances associated with the
presence of guided resonances in periodic structures.
A photonic crystal slab consists
of a two-dimensional periodic index contrast introduced
into a high-index guiding layer Fig.~\ref{ph_fig9}(a).  Such modulated
structures support in-plane guided modes that are completely confined
by the slab without any coupling to external radiations~\cite{rmssw:apl:92}.
In addition to in-plane waveguiding, the slabs can also interact
with external radiations in a complex and interesting way~\cite{sfjdj:PRB:02,kk:PRB:03,swjdj:josa:03}.
Of particular importance is the presence of guided resonances
in the structures.  Guided resonances can provide an efficient
way to channel light from within the slab to the external
environment. In addition, guided resonances can significantly
affect the transmission and reflection of external incident light,
resulting in complex resonant line shapes which can be linked
to Fano resonances.

\textcite{sfjdj:PRB:02} calculated the transmission
and reflection coefficients at
various $k$ points for the structure shown in Fig.~\ref{ph_fig9}(a).
The calculated spectra for s-polarized incident waves are shown
in Figs.~\ref{ph_fig9}(b,c). The spectra consist of sharp resonant
features superimposed upon a smoothly varying background.
The background resembles Fabry-Perot oscillations
when light interacts with a uniform dielectric slab. To
clearly see this, the background is fitted to the spectra of a
uniform slab, which are shown as dashed lines in Figs.~\ref{ph_fig9}(b,c). The
uniform slab has the same thickness as the photonic crystal.
Resonances can be described by employing the Fano-type formulas,
with the effective dielectric constant as the only fitting paramter. 
The fitting
agrees very well with the numerical simulations [see also ~\cite{kk:PRB:03}].

By introducing a nonlinear layer into the slab with a periodic lateral structure, we can generate
a bistable transmission for significant intensity ranges due to Fano resonances, and achieve
a strong frequency-dependent transparency variation related to the transfer via guided modes.
A self-consistent simulation tool which allows for the computation of multivalued transmission
has been developed by~\textcite{vljpv:PRB:04}. It explained the peculiar shape of the
hysteresis loops associated with nonlinear Fano resonances.

\begin{figure}
\includegraphics[width=\columnwidth]{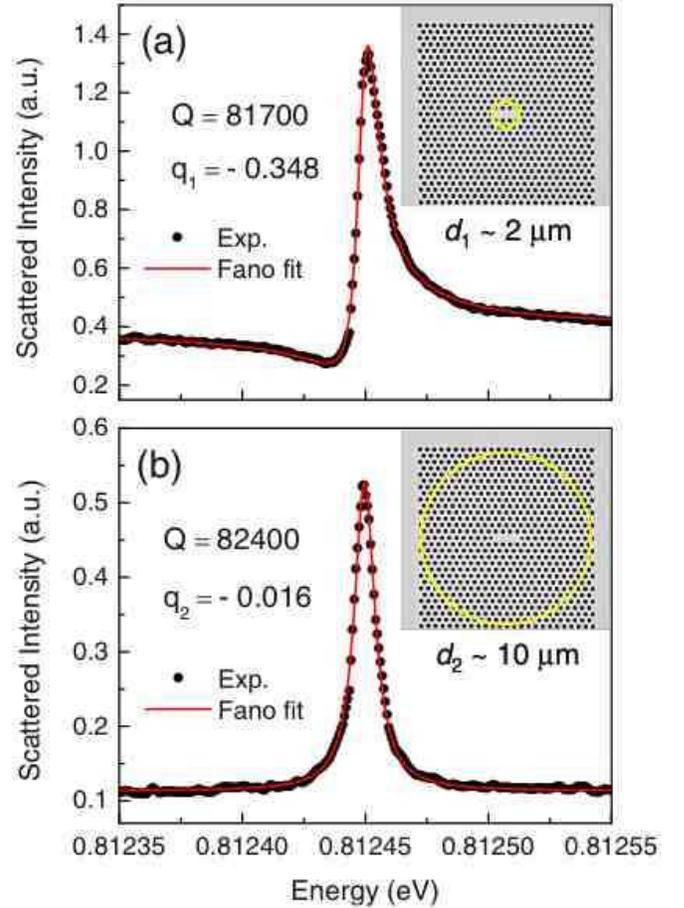}
\caption{\label{fig:PC_slab}
(Color online)
Measured scattering spectra (dots) and fitting by the Fano fromula (solid lines) of a photonic crystal nanocavity for two 
different excitation conditions: (a) a tightly focused, and (b) a slightly defocused laser beam of diameters $d_1$ and $d_2$, 
respectively, indicated by circles. Note here, that the actual profiles are inverted ones because of the use of crossed 
polarized detection.
From~\textcite{mgslpmblcalotfk:apl:09}. }
\end{figure}

Complex resonant line shapes due to Fano resonances were observed experimentally
in several settings~\cite{cgdfbldsmrmdjmmjsbje:OE:06,rhsjnmrfmdejahwdh:APL:07,zqhychpzwz:apl:08,hyhpzqzwz:el:08,czqhyhpzwz:oe:09}. 
In particular, \textcite{cgdfbldsmrmdjmmjsbje:OE:06} 
observed Fano resonances in the optical transmission spectrum of a
chalcogenide glass photonic crystal membrane and demonstrated, for the first time, 
the suppression of optical transmission by over 40 dB, the strongest reported so far, 
and a remarkable result for a dielectric structure with a thickness of only 330 nm. 
These results will allow further progress towards the engineering of very
sharp resonances and, combined with the large intrinsic nonlinearity of the chalcogenide
glasses, should allow for the observation of optical bistability in a photonic-crystal mirror.

Recently, it was experimentally demonstrated that the shape of the Fano resonance in the light scattering by a high-Q 
planar photonic crystal nanocavity can be controlled by varying the waste of the Gaussian beam~\cite{mgslpmblcalotfk:apl:09}. 
For a tightly focused beam with a spot diameter $d_1\approx2\mu$m a strong asymmetric Fano resonance was observed with the 
asymmetry parameter $q_1=-0.348$ [see Fig.~\ref{fig:PC_slab}(a)]. On the other hand, for a slightly defocused Gaussian beam 
with the spot diameter $d_2\approx10\mu$m a symmetric Fano resonance was observed with $q_2=-0.016$ [see Fig.~\ref{fig:PC_slab}(b)]. 
In this geometry the light reflected from the nanocavity mimics the scattering through a discrete level, while the light 
reflected from the photonic crystal pattern, can be considered as the scattering to the continuum. The interference of these 
two reflected components leads to the Fano resonance. The variation of the Fano profile with the increase of the 
excitation area can be understood as an enhancement of the scattering to the continuum, leading to the decrease of the asymmetry 
parameter $q$. Indeed, the variation of the asymmetry parameter $q_1/q_2\sim22$ is proportional to the variation of the excitation 
areas $(d_2/d_1)^2\sim25$.
Thus, by changing the excitation conditions it is possible to tune the Fano resonance in the scattering by photonic crystal nanocavity.

\subsection{Light scattering by spherical nanoparticles}

\begin{figure}[t]
\begin{center}
\includegraphics[width=\columnwidth]{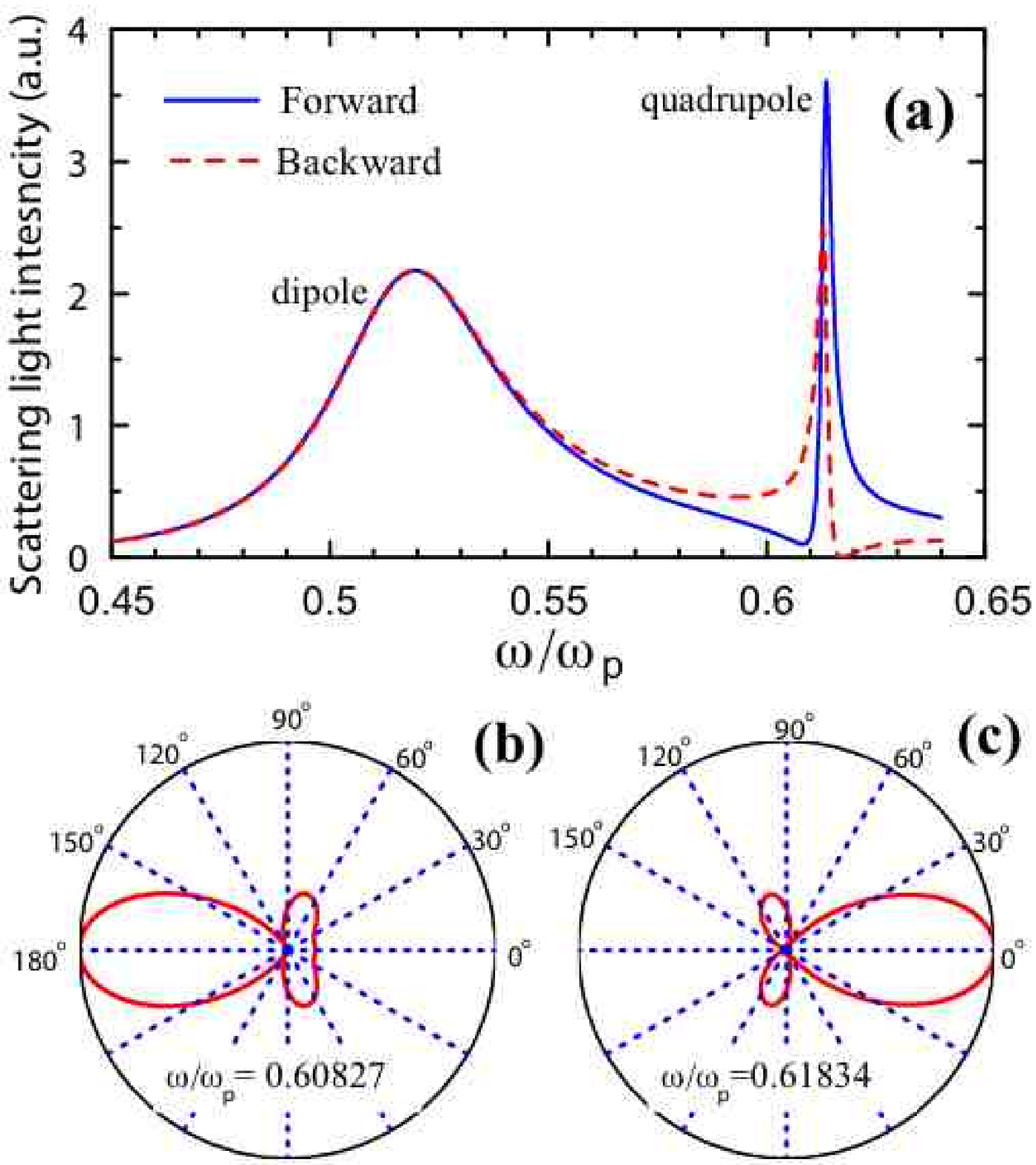}
\end{center}
\caption{(Color online) 
Exact Mie solution of the light scattering by a plasmonic nanoparticle.
The radius of the nanoparticle is much smaller than the light wavelength $a/\lambda=0.083$.
(a) Frequency dependence of the scattering light intensity in the vicinity of the dipole and quadrupole resonances. In the 
latter case both forward (solid lines) and backward (dashed lines) scattering profiles exhibit asymmetric Fano resonances.
(b,c) The angular dependence of the light scattering in the vicinity of the quadrupole resonance. The plasmonic frequency 
is normalized to $\omega_p a/c=1$.
Adapted from ~\textcite{bsltcclpmitzbwlccjrsjh:ipg:08}.}
\label{ph_fig10}
\end{figure}

Light scattering by an obstacle is one of the
fundamental problems of electrodynamics, see, e.g.,
monographs~\cite{Hulst81,BH,BW}. 
It was first described by Lord Rayleigh and is characterized by a sharp increase in scattering intensity with increasing the  
light frequency~\cite{lr:pm:71,lr:pm:71b,lr:pm:71c}. It is used to explain why we can enjoy the blue sky during day time (the intensely 
scattered blue component of the sunlight), and scarlet sunrises and sunsets at dawn and dusk (the weakly scattered red component). 
Lord Rayleigh's studies were generalized by Gustav Mie who obtained the complete analytical solution of Maxwell's equations 
for the scattering of electromagnetic radiation by a spherical particle valid for any ratio of diameter to the wavelength~\cite{gm:AP:08}.

A common assumption is that the general Mie solution transforms into that of Rayleigh when particles are small.
However, recent studies of resonant
scattering by small particles with weak dissipation rates~\cite{mbvfnz:oe:05,mitbsl:prl:06} 
have revealed new and unexpected features, namely giant optical 
resonances with an inverse hierarchy (the quadrupole resonance is much stronger than the dipole one, etc.) , a complicated near-field
structure with vortices, unusual frequency and size dependencies, which allow
to name such a scattering {anomalous}.  \textcite{mitsfaemavgysk:prl:08}
revealed that the physical picture of this anomalous scattering
is analogous to the physics of Fano resonances. This analogy sheds 
new light to the phenomenon. It allows to employ powerful methods
developed in the theory of the Fano resonances (such as, e.g., the
Feshbach-Fano partitioning theory) to describe the resonant light
scattering. It also easily explains  certain features of the
anomalous scattering and related problems, namely sharp changes in
the scattering diagrams upon
small changes in $\omega$  (see Fig.~\ref{ph_fig10}). 
\textcite{mitsfaemavgysk:prl:08} analytically obtained an asymmetric profile of the resonance lines by analyzing the exact 
Mie solution of the light scattering problem by a spherical nanoparticle~\cite{aemsfavgbslyskmit:opn:08}.

Figure ~\ref{ph_fig10} demonstrates light scattering by a potassium colloidal nanoparticle immersed in a KCl crystal, calculated 
with a realistic
dependence $\epsilon(\omega)$ and fitting actual experimental data~\cite{mitsfaemavgysk:prl:08,bsltcclpmitzbwlccjrsjh:ipg:08}.
A slight variation of the incident light frequency in the vicinity of the quadrupole resonance drastically changes the scattering 
pattern (see Fig.~\ref{ph_fig10}), resulting in asymmetric Fano-like profiles for intensities of the forward and backward scattered 
light. In this case, excited localized plasmons (polaritons) are equivalent to the discrete levels in Fano's approach, while 
the radiative decay of these excitations is similar to the tunneling to the continuum. In general, it may lead to a significant 
suppression of the scattering along any given direction.
Note, that in accordance with the theoretical expression obtained from the Mie formula, the points of destructive interference 
for the forward
and backward  scattering lie on different sides of
the corresponding resonant peaks.

\subsection{Plasmonic nanocavities and tunable Fano resonance}
Recent progress in the fabrication and visualization of nano-sized structures gave rise to a novel and rapidly 
emerging field of nanoplasmonics. The optical properties of metals are governed by coherent oscillations of conduction-band 
electrons, known as plasmons~\cite{bdpd:pr:51}. The interaction between light and metallic nanoparticles is mostly dominated 
by charge-density oscillations on the closed surfaces of the particles, called localized surface plasmon resonances (LSPs).
The studies of LSPs in nobel-metal nanoparticles, such as gold and silver, extended applications from various 
surface-enhanced spectroscopies~\cite{mm:rmp:85} to novel nanometer optical devices and waveguides~\cite{wlbadtwe:Nature:03,eo:s:06}.
One of the most important properties of LSPs is the possibility of strong spatial localization of the electron oscillations, 
combined with their high frequencies varying from UV to IR ranges. LSPs have the ability to strongly scatter, absorb, and squeeze 
light on nanometer scales, producing huge enhancement of electromagnetic field amplitudes.
Such unique properties of nanomaterials are essential for the development of novel material functions with potential technological 
and medical applications with specific optical, magnetic, and reactivity properties. 

\begin{figure}[t]
\begin{center}
\includegraphics[width=\columnwidth]{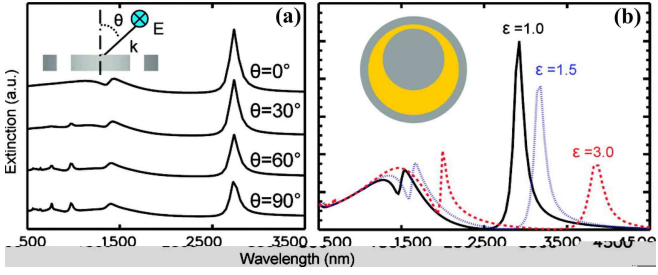}
\end{center}
\caption{(Color online)
A metallic nanostructure consisting of a disk inside a thin ring supports superradiant and very narrow subradiant modes. 
Symmetry breaking in this structure enables a coupling between plasmon modes of differing multipolar order, resulting in a 
tunable Fano resonance:
(a) extinction spectra as a function of incident angle $\theta$;
(b) the impact of filling the cavity with a dielectric material on the extinction spectrum or permittivity $\epsilon=1$ (solid line), 
$\epsilon=1.5$ (dashed line), and $\epsilon=3$ (dotted line).  Adapted from \textcite{fyspdsamnjhpn:nl:08}.
}
\label{Plasmon_fig1}
\end{figure}

Plasmonic nanostructures can be considered as a physical realization of coupled oscillator systems at the nanoscale. The energies 
and linewidths of the LSPs depend mostly on the nanoparticle geometries, such as size and shape. Thus, the spectral tunability 
of LSPs has been widely investigated. As it was suggested by ~\textcite{fpnmtbsam:prcmmp:07}, promising geometries for 
fine tuning are rings and disks. In such structures the dipole-like resonance can be tuned into the near-infrared 
region by changing the width of the metallic ring, for example.
One of the important issues of nanoplasmonics is the effect of symmetry breaking, which allows to excite higher-order multipolar 
modes leading to larger electromagnetic field enhancements. The symmetry breaking can be easily achieved in metallic ring/disk 
cavity structures by displacing the disk with respect to the center of the ring.
The plasmon resonances of a ring/disk cavity system can be understood in terms of the interaction or 
hybridization of the single ring and disk 
cavity plasmons.
This hybridization leads to a low energy symmetric plasmon and high energy anti-symmetric plasmon~\cite{fpnmtbsam:prcmmp:07}. 
The latter one is superradiant, i.e. it strongly radiates because disk and ring dipolar plasmons are aligned and oscillate in phase. 
The low energy  symmetric plasmon is subradiant because of opposite alignment of dipolar moments. It turns out that in a symmetry-broken 
structure, the quadrupole ring resonance couples to the 
superradiant high energy anti-symmetric disk-ring dipole mode ~\cite{fyspdsamnjhpn:nl:08}. The direct coupling interferes 
with the dispersive coupling between the quadrupolar ring mode and the supperradiant mode, resulting in a Fano resonance in the 
extinction spectrum (see Fig.~\ref{Plasmon_fig1}). By varying the incident angle, the shape of the Fano resonance can be altered 
from asymmetric to a symmetric one. 

Other examples of nanoplasmonic structures supporting the asymmetric Fano resonance are metallic nanoshells near a 
metallic film~\cite{fnzlnjhpn:prb:07}, and heterogeneous dimers composed of gold and silver nanoparticles~\cite{gbirebcjnffvpb:prl:08}. 
Both structures show up with a highly tunable plasmonic Fano resonance, accompanied by large local electric field enhancement. 
Thus, the strong response of LSP resonances may be effectively used for biological and medical sensing applications.

A novel type of nonlinear Fano resonance has been found in hybrid molecules composed of semiconductor and metal 
nanoparicles~\cite{wzaoggwb:PR:06}. The latter ones support surface plasmons with a continuous spectrum, while the former 
ones support discrete interband excitations. Plasmons and excitons become strongly coupled via F\"orster energy transfer. 
At high light intensities, the absorption spectrum demonstrates a sharp asymmetric profile, which originates from the coherent 
interparticle Coulomb interaction, and can be understood in terms of a nonlinear Fano resonance.

\subsection{Extraordinary transmission of light through metallic gratings}
The scattering by metallic gratings was the subject of extensive research for over a century. One of the important early 
achievements of the optics of metallic gratings was the discovery and understanding of Wood's 
anomalies~\cite{rww:PRSLA:02,lr:PRSA:07,rww:PR:35}. One type of anomaly is due to ethe xcitation of surface plasmon-polaritons 
propagating on the metallic surface. Another one is the diffractive anomaly, when a diffracted order becomes tangental 
to the plane of the grating. It is characterized by a rapid variation of the diffracted order intensity, corresponding to the 
onset or disappearance of a particular spectral order~\cite{rww:PR:35}. This resonant behaviour of the Wood's anomaly can be 
understood in therms of the coupling of the incoming waves with the surface-bound states of periodic 
arrays~\cite{uf:PR:36,uf:PR:37,uf:ap:38,uf:JOSA:41,ahaao:AO:65}. 
Thus, by considering a surface-bound state as a discrete level and scattered waves as a continuum, Wood's anomaly 
can be interpreted as a Fano resonance~\cite{cbscfpnbjp:oe:09}.

\begin{figure}[t]
\begin{center}
\includegraphics[width=\columnwidth]{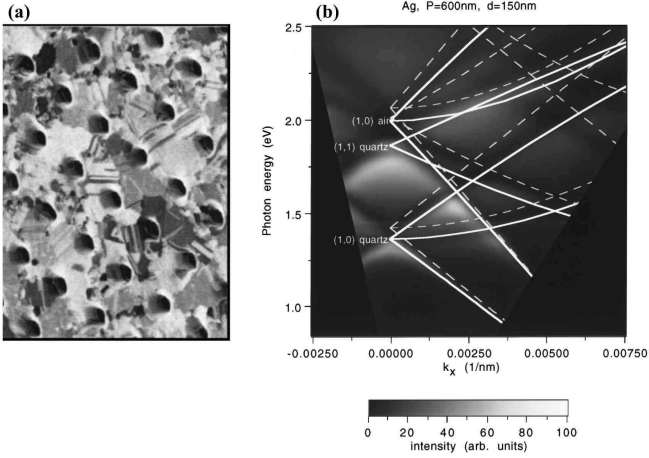}
\end{center}
\caption{
Light scattering by metallic gratings.
(a) Focused ion beam image of a two-dimensional hole array in a polycrystalline silver film.
(b) Observed transmission intensity as a function of photon energy and $k_x$ with predicted energy dispersion of surface 
plasmon-polaritons (solid) and loci of Wood's anomaly (dashed lines). From ~\textcite{hfgttdegtwhjl:PRB:98}.}
\label{fig:Wood1}
\end{figure}

It was demonstrated that for a periodic thin-film metallic grating, formed from a two-dimensional array of holes, the 
transmitted fraction of the incident light can exceed the open fraction of the array for certain 
wavelengths ~\cite{twehjkhfgttpaw:Nature:98,hfgttdegtwhjl:PRB:98}. The enhancement in the transmitted zero-order beam is 
reported to be several orders of magnitude larger than that from pure metallic slab without holes. This phenomenon has been 
called {extraordinary transmission} through periodic arrays of subwavelengths holes in metallic films.

The common understanding of the extraordinary transmission is due to a resonant excitation of surface plasmon-polaritons by 
incoming radiation~\cite{hfgttdegtwhjl:PRB:98,klmkjkksefbsnfhlk:PRB:05}. In addition to the resonant enhancement of the 
transmission the resonant suppression was observed as well.
It was demonstrated that these transmission minima correspond exactly to loci of Wood's anomaly 
(see Fig.~\ref{fig:Wood1})~\cite{hfgttdegtwhjl:PRB:98}.  
According to experimental observations, each extraordinary transmission is accompanied by resonant suppression of transmission resulting in 
asymmetric lineshapes, which can be perfectly fitted by the Fano formula~\cite{fjgda:RMP:07}. Moreover, it was theoretically 
demonstrated by~\textcite{issaynevbalavk:prcmmp:09} that periodically modulated ultrathin metal films may exhibit resonant 
suppression of the transmittance, emphasizing the Wood's anomaly effect. Thus, the extraordinary resonant scattering of light 
by modulated metal film can be described in terms of the Fano resonance, revealing the interference nature of the phenomenon.

\textcite{akarzkmgsad:apl:09} suggested to use active layers to simultaneously enhance both transmittance and reflectance at the 
resonance in subwavelength periodic planar bimetallic grating by exciting gain-assisted surface plasmons.

\subsection{Resonant four-wave mixing induced autoionization}

\begin{figure}[t]
\begin{center}
\includegraphics[width=0.8\columnwidth]{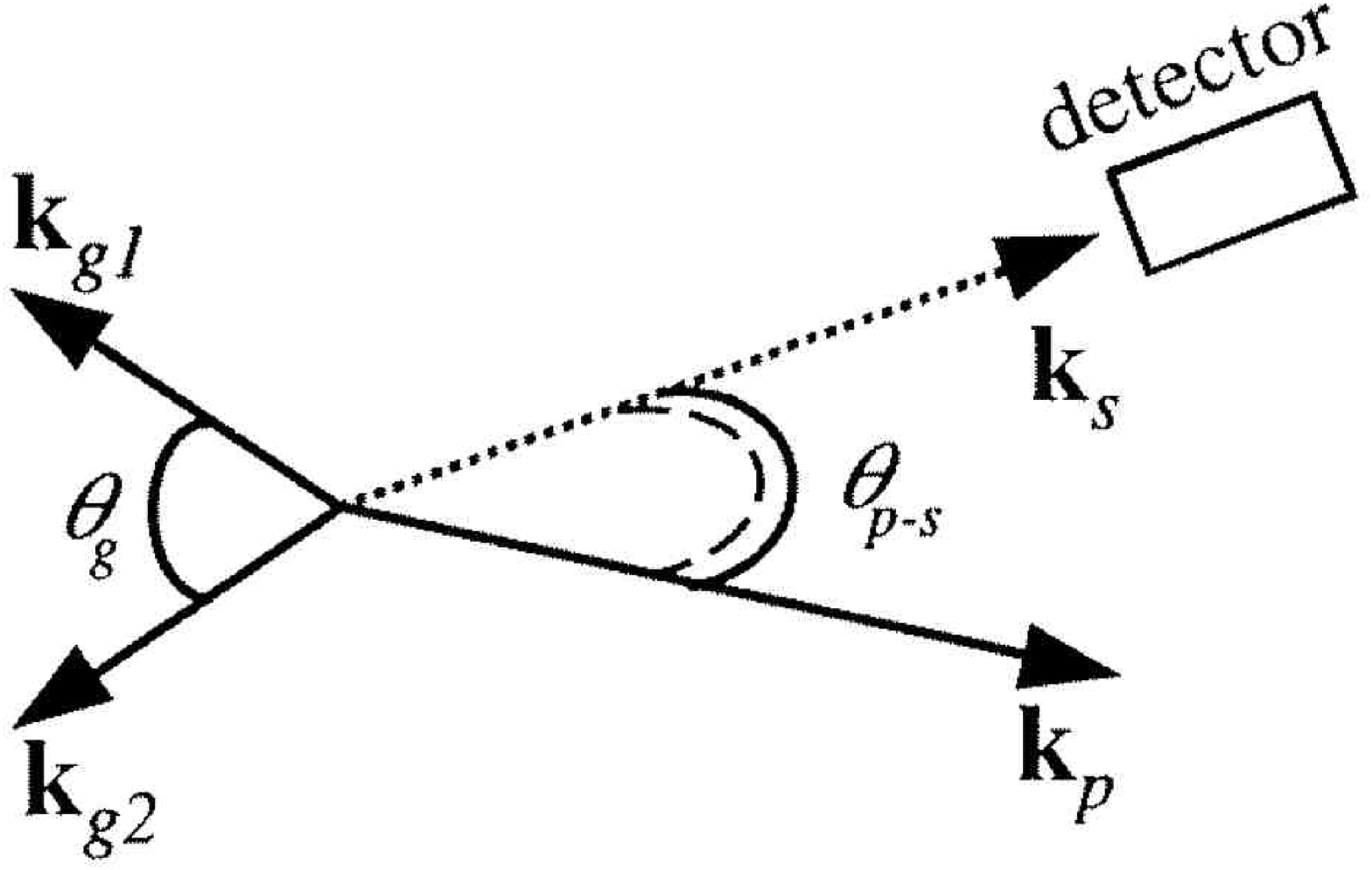}
\end{center}
\caption{Planar wavevector diagram illustrating the phase-matching condition for RFWM. From~\textcite{fdtefm:jpb:99}.}
\label{fig:RFWM1}
\end{figure}
Four-wave mixing involves the interaction of three laser beams to 
produce a nonlinear polarization via the cubic electric susceptibility $\chi^{(3)}$. The induced polarization acts as the 
source of a fourth coherent light beam, detected as the signal.
Four-wave mixing can be considered as the formation and scattering from laser-induced gratings. The grating is formed by two 
laser beams, called grating beams, with wavevector $\mathbf{k}_g$. The third probe beam with the wavevector $\mathbf{k}_p$ is 
then scattered off the laser-induced grating and produces the fourth scattered beam, which is detected as the four-wave mixing 
signal. Due to energy conservation, the frequency of the signal beam must be equal to the frequency of the probe beam 
$\omega_s\equiv\omega_p$. Momentum conservation results in a phase-matching condition for the signal wavevector 
\begin{eqnarray}
|\mathbf{k}_s|=|\mathbf{k}_{g_1}-\mathbf{k}_{g_2}+\mathbf{k}_p|=\omega_p/c\;,
\end{eqnarray}
and the Bragg-scattering angular condition 
\begin{eqnarray}
\frac{\omega_p}{\omega_g}=\frac{\sin(\theta_g/2)}{\sin(\theta_{p-s}/2)}\;,
\end{eqnarray}
where $\theta_g$ is the angle between two grating beams, and $\theta_{p-s}$ is the angle between the probe and signal beams 
(see Fig.~\ref{fig:RFWM1}) .

\begin{figure}[t]
\begin{center}
\includegraphics[width=\columnwidth]{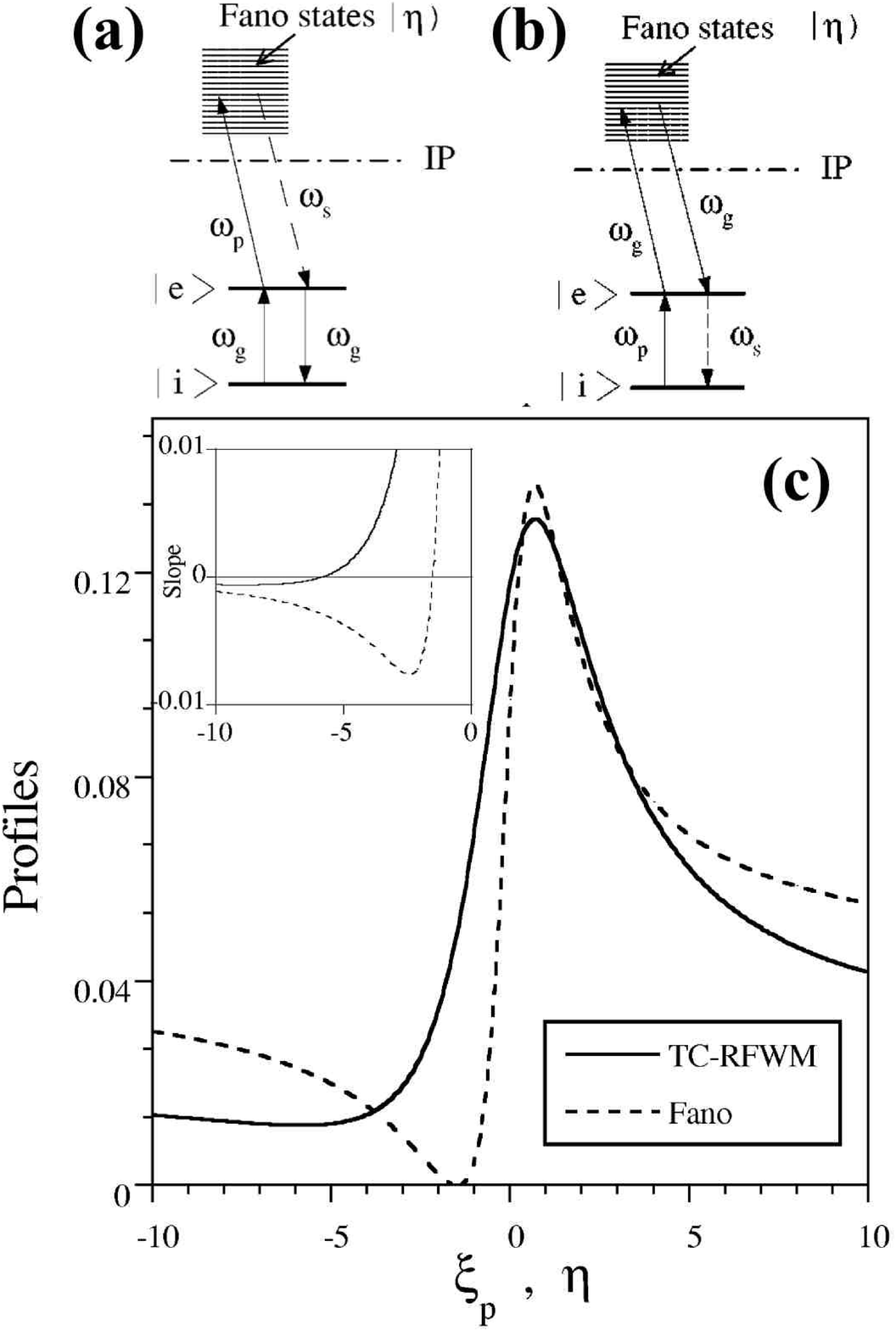}
\end{center}
\caption{Two-color resonant four-wave mixing.
(a) Nonparametric and (b) parametric TC-RFWM process.
The autoionizing level (Fano state) above the ionization potential is indicated by $|\eta\rangle$. 
$|i\rangle$ and $|e\rangle$ 
are ground and intermediate states, respectively.
(c) TC-RFWM and Fano profiles. Inset: the slopes of two profiles.
Note the separation between the slope zeros which correspond to profile minima. From~\textcite{fdtefm:PRA:98}.}
\label{fig:RFWM2}
\end{figure}

In general, the four-wave mixing process can take place in any material. 
When the frequency of the incident laser beams matches the transition resonances of the medium, a drastic enhancement of the 
signal intensity can be observed. Such processes are called resonant four-wave mixings (RFWMs), and they are used as spectroscopic and 
diagnostic tools for probing stable and transient molecular species. 
~\textcite{jaajjw:prl:74} studied experimentally four-wave mixing involving an autoionizing resonance in alkali-metal atomic 
vapor. In their experiment a two-photon transition between two bound states of the metal was excited, followed by a single-photon 
absorption to the autoionizing level. The detected signal demonstrated a characteristic asymmetric response. Using the Fano 
formalism, the authors derived an expression for the line-shape and fitted it with the Fano formula~\cite{jaajjw:prl:74}, 
which allows to obtain the width and asymmetry parameter for the autoionizing 
states~\cite{lablb:prl:75,mcla:jpb:82,mcla:jpb:82b,gapz:pra:83,gsapal:pra:83,slhgsa:pra:87,tmaspthvkm:PRB:95}. Thus, 
this form of RFWM can be considered as one of the techniques to study autoionizing levels.

A double resonance version of RFWM is called two-color RFWM (TC-RFWM) and takes place, 
when two optical fields have frequencies in resonance with 
two different transitions. It yields a variety of excitation schemes, which are very useful for high-resolution spectroscopy. 
In Figure~\ref{fig:RFWM2} possible TC-RFWM excitation schemes are shown, where the grating beams are in resonance with the 
lower transition and the probe is tuned to the upper transition [see Fig.~\ref{fig:RFWM2}(a)], and vice-versa 
[see Fig.~\ref{fig:RFWM2}(b)]. Because of the presence of autoionizing states in overall the FWM process, in both cases  TC-RFWM 
exhibits asymmetric profiles, which can be approximated by the Fano formula [see Fig.~\ref{fig:RFWM2}(c)]. 
Unlike the Fano profile, the TC-RFMW spectral lines have no exact zeroes. This
can be explained 
using the dephasing during nonlinear 
parametric conversions, which is a key difference to the usual Fano resonance case.
Nevertheless, TC-RFMW provides with an efficient way to coherently control the signal lineshape~\cite{efmftjmgstp:jcp:98}.

\section{Charge transport through quantum dots}


\begin{figure}
\includegraphics[width=\columnwidth]{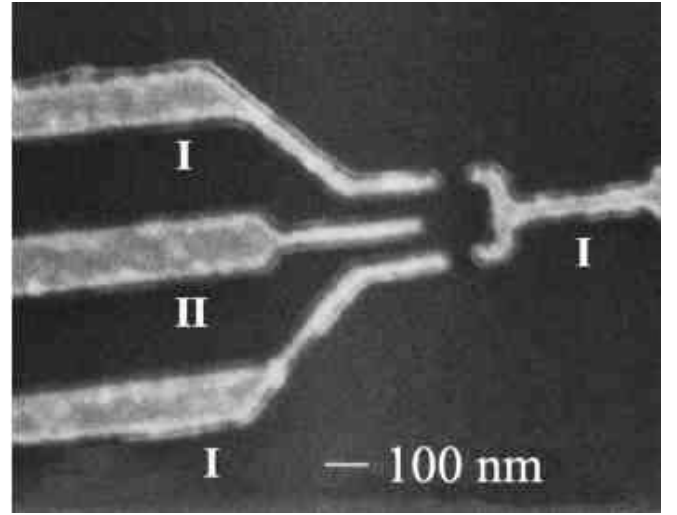}
\caption{\label{fig:kondo1}
Electron micrograph of a single electron transistor 
based on a GaAs/AlGaAs heterostructure.  The
split gates (I) define the tunnel barriers and the additional
gate electrode (II) adjusts the potential energy on the quantum dot.
From~\textcite{jgdggshmakhsdmum:PRB:00}.
}
\end{figure}

In recent decades charge transport through 
quantum dots (QD) has been extensively studied 
both theoretically and 
experimentally \cite{altshuler91,kma:rmp:92,koch92,rhlpkjrpstlmkv:rmp:07,smrmm:rmp:02}.
One of the reasons of that interest is the further miniaturization
of electronic device components.
A comprehensive picture of a big variety of
underlying physical phenomena has emerged 
(see e.g. \textcite{ilapwblig:pr:02,ya:rmp:00} and references therein).
The finite size of the dot is responsible for a dense but discrete
set of single particle levels.
Confinement of electrons in small quantum dots leads 
to the necessity of taking into account their Coulomb repulsion.
As a result, 
at temperatures below the charging energy 
the Coulomb blockade emerges \cite{ilapwblig:pr:02,ya:rmp:00}.
At even lower temperatures 
the phase coherence of the excitations in the quantum dot
is preserved during the scattering, and additional
interference phenomena appear, depending on the coupling strength
to the leads. 
In view of the enormous literature available, 
we will briefly introduce the main physics, 
and focus 
on results which are directly related to the finding of destructive
interferences and Fano resonances.

\subsection{From a single electron transistor to quantum interference}

A quantum dot is a small confinement region for electrons
(typically almost two-dimensional) with leads coupled to it.
The manufacturing of a huge variety of geometries is easily possible.
In the simplest case two leads are used, and a voltage $V$ is applied,
resulting in a current of electrons which enter the dot through one lead, and eventually
exit into the second lead. Various gate voltages can be additionally applied,
e.g. $V_g$ which controls the energy of the electrons in the dot relative to the leads,
and others which control the strength of the coupling between the leads and the dot.
Here we will consider only situations where the applied voltage $V$ between the two leads
is small so that the energy $eV$ is smaller than all other relevant 
energy scales. This is also called the equilibrium case,
at variance to the non-equilibrium case which is also frequently studied.

\begin{figure}
\includegraphics[width=0.7\columnwidth]{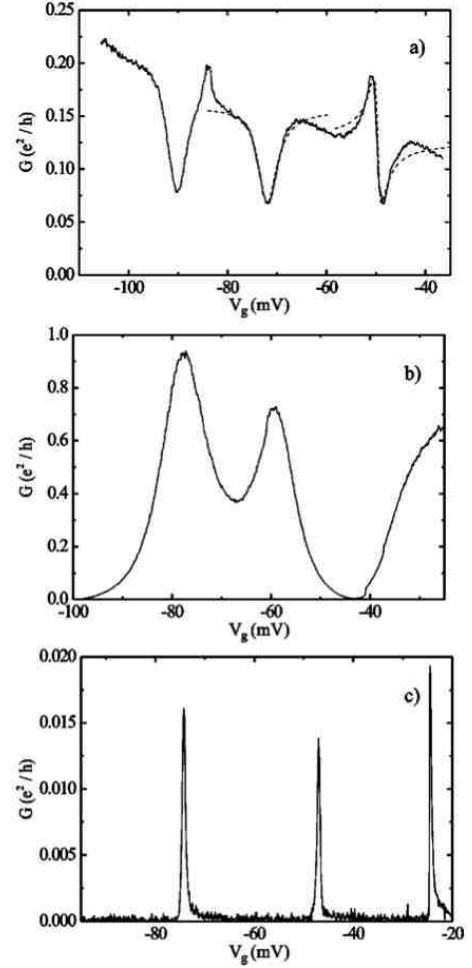}
\caption{\label{fig:kondo2}
Conductance versus gate voltage.
Comparison of conductance measurements in the (a) Fano regime, (b) intermediate
regime, and (c) Coulomb blockade regime. From (c) to (a) the lead-dot coupling 
increases. Fits to the Fano formula (\ref{eq:Fano}) are shown for the center and
right resonances in (a). The respective asymmetry parameters are $q=-0.03$ and $q=-0.99$. From \textcite{jgdggshmakhsdmum:PRB:00}.
}
\end{figure}

Let us consider a closed dot with linear size $L$, when the leads are decoupled.
If one neglects the contribution from Coulomb interaction, 
the spectrum of many body states in a quantum dot can be obtained
from the solution of the single particle problem.
The single particle level spacing $\Delta_{sp} = \pi \hbar^2/m^* L^2$ \cite{ya:rmp:00}.
The effective mass of an electron in GaAs is rather low: $m^* = 0.067 m_e$ \cite{ya:rmp:00}.
For $L=100nm$ one obtains $\Delta_{sp} \approx 2K$, while for $L=500nm$ the spacing is reduced
to $\Delta_{sp} \approx 90mK$.
Adding one electron to the closed dot therefore leads to an energy increase of
the order of $\Delta_{sp}$. Now take Coulomb interaction into account.
If the number of electrons in the dot is $N$, then the charging energy
of adding one additional electron is $E_c \sim N e^2/L$. Therefore,
for large values of $N$, and not too small values of $L$, $E_c \gg \Delta_{sp}$.
Note that typical dot sizes are of  the order $100nm - 1\mu m$. 
$N$ can strongly vary, with values $N\sim 10^2-10^3$.
Characteristic values of the charging energy are in the range $E_c \sim 100 - 400 K$ ($12 - 50 meV$).
Therefore, for all practical purposes, $E_c \gg \Delta_{sp}$.

The number of electrons in a quantum dot is defined by
minimizing the energy of the dot with respect to $N$.
This energy is given by \cite{ya:rmp:00}
\begin{equation}
E(N) = -NeV_g + N^2 e^2 / 2C\;,
\label{dotenergy}
\end{equation}
where $C$ is the total capacitance between the dot and its surroundings.
Apart from special values of the gate voltage,
there will be a given electron number $N$ with smallest energy, and 
changing the number of electrons will cost an amount about one charging energy $E_c$. 
For particular values of the gate voltage $V_g^{(n)}$ however degeneracies between $E(N)$
and $E(N+1)$ appear. 

Consider an experimental geometry shown in Fig.\ref{fig:kondo1}. 
If the coupling to the leads is weak enough and the temperature $kT  < E_c$ the Coulomb blockade regime sets in. 
As long as $V_g \neq V_g^{(n)}$ the charging energy prevents lead electrons
from entering the dot, and the conductance $G$ is practically zero. 
However, when $V_g = V_g^{(n)}$, the degeneracy between $N$ and $N+1$ electron states on the dot sets in.
Therefore, electrons can pass through the dot one by one, and the conductance takes
the universal value $G=2e^2/h$ (here the factor 2 accounts for spin degeneracy).
Note that the Coulomb interaction is treated in a mean-field type way, therefore
no phase coherence of dot electrons is required. 

Lowering the temperature further, the phase coherence
of the dot electrons becomes essential [see e.g. \cite{ymhdsdmhs:s:00,hakkassky:prl:04}]. Note that 
the typical electron mean free path can be of the order of $10\mu m$, one-two orders of magnitude
larger than the dot size.
It may also be possible to reduce {\sl decoherence} effects within some suitable range
by {\sl increasing} the coupling of the dot to the leads, which may lead to a shorter residence time
of electrons inside the dot and therefore to less scattering.
With the option of having several channels which electrons can use to pass through the dot,
phase coherence will lead to interference effects, and therefore to possible Fano resonances.

If a magnetic field is added, orbital and spin effects have to be considered as well.
The Zeeman energy $E_Z = g \mu_B H$ sets another temperature scale.
Depending on the Lande factor $g$, which can vary strongly from sample to sample,
the corresponding Zeeman energy $E_Z$ is of the order of $ 100-200 mK$
for $B=1T$. Allowing the electrons to traverse the dot along different pathes,
an Aharonov-Bohm phase shift $\phi$ occurs due to a nonzero magnetic flux penetrating the
area $S$ enclosed by them \cite{bladkailpal:PRB:80}: 
$\phi=\frac{e}{h}BS$. With $S=L^2$ we find for
$L=100nm$ that $\phi/2\pi = 0.38 B/T$, and for $L=1\mu m$  that $\phi/2\pi = 38 B/T$.

Therefore, for $L=100nm$ and $B=1T$ it follows $E_Z \ll \Delta_{sp}$. 
Then at low temperatures $kT < E_z$ of the order of $T \sim 50-100mK$ 
and at a magnetic field $B \sim 1T$
the Coulomb blockaded dot
has a well defined spin: either $|S_z|=1/2$, or $S_z=0$. 
Changing the gate voltage and reaching the next
degeneracy $E(N)= E(N+1)$, an electron with a well-defined spin is allowed to enter the dot -
either spin up or spin down. The allowed spin value alternates as one further tunes the gate voltage
to the next degeneracy. If the phase coherence of electrons is preserved during the scattering,
one may again expect interference phenomena - but this time, depending on the chosen value of $V_g$,
only electrons with spin up (respectively spin down) will interfere along different channels.  
Increasing the coupling to the leads may cause spin-selective destructive interference for a given
spin species, while the other spin species is freely passing through.
The orbital effect of the magnetic field leads to an additional phase shift of the order of $0.8 \pi$,
independent of applied gate voltages.

For $L=1\mu m$ the single particle spacing $\Delta_{sp} \approx 20mK$. Therefore at $B=1T$ it follows
$E_Z \gg \Delta_{sp}$. Then at temperatures $T \sim 50-100mK$ the Coulomb blockaded dot is magnetized,
but electrons which enter the dot can have any spin, preventing spin-selective destructive interference.
The orbital effect of the magnetic field is huge with a $2\pi$ phase shift every $25 mT$ upon changing the
magnetic field.

Before proceeding, let us briefly mention related studies of the Kondo effect in transport through quantum dots.
In the Coulomb blockade, the number of electrons on the dot is well defined, and either even or odd.
Assuming a ground state only, the total electronic spin is either 1/2 (odd number of electrons)
or zero (even number). In the absence of a magnetic field and for odd numbers of electrons,
the whole dot could be viewed as some magnetic impurity with spin 1/2, which scatters conduction
electrons passing from one lead to another. That calls for an analogy with the well known Kondo
effect which is observed in the low temperature properties of the conductivity of electrons
in metals with magnetic impurities \cite{hewson93}. 
The resistivity in metals usually drops with lowering the temperature, since the number of phonons, which 
are responsible for electron scattering due to electron-phonon interaction, decreases.
At around $30 K$ a minimum in the resistivity appears for some metals, and subsequently the resistivity increases again
with further lowering the temperature. This increase is due to scattering of electrons by magnetic impurities,
and originates from an exchange interaction of the conductance electron spin with the spin of the
magnetic impurity. The exchange interaction sets an energy and temperature scale (the Kondo temperature $T_K$), which is typically
of the order of $T_K \sim 100 mK - 1K$, similar to the Zeeman energy of an electronic spin 1/2 in a magnetic field of 1 Tesla.
For temperatures $T < T_K$ the impurity spin is screened by a cloud of renormalized conduction electrons.
The Kondo temperature depends sensitively on the coupling strength (hybridization) $\Gamma$ 
between the conduction electrons and the magnetic
impurities. For weak coupling $T_K$ is exponentially small in $-1/\Gamma$. 
This analogy stirred ideas to observe the Kondo effect in the conductance of electrons through quantum dots.
For that low temperatures have to be used, and the coupling of the leads to the dot has to be increased
(in order to increase $T_K$). An enormous amount of theoretical studies was performed \cite{ilapwblig:pr:02}.
Experimental results showed a deviation from the Coulomb blockade regime for strong lead-dot coupling (see below).
The relation to theoretical models based on Kondo mechanisms is still debated [see e.g. \cite{ilapwblig:pr:02,ymhdsdmhs:s:00}].  

\begin{figure}
\includegraphics[width=0.7\columnwidth]{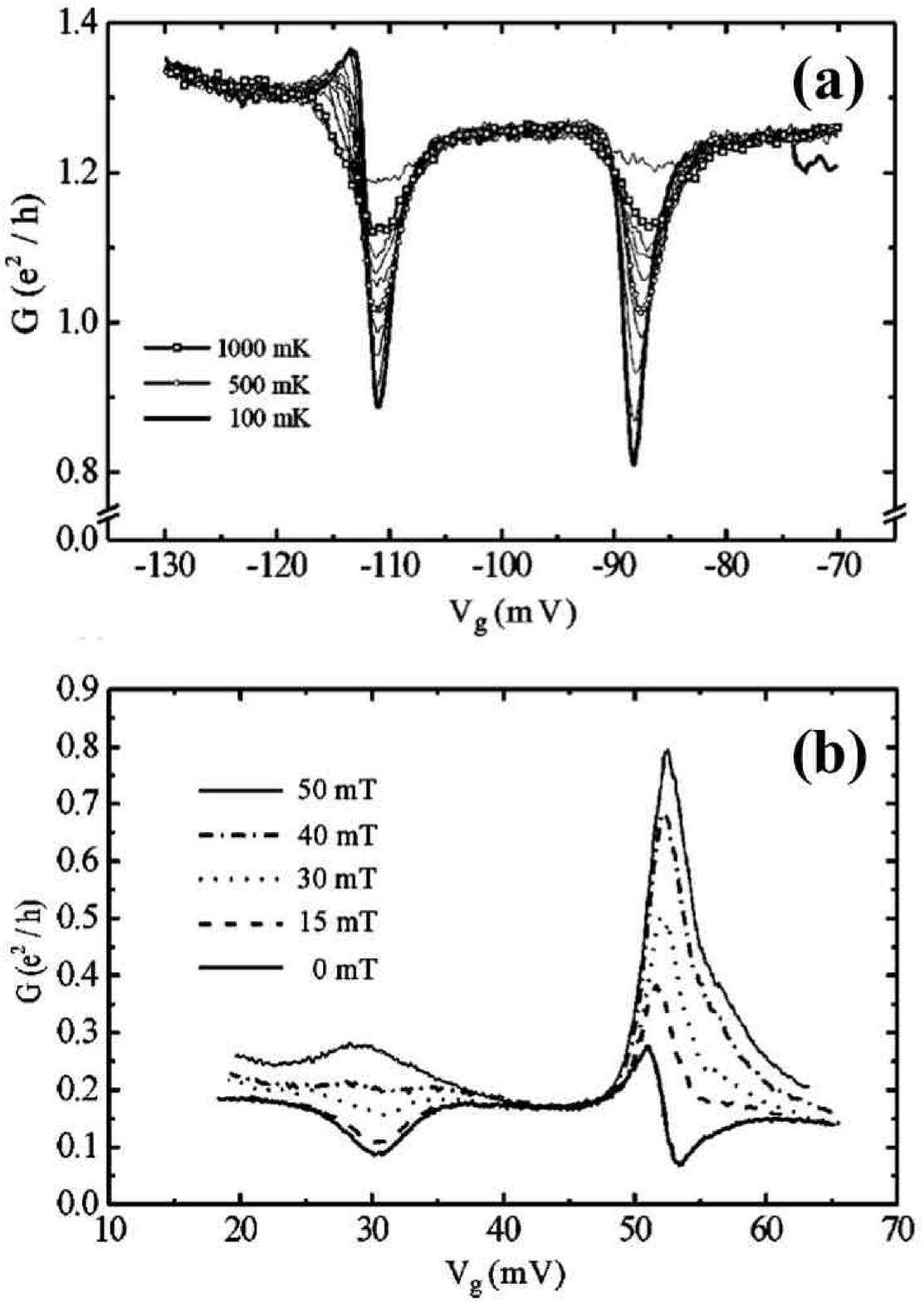}
\caption{\label{fig:kondo3}
Conductance versus gate voltage.
(a) Temperature dependence of the conductance for two Fano resonances.
(b) Conductance as a function of the gate voltage for various magnetic fields applied
perpendicular to the two-dimensional electron gas. Adapted from \textcite{jgdggshmakhsdmum:PRB:00}.
}
\end{figure}

\subsection{From Coulomb blockade to Fano resonances}

A number of experimental studies report on the observation of Coulomb blockade in
various quantum dot realizations on the basis of AlGaAs heterostructures 
\cite{jsjwkekvk:pb:98,csmothkp:s:98,gdshmdadmukma:n:98,gdgjkmashmdmu:prl:98,jgdggshmakhsdmum:PRB:00,dgjghsdmummak:mse:01,kkhaskyi:PRL:02}.
The charging energies are in the range $E_c \sim 100-300 K$.
Temperatures were as low as $30 mK$, applied magnetic fields up to 1T, and higher. 
Therefore, the Zeeman energy $E_Z$ is 2-3 orders of magnitude lower than the charging energy $E_c$.
The Coulomb blockade is usually observed in the case of weak coupling between the leads and the dot.
In Fig. \ref{fig:kondo2} the results of \textcite{jgdggshmakhsdmum:PRB:00} are shown,
which correspond to the setup in Fig. \ref{fig:kondo1}. 
For weak lead-dot coupling (c) the Coulomb blockade regime is nicely observed (temperatures 
are around $100 mK$, and the drain source voltage $V_{ds} \approx 5 \mu V \ll V_g$).
With increasing coupling the sharp peak structure is smeared out (b), which has been
discussed in relation to the Kondo effect. Even further increasing the coupling,
Fano resonances are observed in the strong coupling case (a). A fitting yields asymmetry parameters
$q=-0.03$ and $q=-0.99$ for the center and right resonances, respectively.
Note that also the peaks in (c) separating Coulomb blockades with different numbers of electrons
on the dot, are clearly asymmetric.
The same authors studied the temperature and weak magnetic field dependence of the Fano profiles in
the strong coupling regime for even larger absolute values of the gate voltage, shown
in Fig. \ref{fig:kondo3}.
\begin{figure}
\includegraphics[width=\columnwidth]{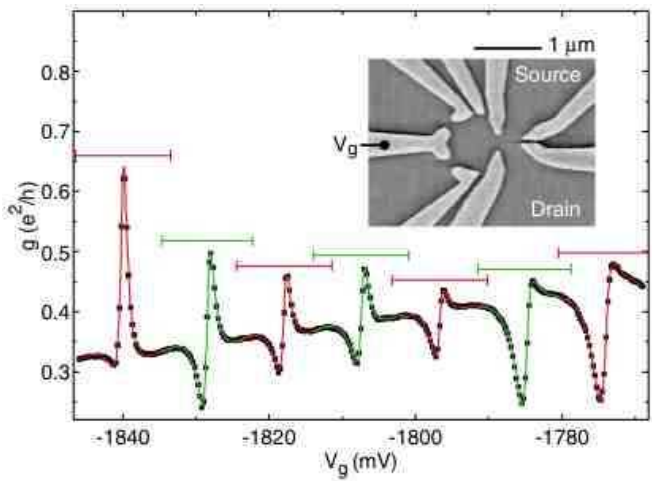}
\caption{\label{fig:kondo4}
Channel conductance data (squares) and fits (curves) vs gate voltage in the Fano regime.
Bars show fitting ranges. Inset: SEM image of a similar sample.
From \textcite{acjcmmmphacg:PRL:04}.
}
\end{figure}

The fitting of the resonances in Fig. \ref{fig:kondo3}(a) yields an almost linear decrease of the
linewidth $\Gamma$ with temperature, reaching values of  $2meV$ at $100 mK$. The depth of the
Fano resonance increases with decreasing temperature, making the Fano resonance sharper and deeper
at low temperatures.
The Fano resonances show very strong dependence on the value of a weak applied magnetic field
Fig. \ref{fig:kondo3}(b). Note that the largest applied fields are at $ 50 mT$, which corresponds
to a Zeeman energy of the order of $10mK$ or less. 

The origin of the observed Fano resonances is interferences of electrons along several channels (paths)
traversing the quantum dot. When the lead-dot coupling is weak, the background conductance is very small
[see Fig. \ref{fig:kondo2}(c)]. Still an asymmetric line shape is observed. The Fano resonance (dip) may
either be hard to be detected with that background, or simply be absent, since essentially only one
path is active. Another possibility is that the antiresonance is extremely narrow (weak coupling to
a dot state).
Since the Fano resonances are well observed at large lead-dot coupling, phase coherence of electrons
passing through the dot is therefore established, and is further increased with lowering the temperature.

The dramatic change of the resonance shape at weak magnetic fields is attributed to a suppression of the
coupling into the dot states \cite{jgdggshmakhsdmum:PRB:00}. 
That leads to an enhancement of the asymmetry parameter $q$, and respectively to
a shifting of the Fano resonance (dip) out of the window of available gate voltages.
The alternative explanation of loosing phase coherence of traversing electrons does not
account for the extremely low-field scale at which the change occurs \cite{jgdggshmakhsdmum:PRB:00}.
In a similar way one can exclude orbital Aharonov-Bohm effects, since the expected phase shifts
are of the order of $\phi \leq 0.12$.

\begin{figure}
\includegraphics[width=\columnwidth]{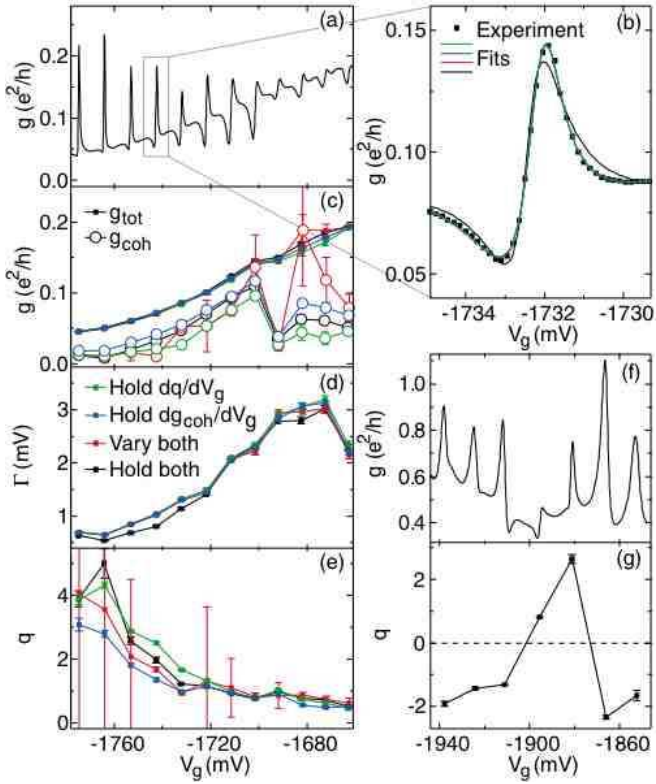}
\caption{\label{fig:kondo5}
Tunable Fano interferometer. (a) Experimental data with 12 Fano resonances.
(b) Fits of one resonance using different fitting parameters.
(c)-(e) $g_{tot}$, $g_{coh}$, $\Gamma$, and $q$ from (a).
(f) Data exhibiting reversals of $q$.
(g) Extracted $q$ values.
From \textcite{acjcmmmphacg:PRL:04}.
}
\end{figure}

\subsection{From Fano to Aharonov-Bohm interferometers}

In the above described experiments, the quantum dot design allowed essentially only
to control the lead-dot coupling. 
To further advance in the tunability of Fano resonances with quantum dots, 
interferometer devices have been manufactured. In addition to a small quantum dot,
which can be traversed by electrons, a second region (second dot, or additional
channel, or additional arm) is coupled in a controlled way. Therefore, the coupling
to a second channel can be tuned systematically. Of course there may be already 
several channels involved in the traversing of electrons through the primary dot.
\begin{figure}
\includegraphics[width=\columnwidth]{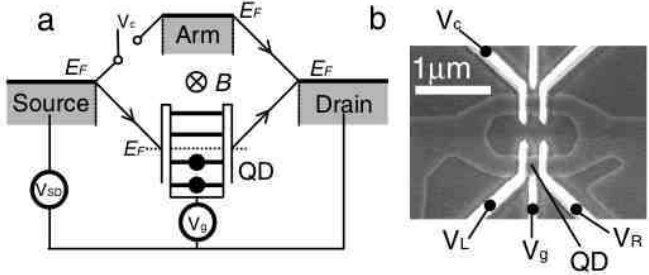}
\caption{\label{fig:kondo6}
An Aharonov-Bohm ring with an embedded QD in one of its arms. (a) Schematic representation of the experimental setup.
(b) Scanning electron micrograph of the fabricated device.
From \textcite{kkhaskyi:PRL:02}.}
\end{figure}

Impressive results have been obtained by \textcite{acjcmmmphacg:PRL:04} 
in designing a tunable Fano interferometer which consists of a quantum dot
and an additional tunnel-coupled channel (see Fig.\ref{fig:kondo4}).

A sequence of several Fano resonances was observed, and well fitted with the Fano formula (\ref{eq:Fano}).
Moreover, \textcite{acjcmmmphacg:PRL:04} performed careful fittings of various resonance shapes
as shown in Fig.\ref{fig:kondo5}. 
In panel (a) another set of resonances is observed. Upon variation of the gate voltage
the asymmetry of the resonance shape clearly changes, as also seen in panel (e).
In addition, also the line width $\Gamma$ is changing (panel (d)). In another gate voltage window
[panel (f)] these changes are even more drastic. Indeed, the fit yields a change of the sign of
$q$ with $V_g$ (panel (g)). Note, that according to (\ref{eq:Fano}) at $q=0$ a symmetric resonant reflection,
with no resonant transmission, is predicted. Indeed, around the value $V_g \approx -1900 mV$
the conductance in panel (f) shows practically a dip only.
\begin{figure}
\includegraphics[width=\columnwidth]{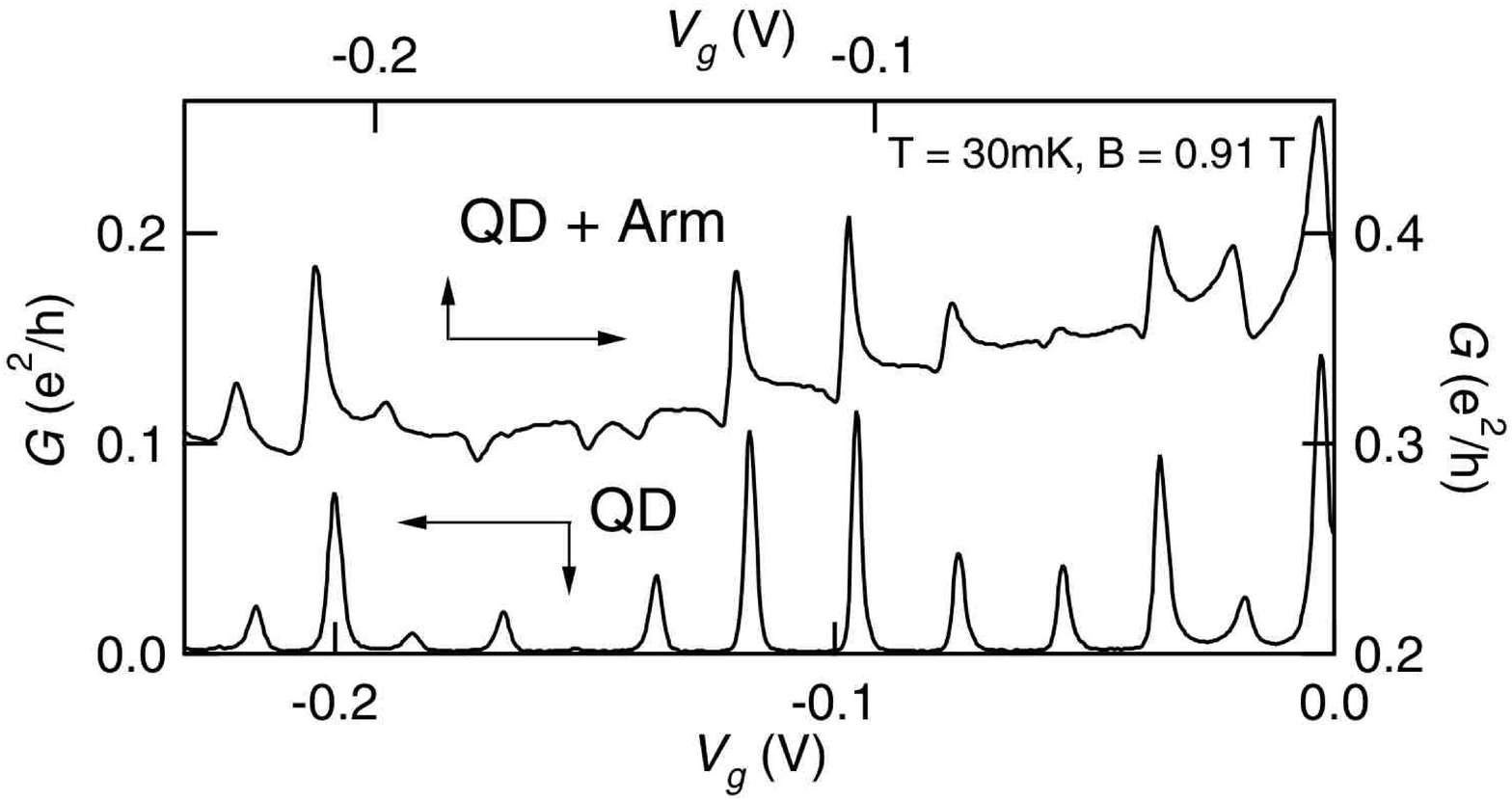}
\caption{\label{fig:kondo7}
Coulomb oscillation at $V_c=-0.12V$ with the arm pinched off, and asymmetric
Coulomb oscillation at $V_c = -0.086 V$ with the arm transmissible. 
Here $T=30mK$ and $B=0.91T$.
Adapted from \textcite{kkhaskyi:PRL:02}.
}
\end{figure}

Yet another twist was taken by \textcite{kkhaskyi:PRL:02} with a qualitatively similar geometry
but an additional magnetic field penetrating the interferometer area and turning it into
an Aharonov-Bohm (AB) device (see Fig.\ref{fig:kondo6}). The current through the quantum dot and the additional arm (channel)
can be controlled independently. Magnetic fields were around 1 Tesla. With the arm switched off,
a series of Coulomb blockade peaks is observed (see Fig.\ref{fig:kondo7}).

When making the arm transmittable, clear interference effects are observed through asymmetric Fano lineshapes
(see Fig.\ref{fig:kondo7}). 
In that system, the discrete level and the continuum are spatially separated, allowing to
control Fano interference via the magnetic field piercing the ring as shown in Fig.\ref{fig:kondo8}.
The line shape changes periodically with the AB period $\sim 3.8 mT$, which agrees
with the expected value using the ring dimension \cite{kkhaskyi:PRL:02}.
As magnetic field $B$ is swept, an asymmetric line shape with negative $q$ continuously changes to
a symmetric one and then to an asymmetric one with positive $q$. 
\textcite{kkhaskyi:PRL:02} argue, that due to the breaking of time reversal symmetry
in the presence of a magnetic field, the matrix elements defining $q$ are not real as usually
assumed, but complex, therefore leading to complex $q$ values. This confirms theoretical investigations 
for the noninteracting single particle AB interferometer case 
\cite{oewaayiyl:JLTP:02,aaoewbihyi:PRB:02,oewaayiylas:PRL:02,aaoewyi:PRL:03,ksnh:pe:05}.

\subsection{Correlations}

An enormous bulk of theoretical literature on various facets of the conductance properties
of quantum dots is available. We will discuss some of these results below.
Let us remind the reader about some characteristic scales. 
The Coulomb energy (charging energy) of quantum dots is of the order of $50 meV$ ($380 K$).
The Kondo temperature in a typical metal with magnetic impurities is of the order of 
$10 \mu eV$  ($100 mK$), comparable to the Zeeman energy of a spin 1/2 electron in a magnetic field of around 1 Tesla. 
Therefore, when operating at temperatures of the order of the Zeeman energy, the charge on a typical
quantum dot is extremely well fixed by the number of electrons.
The next question is whether a conductance electron, when penetrating the quantum dot, is able to efficiently
interact with an excess spin 1/2 particle for odd electron numbers, or whether it will
usually follow a path which avoids strong exchange interaction. 
These, partly open, issues make it sometimes to hard to judge the relevance of many interesting
theories.


The simplest model, which keeps the effect of Coulomb interactions and correlations, uses
exactly one level from the quantum dot, adds links to leads (left and right), and takes 
Coulomb interaction of spin up and spin down electrons into account - but only on the dot (see Fig.\ref{fig:kondo9_picasso}a).
The resulting Hamiltonian has the following form:

\begin{figure}
\includegraphics[width=0.8\columnwidth]{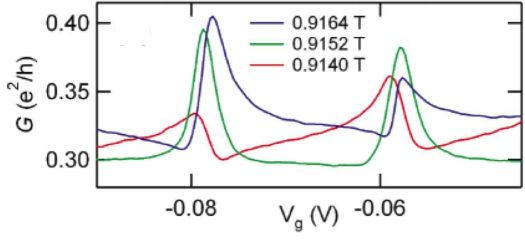}
\caption{\label{fig:kondo8}
Conductance of (a) two Fano peaks at $30 mK$ and selected magnetic fields,
(b) one Fano peak vs $V_g$ and $B$,
(c) Same as (b) but for larger windows of $V_g$ variations. The white line
represents the AB phase as a function of $V_g$.
From \textcite{kkhaskyi:PRL:02}.
}
\end{figure}

\begin{eqnarray}
H_s = H_D + H_W\;,\; H_D = \epsilon_d \sum_{\sigma} n_{\sigma} + U n_{\uparrow}n_{\downarrow}\;, 
\label{dot-anderson1} \\  
H_W = \sum_{k\sigma r} \epsilon_{kr} c^\dagger_{k\sigma r} c_{k\sigma r} + 
(V_r c^\dagger_{k\sigma r} d_{\sigma} + H.c.)\;. 
\label{dot-anderson2}
\end{eqnarray}
Here $n_{\sigma} = d^\dagger_{\sigma} d_{\sigma}$ measures the number of 
electrons on the quantum dot level, which interact with each other with strength $U$.
The left and right leads are denoted by $r=L(R)$.
The level energy $\epsilon_d$ is measured from the Fermi energy of the leads.
The lead states are chosen in the momentum representation. 
All fermionic creation and annihilation operators $c,c^\dagger,d,d^\dagger$ obey the standard
anticommutation relations.

\subsection{Interference}

\begin{figure}
\includegraphics[width=0.7\columnwidth]{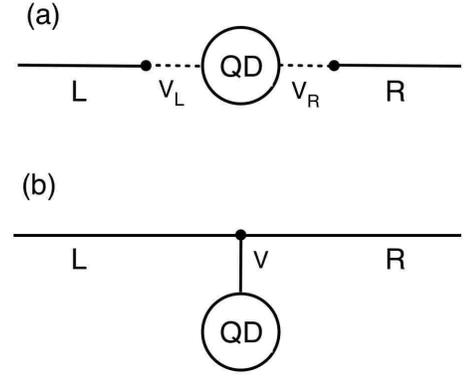}
\caption{\label{fig:kondo9_picasso}
Schematic representation of (a) a serial model of leads and a quantum dot (\ref{dot-anderson1})-(\ref{dot-anderson2}), and
(b) a T-shape model of leads and a side-coupled quantum dot (\ref{dot-sidedot}).
}
\end{figure}

There are many ways to incorporate interference and multiple pathes, in order to reach Fano resonances.
One of the simplest ones is a T-shaped scheme, which is a small change of the above model
by {\sl side-coupling} the quantum dot to the quantum wire (leads) (see Fig.\ref{fig:kondo9_picasso}b):
\begin{eqnarray}
H_T&=&-t \sum_{n,\sigma} 
(c^\dagger_{n,\sigma} c_{n-1,\sigma}+c^\dagger_{n,\sigma} c_{n+1,\sigma})
+\sum_{\sigma}\epsilon_{d,\sigma} n_{\sigma}\nonumber \\
&+&
\sum_\sigma(V d^{\dagger}_{\sigma}c_{0,\sigma}+V^\star c^{\dagger}_{0,\sigma}d_{\sigma} )
+ U n_{\uparrow}n_{\downarrow}\;.
\label{dot-sidedot}
\end{eqnarray}
The lead states are chosen in the coordinate representation.
Interference is possible because electrons can directly pass from the left to the right,
but can also visit the side dot and exit again. These two pathes are enough for destructive
interference.

Another possibility is to extend the serial dot scheme (\ref{dot-anderson1})-(\ref{dot-anderson2}) 
by adding a direct path (arm) for
electrons to transit from the left to the right leads \cite{whjkhs:PRL:01}:
\begin{equation}
H_{AB} = H_s + H_{a}\;,\; H_{a} = \sum_{kq\sigma}W{\rm e}^{i\phi } c^\dagger_{k\sigma R}
c_{q \sigma L} + H.c.\;.
\label{dot-hofstetter}
\end{equation}
The phase $\phi$ models a magnetic flux which is encompassed by the loop of the direct path
and the path via the quantum dot.

The Hamiltonians (\ref{dot-anderson1})-(\ref{dot-anderson2}),(\ref{dot-sidedot}) belong to the class of Anderson
Hamiltonians \cite{pwa:PR:61}.
Thus the thermodynamic properties of both models are similar, e.g. the average number of (spin up and spin down)
electrons on the dot $\left\langle n_{\sigma} \right\rangle$. 
However, the transport properties depend crucially on the chosen geometry~\cite{hgltxxqwzbsly:PRL:04}.
Note that changing the dot level $\epsilon_d$ is qualitatively similar to varying the
gate voltage of a quantum dot. The dot level is capable of accepting at most one spin up and one spin down electron.

\textcite{pbwamt:JPC:83}
obtained analytical results for $\left\langle n_{\sigma} \right\rangle$
assuming a linearized spectrum of lead electrons, which is not a crucial constraint,
as long as the lead electron bands are partially filled (ideally half-filling),
and as long as the temperature is much smaller than the distance from the Fermi energy to the band egdes.
In addition, there exist various numerical methods to compute $\left\langle n_{\sigma} \right\rangle$
approximately.

With standard scattering matrix approaches, as well as using the Friedel sum rule
~\cite{dcl:PR:66,hewson93},
the conductance of the serial dot scheme (\ref{dot-anderson1})-(\ref{dot-anderson2}) 
at zero temperature can be expressed in the following way~\cite{ligmer:JETP:88,tknpal:PRL:88}:
\begin{equation}
g_{\sigma} = \left( \frac{2V_LV_R}{V_L^2+V_R^2} \right) ^2 \sin^2 \pi \left\langle n_{\sigma} \right\rangle \;.
\label{dot-serialconductance}
\end{equation}

\begin{figure}
\includegraphics[width=0.8\columnwidth]{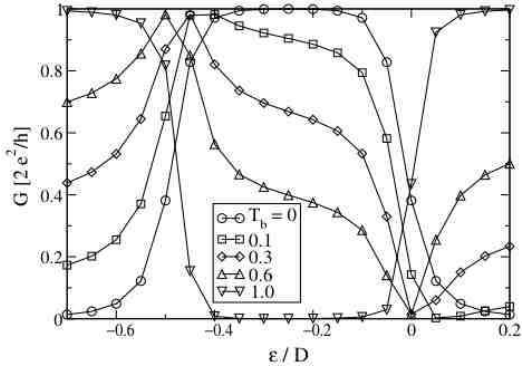}
\caption{\label{fig:kondo10}
Conductance as a function of $\epsilon_d$ for different values of background
transmission $T_b$. The AB phase $\phi=0$.
From \textcite{whjkhs:PRL:01}.
}
\end{figure}

\textcite{whjkhs:PRL:01} studied Fano resonances in transport through the the AB interferometer model (\ref{dot-hofstetter})
at zero temperature. The schematic view of the AB interferometer is similar to Fig.\ref{fig:kondo6}(a).
For zero AB phase $\phi=0$, and the direct path being switched off $W=0$, there are three
states of a Coulomb blockade to be expected upon variation of the gate voltage $\epsilon_d$:
the dot contains either zero, one, or two electrons, with sharp transitions between them.
We remind again, that the empty dot is almost not conducting (Coulomb energy too large),
and the dot filled with two electrons as well (Pauli principle).
When there is one electron on the dot, a second can enter while the first leaves. Despite of applying a magnetic field,
model (\ref{dot-hofstetter}) is invariant under spin reversal (because the bare dot levels are not
Zeeman splitted). 
This may be not easy to be achieved in an experiment.
Therefore, when there is one electron on the 
dot, it can have either spin up or spin down, and on average $\left\langle n_{\sigma} \right\rangle=1/2$
in that case. For $\epsilon_d > 0 $ (Fermi energy is placed at zero) the dot is empty, and the conductance is
zero. When $-U < \epsilon_d < 0$, one electron can enter the dot, but not two.
Then additional electrons can tunnel through, giving maximal conductance.
Finally, for $\epsilon_d < -U$, two electrons occupy the dot, and the conductance
is zero again. 
This broad region of almost perfect conductance is due to spin exchange processes on the quantum dot level,
and can be therefore related to the discussed  above Kondo effect. 
Indeed, in Fig.\ref{fig:kondo10}
this is observed for $T_b=0$ with $T_b=4x/(1+x)^2$ being the background transmission probability,
where $x=\pi^2 W^2 N_LN_R$, and $N_{L,R}$ is the density of states in the left (right) lead.
With increasing $T_b$ the curves change dramatically. Most importantly, a Fano resonance is appearing
in the studied energy window, qualitatively similar to experimental observations \cite{kkhaskyi:PRL:02}.
For the considered model the resonance location is shifting towards $-U/2$, and its width tends to
$-U$, as $T_b$ further increases Fig.\ref{fig:kondo10}, 
A variation of the AB phase $\phi$ in some intermediate $T_b$ regime yields the possibility to change
the sign of the asymmetry parameter $q$.

\subsection{Spin filters}

When a magnetic field is applied to the AB interferometer setup in Fig.\ref{fig:kondo6}(a),
it is reasonable to consider also its action on the quantum dot region itself, which
leads to a Zeeman splitting of the dot level.
This is incorporated in the side dot model (\ref{dot-sidedot}) with specifying
\begin{equation}
\epsilon_{d,\uparrow}=\epsilon_d + \Delta/2\;,\;
\epsilon_{d,\downarrow}=\epsilon_d - \Delta/2\;,
\label{dot-zeeman}
\end{equation}
where $\Delta$ is the Zeeman energy up to which the single particle level is splitted for spin down
and spin up electrons. It is easy to incorporate the AB phase shift as well, we will discuss it below.

For $U=0$ (\ref{dot-sidedot}) is reduced to the Fano-Anderson model 
(\ref{eq:fano_model4}), and
the transmission is computed within the one-particle picture for an electron moving
at the Fermi energy $\epsilon_F$:
\begin{eqnarray}
-\epsilon_F \phi_i&=&t(\phi_{n-1}+\phi_{n+1})+V^\star\varphi\delta_{n0},\\
-\epsilon_F \varphi&=&-\epsilon_{d,\sigma}\varphi+V\phi_0,
\label{dot-sf-singleparticle}
\end{eqnarray}
where 
$\phi_n$ refers to the amplitude 
of a single particle at site $n$ in the conducting channel
and $\varphi$ is the amplitude at the side dot. 
With the help of the Friedel sum rule~\cite{dcl:PR:66,hewson93} one arrives at
\cite{metkhsfaemmt:EPJB:04}
\begin{equation}
\label{dot-sf-cos}
g_\sigma=\cos^2\pi\left\langle n_{\sigma} \right\rangle\;.
\end{equation}
This relation has a geometric origin and actually 
holds for arbitrary $U$ (at zero temperatures).
For a nonzero magnetic field $\Delta \gg \Gamma$ 
the two Fano resonances for spin up and spin down 
electrons are energetically separated. Therefore,
the current through the channel is completely 
polarized at $\epsilon_F=\epsilon_{d,\uparrow}$ and
$\epsilon_F = \epsilon_{d,\downarrow}$. 
The AB phase can be easily included into the model (\ref{dot-sidedot}) similar to (\ref{dot-hofstetter}).
Remarkably it will not change the position of the resonances (cf. also (\ref{eq:fano_model7})),
since the position of the Fano resonance is entirely determined by the matching condition between the dot level(s)
and the Fermi energy.

The obtained spin filter will operate at temperatures
$kT \ll \Delta$. For a field of a few Tesla that implies temperatures less than $100mK$.
While that is possible in principle, two more problems appear. 
First, to control such a spin filter, one would have to control the gate voltage on the 
scale of $\mu eV$ (because the spin polarized Fano resonances are separated in the gate voltage
by the same amount of the Zeeman energy).
Second, as discussed
above, Coulomb interactions have to be taken into account. 

\begin{figure}
\includegraphics[width=\columnwidth]{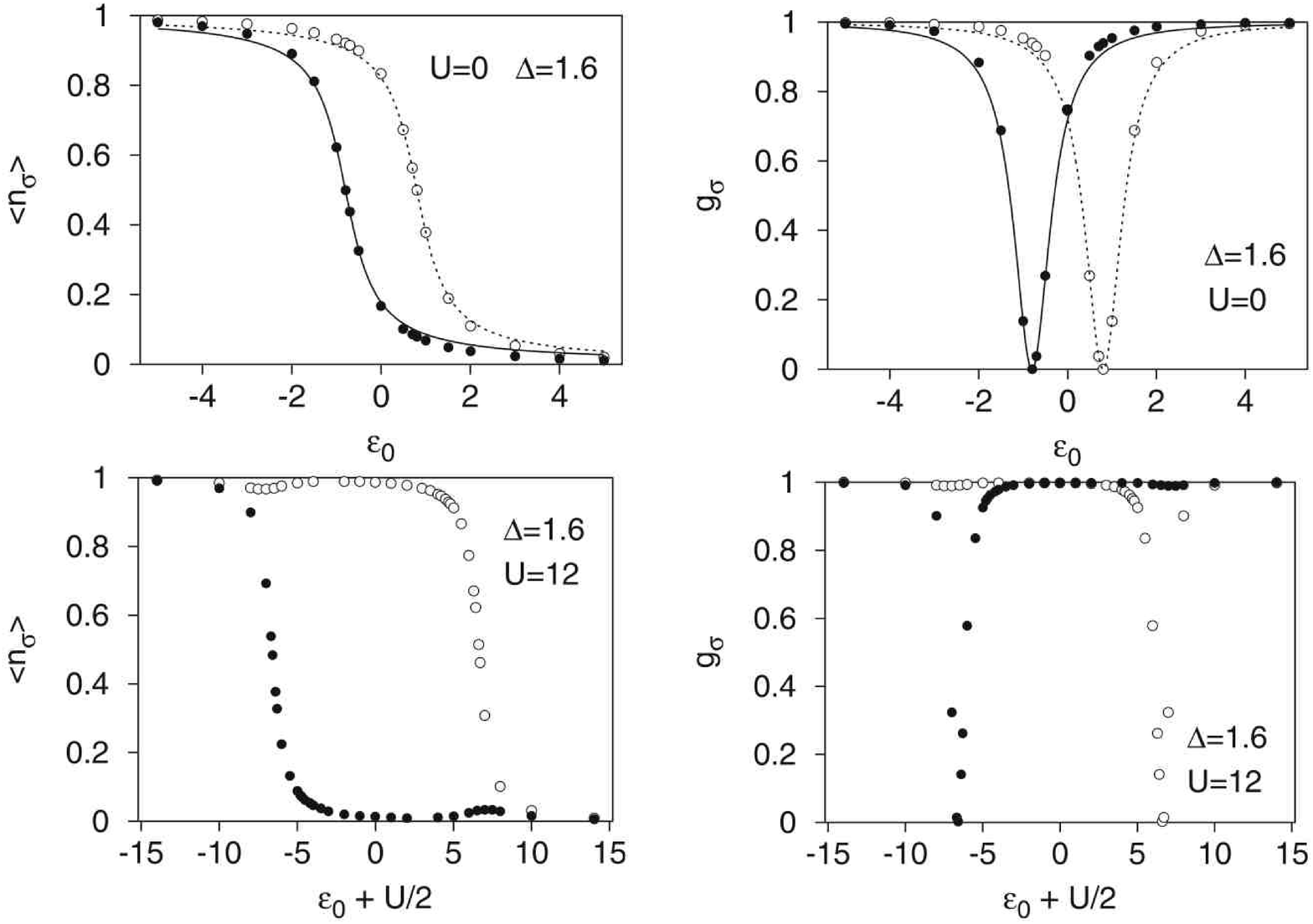}
\caption{\label{fig:kondo11}
Left plots: $\langle n_{\sigma} \rangle$
versus $\epsilon_d$ 
for a finite splitting $\Delta=1.6$, $U=0$ (top) and $U=12$ (bottom).
The black (white) dots represent the numerical results for the
spin up (spin down) occupation number. 
Right plots: $g_\sigma$
versus $\epsilon_d$ for 
finite splitting $\Delta=1.6$, $U=0$ (top) and $U=12$ (bottom). 
The conductance is computed numerically for spin up (black dots)
and spin down (white dots) electrons. 
The solid and the dashed 
lines on the top figures represent the exact results of Eq. (\ref{dot-sf-cos}).
Adapted from \textcite{metkhsfaemmt:EPJB:04}.
}
\end{figure}

For nonzero $U$ and $\Delta$, the results for the mean number of particles on the dot, and
for the spin-resolved conductance, have been obtained by \textcite{metkhsfaemmt:EPJB:04},
and are shown in Fig.\ref{fig:kondo11}.
The main outcome is, that the presence of a strong Coulomb interaction is shifting the 
two Fano resonances for spin up and spin down electrons further apart.
Therefore,
the current through the channel is completely 
polarized at $\epsilon_F=\epsilon_{d,\uparrow}+U$ and
$\epsilon_F = \epsilon_{d,\downarrow}$. For $U \gg \Delta$ the distance between the
two spin polarized Fano resonances is of the order of the charging energy (and not the Zeeman energy).
At the same time, the Kondo regime is completely suppressed.
For $\epsilon_F < \epsilon_{d,\downarrow}$, the dot level is empty, and electrons pass directly from the left
to the right lead (background transmission). For $\epsilon_F = \epsilon_{d,\downarrow}$
the dot is opening for spin down electrons. A Fano resonance appears, and its width is determined
solely by $\Gamma=2|V|^2/|v_F|$ where $v_F=d\epsilon/dq|\epsilon_F$ is the Fermi velocity. 
For $\epsilon_{d,\downarrow} < \epsilon_F < \epsilon_{d,\uparrow}+U$ the dot level is filled with one spin down
electron, and does not contribute to the conductance, leading to direct transmission from left to right leads.
For $\epsilon_F=\epsilon_{d,\uparrow}+U$ the dot is opening for spin up electrons.
A Fano resonance appears, with the same width as for the previous case.
Finally, for $\epsilon_F > \epsilon_{d,\uparrow}+U$, the dot is filled with two electrons and
does not contribute to the conductance, leading to direct transmission from left to right leads.

For typical quantum dots with $L \approx 100nm$ and $B\approx 1T$, the spin filter effect is expected to be active for temperatures
below $100 mK$, with a distance between the spin polarized Fano resonances of the order of $20-50meV$.
To observe it, one needs to monitor experimentally the spin-resolved flow of electrons
with a spatial resolution less than the dot dimension.

\subsection{Perspectives}

\textcite{sagybl:PRB:93} obtained resonant reflection and transmission within a generalized
description of a conducting channel (with several transverse modes) with a single impurity.

Extensions of the theoretical models in order to include many dot levels were performed by
\textcite{psatbrb:PRL:04} for very large ($0.1eV$) charging energies. Two dots with rather small
charging energies ($1meV$) were discussed by \textcite{ps:SSC:03}.
A series of authors considered the limit $U\rightarrow \infty$ \cite{brbps:PRL:01,kkscs:PRL:5619,
kksycjjkscs:PRB:113304}.
It remains to be clarified, whether such models can be used to discuss temperature effects on
transport properties through quantum dots.

\textcite{mlcb:PRB:06} extended the spin filter model by including spin-orbit interactions and
extending the side dot into a side ring with many levels.
Estimates of Kondo temperatures, and general temperature effects, have been discussed
by \textcite{aaalas:PRB:04}.
\textcite{amlaaa:PRL:08} included Rashba spin-orbit coupling into the consideration of AB interferometers
(see also \textcite{dsls:PRB:06,lsds:JPCS:07,fcjlllls:JAP:07,wgyzylfnktl:PLA:08}).
Spin inversion devices in a quasi-two-dimensional semiconductor wave guide under
sectionally constant magnetic fields and spin-orbit interactions were discussed by
\textcite{jlcpp:EPL:08}.

Experimental progress was reported by \textcite{nnjkllkpwahrb:PRL:07} through contacting
the tip of a low-temperature scanning tunneling microscope with individual cobalt atoms
adsorbed on Cu(100), where Fano resonances have been observed.

Single-molecule devices attracted attention recently. There one sandwiches various molecules
between gold electrodes and studies their conductance properties.
Impressive Fano resonances (with the background transmission dropping by several orders of magnitude)
were reported recently by \textcite{cmfvmgscjl:PRB:09}.
The additional influence of Andreev reflection at low temperatures, when the metallic contacts
turn superconducting, was studied by \textcite{akigcjl:PRB:09}.

Since Fano resonances rely on phase coherence of electrons traversing the structure along different paths,
several authors investigated the influence of phonons on decoherence in quantum dots 
\cite{hmplefftem:CP:02,
leffthmpem:EPL:06}.
\textcite{aacxwpwb:PRL:01} studied the possibility to extract phase decoherence properties
from measurements on the $q$-factor of the Fano resonance.


During the last decades, carbon nanotubes have been studied extensively because of 
their unconventional properties~\cite{carbon_book}. For applications to nanoscale electronic
devices, researchers have fabricated various forms of carbon nanotubes
to engineer their physical properties, including new morphologies such
as X- and T-shaped junctions~\cite{mthtfbjcpma:s:00}. These
developments offer interesting opportunities to study phase coherent
transport in novel geometries. 
Carbon nanotubes are excellent objects for observing phase coherence phenomena and Fano effects, 
and there are many theoretical studies and experimental signatures of the Fano effect 
in different types of carbon 
nanotubes~\cite{jkjrkjljwphmsnkkkkyjk:PRL:03,wyllhhzwpssx:PRL:03,zzdadrsrvc:el:04,bbcs:PRB:04,gksbltskji:PRB:05,zzvc:prb:06,fhhyxyjd:PRB:06}. 
%
%
%
In particular, Fano resonances are very pronounced in the 
transport properties of multiply connected carbon nanotubes
where a single tube is branched 
off into two smaller arms and then they merge into one.
Both $\pi$-bonding and $\pi^*$ ($\pi$ anti)-bonding electron
transport channels show resonant Fano tunneling through discrete
energy levels in the finite arms~\cite{gksbltskji:PRB:05}. 

There are many other systems where Fano resonances was observed and studied in details, e.g. for the resonant phonon transport between two crystalline media in the presence of a weakly bounded intermediate layer due to nonlocal interaction~\cite{yak:pss:97,yakafess:ltp:08,yak:pu:08}.

\section{Conclusions}
This Review offers a bird-eye view on the Fano resonances in various physical systems. All examples presented here share the 
same basic feature - coexistence of resonant and nonresonant paths for scattering wave to propagate. It results in constructive 
and destructive interference phenomena and asymmetric lineshapes, first quantitatively described by Ugo Fano. 
It turns out to be a very common situation in any complex system describing wave propagation, either on a classical
footing, or on a quantum mechanical one. 
This makes the Fano resonance a very generic phenomenon. 
The characteristic fingerprints of the Fano resonance are usually assumed to be related to
an asymmetric profile of a cross-section or transmission as a function of some relevant control parameters.
A detailed study of the problem shows, that symmetric profiles are allowed as well,
and therefore a Fano resonance is indicating its presence whenever a resonant suppression of forward scattering (transmission)
is observed. It is intimately related to the presence of a quasi-bound state resonantly interacting with a continuum
of scattering states. The pinning down of such a bound state may or may not be an obvious
undertaking, depending on the given physical setting. In particular, such quasi-bound states can be generated
by geometrical means, and in more complicated settings by many body interactions.
We focussed here on the study of Fano resonances in light propagation through artificial nanoscale optical devices,
and in charge transport through quantum dots. Several other potential applications were discussed as well,
touching such areas as superconductivity, Bose-Einstein condensates in optical lattices, among others.

Despite being interference in nature the Fano resonance we should pointed out here that it is quite different to other interference 
phenomena, such as, for instance,  double slit experiment or weak localization in disordered media~\cite{vfg:CLARENDON:2005}.
The latter two share the common feature of interference between two open channels (or broad continuums) represented by similar  
diffraction pattern of the slits in the first case, or  identical length of the two counter-propagating paths along a loop in the 
second. The phase of a scattering wave varies relatively slowly along a continuum. 
Therefore, for nearly identical continua the phase accumulation during propagation along two paths will be practically the same. 
The constructive/destructive interference takes place when the sum of these two phases become equal to zero or $\pi$, and, in 
general, are very well separated from each other. In the case of a Fano resonance the situation is quite different. Along the 
discrete level path the phase undergoes sharp variation (in comparison with the continuum) with a consequent change of its sign. It results 
in a very strong asymmetric profile where constructive and destructive interferences are located very close to each other.
Several detailed examples considered in this Review demonstrate that systems which support Fano resonance can be 
mapped onto the Fano-Anderson model (\ref{eq:fano_model1}). This model is very simple and  provides with a core understanding of 
the phenomenon. It can be considered as a guideline for explanation of the Fano resonance in a particular system.

\begin{acknowledgments}
We thank A. Dyugaev, Yu. Ovchinnikov, M. Titov and M. Rybin for very useful discussions and careful reading of parts of the manuscript.
The work has been supported by the Australian Research Council through the Discovery and Centre of Excellence projects.

\end{acknowledgments}

\bibliographystyle{apsrmplong}
\bibliography{fano_rpm_all_minimal}

\end{document}